%% file: JOURNAL.tex
\RequirePackage{fix-cm}
\documentclass[smallextended,natbib]{svjour3}       % onecolumn (second format)
\smartqed  % flush right qed marks, e.g. at end of proof

\renewcommand\cite{\citep} % paranthesis around citation inthe natlib style

\usepackage{tabularx}
\usepackage{multirow}
\usepackage{booktabs} % For formal tables
\usepackage{array}
\usepackage{xspace}
\usepackage{graphicx,wrapfig}

\usepackage{balance}
\usepackage{subcaption}
\usepackage{enumitem}

\usepackage{url}
\usepackage[group-separator={,}]{siunitx}
\usepackage[skins]{tcolorbox}

\usepackage{ifthen}
\newboolean{showcomments}
\setboolean{showcomments}{true}
\ifthenelse{\boolean{showcomments}}
{ \newcommand{\mynote}[2]{
    \fbox{\bfseries\sffamily\scriptsize#1}
    {\small$\blacktriangleright$\textsf{\emph{#2}}$\blacktriangleleft$}}}
{ \newcommand{\mynote}[2]{}}

\newcommand{\toolname}{{\em FaRM}\xspace}

\newcommand{\toolnameB}{{\em FaRM*}\xspace}

\newcommand{\toolnameC}{{\em PredKillable}\xspace}

\begin{document}

\title{Selecting Fault Revealing Mutants%\thanks{Grants or other notes
}


\author{Thierry Titcheu Chekam  \and  \\
				Mike Papadakis  \and 
        		Tegawend\'e Bissyand\'e  \and
        		Yves Le Traon  \and 
        		Koushik Sen  %etc.
}


\institute{Thierry Titcheu Chekam \at SnT Centre, University of Luxembourg  \\ \email{thierry.titcheu-chekam@uni.lu} 
				\and Mike Papadakis \at SnT Centre, University of Luxembourg   \\ \email{michail.papadakis@uni.lu}
				\and Tegawend\'e F. Bissyand\'e  \at SnT Centre, University of Luxembourg  \\\email{tegawende.bissyande@uni.lu}
				\and Yves Le Traon \at SnT Centre, University of Luxembourg    \\\email{yves.letraon@uni.lu}
				\and Koushik Sen \at University of California, Berkeley    \\\email{ksen@berkeley.edu} \\
%\and
  %            second address
 %  \\
}

\date{Received: date / Accepted: date}

\maketitle

\begin{abstract}
%Mutation testing is considered as one of the strongest test criteria to date. However, it is expensive due to large numbers of mutants involved. There have been many attempts to reduce the mutants' number. Unfortunately, recent studies have proven them unsuccessful.  
Mutant selection refers to the problem of choosing, among a large number of mutants, the (few) ones that should be used by the testers. In view of this, we investigate the problem of selecting the fault revealing mutants, i.e., the mutants that are most likely to be killable and lead to test cases that uncover unknown program faults. We formulate two variants of this problem: the fault revealing mutant selection and the fault revealing mutant prioritization. We argue and show that these problems can be tackled through a set of `static' program features and propose a machine learning approach, named \toolname, that learns to select and rank killable and fault revealing mutants.  %Therefore, %instead of selecting mutants that, when they are killed result in the killing of other mutants, we aim at selecting those mutants that are most likely to reveal real faults. 
Experimental results involving 1,692 real faults show the practical benefits of our approach in both examined problems. Our results show that \toolname achieves a good trade-off between application cost and effectiveness (measured in terms of faults revealed). We also show that \toolname outperforms all the existing mutant selection methods, i.e., the random mutant sampling, the selective mutation and defect prediction (mutating the code areas pointed by defect prediction). In particular, our results show that with respect to mutant selection, our approach reveals 23\% to 34\% more faults than any of the baseline methods, while, with respect to mutant prioritization, it achieves higher average percentage of revealed faults with a median difference between 4\% and 9\% (from the random mutant orderings). 
\end{abstract}

\input{samplebody-conf}

\balance
\bibliographystyle{spbasic}
\bibliography{sigproc}

\end{document}

%% file: samplebody-conf.tex
\input{introduction}

\input{Background}

\input{approach}

\input{RQs}

\input{Experiment}

\section{Results}
\label{sec:results}

\input{predictions}

\input{comparison}

%\section{Results on Large Programs}

%\section{Discussion}
%\label{sec:discussion}
\input{threats}
\input{RelatedWork}

\section{Conclusions}
\label{sec:conclusion}

The large number of mutants involved in mutation testing has long been identified as a barrier to the practical application of the method. 
 Unfortunately, the problem of mutant reduction remains open, despite significant efforts within the community. To tackle this issue, we introduce a new perspective of the problem: the fault revelation mutant selection. We claim that valuable mutants are the ones which are most likely to reveal real faults, and we conjecture that standard machine learning techniques can help in their selection. In view of this, we have demonstrated that  some simple `static' program features capture the important properties of the fault revealing mutants, resulting in uncovering significantly more faults (6\%-34\%) than randomly selected mutants. 

Our work forms a first step towards tackling the fault revelation mutant selection with the use of machine learning. As such, we expect that future research will extend and improve our results by building more sophisticated techniques, augmenting and optimizing the feature set, by using different and potentially better classifiers, and by targeting specific fault types. To support such attempts we make our subjects (programs \& tests), feature, kill and fault revelation matrices publicly available.

\section*{Acknowledgements}
Thierry Titcheu Chekam is supported by the AFR PhD Grant of the National Research Fund, Luxembourg. Mike Papadakis is funded by the CORE Grant  of National Research Fund, Luxembourg, (C17/IS/11686509/CODEMATES).

%% file: introduction.tex
\section{Introduction}

Mutation testing has been shown to be one of the most effective techniques with respect to fault revelation~\cite{ChekamPTH17}. Researchers typically use mutation as an assessment mechanism (measuring effectiveness) for their techniques~\cite{PapadakisSurvey}, but %Empirical research has also shown that mutation-based test suites are capable of subsuming almost every other test criterion. 
it can be used as every other test criterion. To this end, mutation can be used to assess the effectiveness of test suites or to guide test generation \cite{0020331, FraserZ12,Goran,Papadakis2018hi,ChekamPTH17}. %Researcher have devised many mutation-based techniques cap
 
Unfortunately, mutation testing is expensive. This is due to the large number of mutants that require analysis. An important cost parameter is the so-called {\em equivalent mutants}, which are mutants forming equivalent program versions \cite{PapadakisJHT15,0020331}. These need to be manually inspected by testers since their automatic identification is not always possible~\cite{Budd:1982:TNC:2697733.2697965}. 

While the problem of the equivalent mutants have been partly addressed by recent methods such as the Trivial Compiler Equivalence (TCE)~\cite{PapadakisJHT15}, the problem of the large number of mutants remains challenging. Yet, addressing this problem will in return contribute to addressing the equivalent mutant problem: any approach that is effective in reducing the large number of mutants, would indirectly reduce the equivalent mutant problem since less equivalent mutants will be available. 

Nevertheless, producing a large number of mutants is impractical. The mutants need to be analyzed, compiled, executed and killed by test cases. Perhaps, more importantly testers need to manually analyse them in order to design effective test cases. The scalability, or lack thereof, of mutation testing, with respect to the number of mutants to be processed, is thus a key factor that hinders its wide applicability and large adoption~\cite{PapadakisSurvey}. 
Consequently, if we can find a lightweight and reasonably effective way to diminish the number of mutants without sacrificing the power of the method, we would then manage to significantly improve the scalability of the method. Since the early days of mutation testing, researchers attempted to find such solutions by forming many mutant reduction strategies~\cite{PapadakisSurvey}, such as selective mutation~\cite{Offutt1993ICSE, WongM95b} and random mutant selection \cite{Acree1979}. 

Our goal is to form a mutant selection technique that identifies killable mutants that are fault revealing, prior to any mutant execution. %When mutants are used as test objectives, 
We consider as fault revealing, any mutant (i.e. test objective) that leads to test cases capable of revealing the faults in the program under test. We argue that such mutants are program specific and can be identified by a set of static program features. In this respect, we need features that are simultaneously generic, in order to be widely applicable, and powerful to approximate well the program and mutant semantics.

We advance in this research direction by proposing a machine learning-based approach, named \toolname, which learns on code and mutants' properties, such as mutant type and mutation location in program control-flow graphs, as well as code complexity and program control and data dependencies, to (statically) classify mutants as likely killable/equivalent and likely fault revealing. This approach is inspired by the prediction modelling line of research, which has recorded high performance by using machine learning to triage likely error-prone characteristics of code~\cite{MenziesGF07,KameiS16}. 

The use case scenario of \toolname is a standard testing scenario where mutants are used as test objectives, guiding test generation. To achieve this, we train on a set of faulty programs that have been tested with mutation testing, prior to any testing or test case design for the particular system under analysis. Then, we predict the killable and fault revealing mutants based on which we test the particular system under analysis. The training corpus can include previously developed projects (related to the targeted application domain) or previous releases of the tested software. In a sense, we train on system(s), say \textit{x}, and select mutants on the system under test, say \textit{y}, where $x \neq y$. 

Experimental results using 10-Fold cross validation on 1,692 + 45 faulty program versions show a high performance of 
\toolname in yielding an adequately selected set of mutants. In particular our method achieves statistically significantly better results than the random, selective mutation and defect prediction (mutating the areas predicted by defect prediction), mutant selection baselines by revealing 23\% to 34\% more faults than any of the baselines. Similarly, our mutant prioritization method achieves statistically significant higher Average Percentage of Faults Detected (APFD)~\cite{HenardPHJT16} values than the random prioritisation (4\% to 9\% higher in the median case). With respect to test execution, we show that our selection method requires less execution time (than random). 

We also demonstrate that our method is capable of selecting killable (non-equivalent) mutants. In particular, by building an equivalent classification method, using our features, we achieve an AUC value of 0.88 and 95\%, 35\% precision and Recall. These results indicate drastic reductions on the efforts required by the analysis of equivalent mutants. A combined approach, named \toolnameB, achieves similar to \toolname fault revelation,  but potentially at a lower cost (lower number of equivalent mutants), indicating the capabilities of our method. 

In summary, our paper makes the following contributions:

\begin{itemize}
\item It introduces the fault revealing mutant selection and fault revealing mutant prioritization problems.
\item It demonstrates that the killability and fault revealing utility of mutants can be captured by simple static source code metrics. 
\item It presents \toolname, a mutant selection technique that learns to select and rank mutants using standard machine learning techniques and source code metrics. 
\item It provides empirical evidence suggesting that \toolname outperforms the current state-of-the-art  mutant selection and mutant prioritization methods by revealing 23\% to 34\% more faults and achieving 4\% to 9\% higher  average percentage of revealed faults, respectively. 
\item It provides a publicly available dataset of feature metrics, kill and fault revelation matrices that can support reproducibility, replication and future research.
 
\end{itemize}

The paper is organized as follows. Section~\ref{sec:background} provides background information on mutation testing,  the mutant selection problem and defines the targeted problem(s). Section~\ref{sec:approach} overviews the proposed approach. Evaluation research questions are enumerated in Section~\ref{sec:rqs},  while experimental setup is described in Section~\ref{sec:setup} and experimental results are presented in Section~\ref{sec:results}. A detailed discussion on the applicability of our approach and the threats to validity are given in Section~\ref{Discussion}, and  related work is discussed in Section~\ref{RelatedWork}. Section~\ref{sec:conclusion} concludes this work.

%% file: Background.tex
\section{Context}
\label{sec:background}

\subsection{Mutation Testing}

Mutation testing \cite{HintTestDataSelection1978} is a test adequacy criterion that sets the revelation of artificial defects, called mutants, as the requirements of testing. As every test criterion, mutation assists the testing process by defining test requirement that should be fulfilled by the designed test cases, i.e., defining when to stop testing. %, and by defining  test suite thoroughness.

Software testing research has shown that designing tests that are capable of revealing mutant-faults results in strong test suites that in turn reveal real faults \cite{FranklWH97, LiPO09, ChekamPTH17, PapadakisSurvey} and are capable of subsuming or almost subsuming all other structural testing criteria \cite{OffuttPTZ96, FranklWH97, 0020331}. 

Mutants form artificially-generated defects that are introduced by making changes to the program syntax. The changes are introduced based on specific syntactic transformation rules, called {\em mutation operators}. The syntactically changed program versions form the mutant-faults and pose the requirement of  distinguishing their observable behaviour from that of the original program. A mutant is said to be {\em killed}, if its execution distinguishes it from the original program. In the opposite case it is said to be {\em alive}.  

Mutation quantifies test thoroughness, or test adequacy \cite{HintTestDataSelection1978, DeMilloO91, Frankl:1998:EST:291252.288298}, by measuring the number of mutants killed by the candidate test suites. In particular, given a set of mutants, the ratio of those that are killed by a test suite is called mutation score. Although all mutants differ syntactically from the original program, they do not always differ semantically. This means that there are some mutants that are semantically equivalent to the original program, while being syntactically different \cite{doi:10.1002/stvr.4370040303, PapadakisJHT15}. These mutants are called equivalent mutants \cite{HintTestDataSelection1978, doi:10.1002/stvr.4370040303} and have to be removed from the test requirement set. 

Mutation score denotes the degree of achievement of the mutation testing requirements \cite{0020331}. Intuitively, the score measures the confidence on the test suites (in the sense that mutation score reflects the fault revelation ability). Unfortunately, previous research has shown that the relation between killed mutants and fault revelation is not linear~\cite{FranklWH97, ChekamPTH17} as fault revelation improves significantly only when test suites reach high mutation score levels.

\subsection{Problem Definition}

Our goal is to select among the many mutants the (few) ones that are fault revealing, i.e., mutants that are most likely to lead to test cases that reveal existing, but unknown, faults. %To be practical, somutant selection needs to be performed statically and prior to any test generation and execution.
This is a challenging goal since only 2\%  (according to our data) of the killable mutants are fault revealing. % fact indicating that the majority of the mutants are ``irrelevant'' to the sought faults and should not be considered by testers. 

\begin{figure}[!t]
	\centering
	\vspace{-.6em}
	\includegraphics[width=0.9\linewidth]{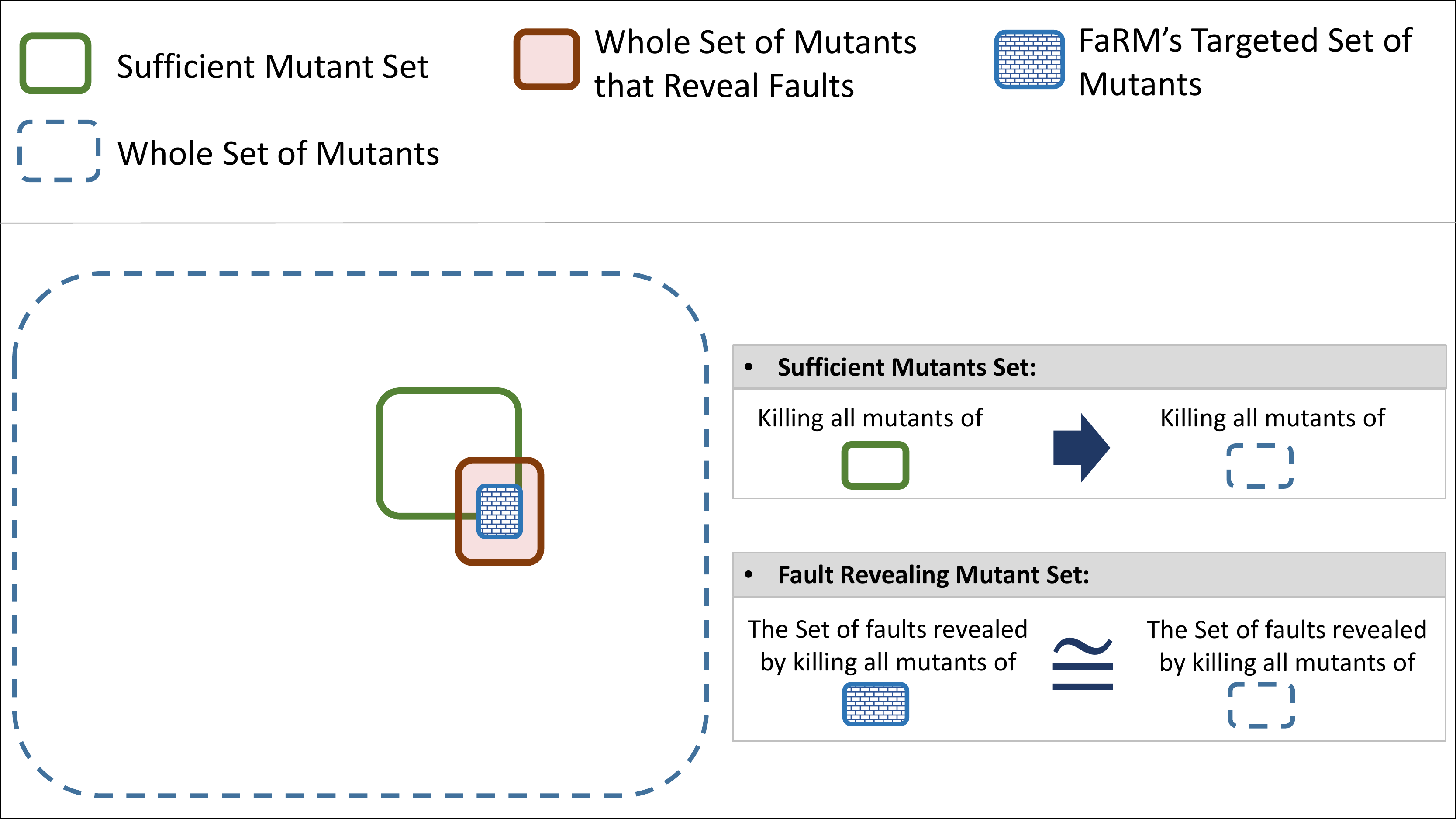}
	\caption{Fault revealing mutant selection. Contrast between sufficient mutant set selection and fault revealing mutant selection. Sufficient mutant set selection aims at selecting a minimal subset of mutants that is killed by tests that also kill the whole set of mutants. Fault revealing mutant selection aims at 	
	selecting a minimal subset of mutants that is killed by tests that reveal the same underlying faults as the tests that kill the whole set of mutants.}
%	\vspace{-1.0em}
		\label{fig:problem}
\end{figure}

The fault revealing mutant selection goal is different from that of the ``traditional'' mutant reduction techniques, which is to reduce the number of mutants \cite{OffuttLRUZ96, WongM95, Cutigi2018ICSTW, PapadakisSurvey}.  Mutant reduction strategies focus on selecting a small set of mutants that is representative of the larger set. This means, that every test suite that kills the mutants of the smaller set, also kills the mutants of the large set. Figure~\ref{fig:problem} illustrates our goal and contrasts it with the ``traditional'' mutant reduction problem.

Previous research \cite{Papadakis2018hi, Papadakis2018Mutation} has shown that the majority of the mutants, even in the best case, are ``irrelevant'' to the sought faults. This means that testers need to analyse a large number of mutants before they can find the actually useful ones (the fault revealing ones), wasting time and effort. According to our data, 17\% of the minimal mutants (ideal mutant reduction) is fault revealing, i.e., subsuming mutants (a set of mutants with minimal overlap that are sufficient for preserving test effectiveness~\cite{JiaH09, KintisPM10, AmmannDO14}), also indicating that  the majority of the mutants, even in the best case, are ``irrelevant'' to the sought faults. We therefore claim that mutation testing should be performed only with the mutants that are most likely to be fault revealing. This will make possible the best effort application of the method.

\begin{figure*}[!t]
	\centering
	\vspace{-1.25em}
	\includegraphics[width=0.9\linewidth]{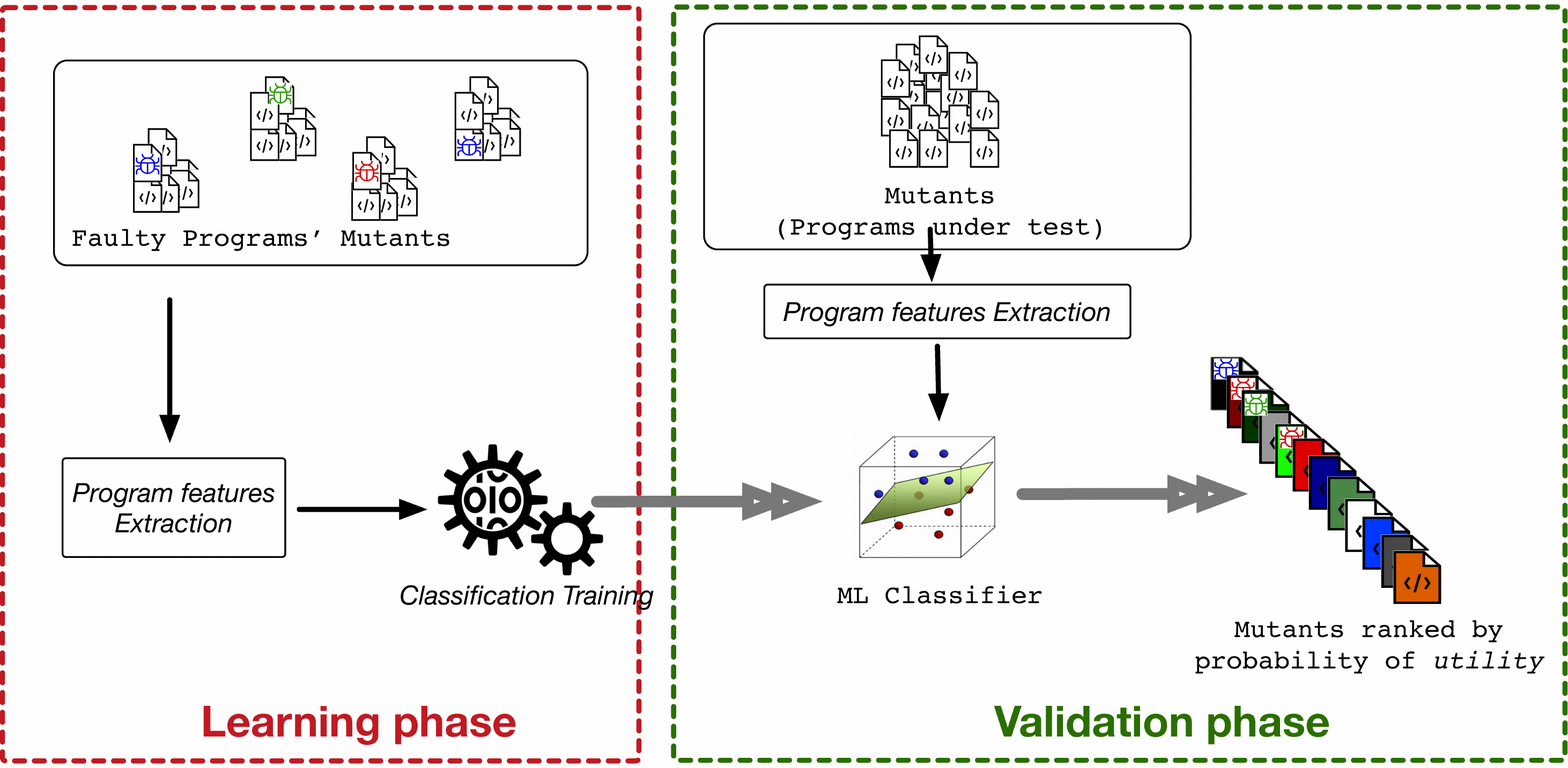}
	\caption{Overview of the \toolname{} approach. Initially, \toolname{} applies supervised learning on the mutants generated from a corpus of faulty program versions, and builds a prediction model that learns the fault revealing mutant characteristics. This model is then used to predict the mutants that should be used to test other program versions. This means that at the time of testing and prior to any mutant execution, testers can use and focus only on the most important mutants. 
}
	\vspace{-1.5em}
		\label{fig:approach}
\end{figure*}

Formally, we consider two aspects of this selection problem: the mutant selection one and the mutant pioritization one. 

The \textbf{fault revealing mutant selection problem} is defined as: 

\noindent
\textbf{Given:} A set of mutants $M$ for program $P$. %and a function, F, from M to real numbers.
 
\noindent
\textbf{Problem:} Subset selection. Select a subset of mutants, $S \in M$, such that $F(S) = F(M)$ and $(\forall m \in S)$, $(F(S - \{m\}) \ne F(M))$.

\noindent
$S$ represents a subset of $M$; $F(X)$ represents the number of faults in $P$ that are revealed by the test suites that kill all the mutants of the set $X$. In practice, the challenge is to approximate well $S$, statically and prior to any test execution, by finding a relatively good trade-off between the number of selected mutants (to minimise) and the number of faults revealed by their killing (to maximize). 

Similarly, the \textbf{fault revealing mutant prioritization problem} is defined as: 

\noindent
\textbf{Given:} A set of mutants, $M$ and the set of permutations of $M$, $PM$ for program P.

\noindent
\textbf{Problem:} Find $Pm' \in PM$ such that $(\forall Pm'') (Pm'' \in PM)$ $(Pm'' \ne Pm')$ $[f(Pm') \geq f(Pm''))]$

\noindent
PM represents the set of all possible mutant orderings of $M$, and $f(X)$ represents the average percentage of faults revealed by the test cases that kill the selected mutants in the given order X (measures the area under the curve representing the faults revealed by the killing of each one of the mutants in the order). The challenge is to statically and prior to any test execution, rank the mutants so that the fault revealing potential is maximized when killing any (arbitrary) number of them. The idea is that fault revelation is maximized whenever the tester decides to stop killing mutants. 

\subsection{Mutant Selection}

In the literature many mutant selection methods have been proposed \cite{PapadakisSurvey, Cutigi2018ICSTW} by restricting the considered mutants according to their types, i.e., applying one or more mutant operators. Empirical studies \cite{KurtzAODKG16, DengOL13}, have shown that the most successful strategies are the statement deletion \cite{DengOL13} and the E-Selective mutant set \cite{OffuttLRUZ96, Offutt1993ICSE}. We therefore compare our approach with these methods. We also consider the random mutant selection \cite{Acree1979} since there is evidence demonstrating that it is particularly effective \cite{ZhangHHXM10,PapadakisM10a}. 

\subsubsection{Random Mutant Selection}

Random mutant sampling \cite{Acree1979} forms the simplest mutant selection technique, which can be considered as a natural baseline method. Interestingly, previous studies found it particularly effective~\cite{ZhangHHXM10,PapadakisM10a}. Therefore, we compare with it.

We use two random selection techniques, named as SpreadRandom and DummyRandom. SpreadRandom iteratively goes through all program statements (in random order) and selects mutants (one mutant among the mutants of each statement), while DummyRandom selects them from the set of all possible mutants. The first approach is expected to select mutants residing on most of the program statements, while the second one is expected to make a uniform selection. 

\subsubsection{Statement Deletion Mutant Selection}

Mutant selection based on statement deletion is a simple approach that, as the name suggests, deletes every program statement (once at a time). To avoid introducing compilation issues (mutants that do not compile) and introduce relatively strong mutants, the statement deletion is usually applied on parts of a statement (deleting parts of expressions, i.e., the expression $a + b$ becomes $a$ or $b$ ). Empirical studies have shown that statement deletion mutant selection is powerful (achieves a very good trade-off between the number of selected mutants and test effectiveness) and has the advantage of introducing few equivalent mutants \cite{DengOL13}. 

\subsubsection{E-Selective Mutant Selection}
E-Selective refers to the 5 operator mutant set introduced by Offutt \etal~\cite{OffuttLRUZ96, Offutt1993ICSE}. This set is the most popular operator set \cite{PapadakisSurvey} that is included in most of the modern mutation testing tools. This set includes the mutants related to relational, logical (including conditional), arithmetic, unary and absolute mutations.  According to the study of  Offutt \etal~\cite{OffuttLRUZ96} this set has the same strengths as a much larger comprehensive set of operators. Although, there is empirical evidence demonstrating that the E-Selective set has lower strengths than a more comprehensive set of operators ~\cite{KurtzAODKG16}, it still provides a very good trade-off beetween selected mutants and strengths \cite{KurtzAODKG16}. 

\subsection{Mutant Prioritization}

Mutant prioritization has received little or even no attention in literature (refer to the Related Work Section \ref{RelatedWork} for details). 
Given the absence of other methods, we compare our approach with the random baselines. We also consider alternative schemes, such as Defect Prediction prioritization.

\subsubsection{Random Mutant Prioritization}

Random mutant prioritization forms a natural baseline for our approach. Comparing with random orderings is a common practice in test case prioritization studies~\cite{RothermelUCH01, HenardPHJT16} and shows the ability of the prioritization method to systematically order the sought elements. Similarly to mutants selection, we applied two random ordering techniques, the SpreadRandom and DummyRandom. SpreadRandom orders mutants by iteratively going through all program statements (in random order) and selects one mutant among the mutants of each statement (statement-based orders), while DummyRandom orders them from the mutant set (uniform orders). 

\subsubsection{Defect Prediction Mutant Prioritization}

Naturally, one of the main attributes determining the utility of the mutants is their location. Thus, instead of selecting mutants based on other properties, one could select them based on their location. To this end, we form a prioritization method that predicts and orders the error-prone code locations, i.e., code parts that are most likely to be faulty. Then, we mutate the predicted code areas and form a baseline method. Such an approach is in sense equivalent to the application of mutation testing on the results of defect prediction. Moreover, such a comparison demonstrates that mutants depend on the attributes (features) we train on not solely on their location.

%% file: approach.tex
\section{Approach}
\label{sec:approach}

 Our objective is to select mutants that lead to effective test cases. In view of this, %instead of selecting mutants that can collaterally kill other mutants~\cite{OffuttLRUZ96}, 
 we aim at selecting and prioritizing mutants so that we reveal most of the faults by analysing the smallest possible number of mutants. 

We conjecture that mutant selection strategies should account for the properties that make them killable and fault revealing. Defect prediction studies~\cite{MenziesGF07,KameiS16}, investigated properties related to error-prone code locations, but not related to mutants. Mutation testing is a behaviour oriented criterion and requires mutants introducing small and useful semantic deviations. Therefore, we propose building a model, which captures the essential properties that make mutants valuable (in terms of their utility to reveal faults).

Figure~\ref{fig:approach} depicts the \toolname{} approach, which learns to rank mutants according to their fault revealing potential (likelihood to reveal (unknown) faults).  
Initially, \toolname{} applies supervised learning on the mutants generated from a corpus of faulty program versions, and builds a prediction model.  
This model is then used to predict the mutants that should be used to test the particular instance of the program under test. This means that at the time of testing and prior to any mutant execution, testers can 
use and focus only on the most important mutants. 

%\begin{figure*}[!t]
%	\centering
%	\vspace{-1.0em}
%	\includegraphics[width=0.9\linewidth]{figures/approach.pdf}
%	\caption{Overview of the \toolname{} approach}
%		\label{fig:approach}
%\end{figure*}

\noindent 
{\bf \em ML-based measurement of mutant utility.}
The selection process in \toolname{} is based on training a predictor for assessing the probability of a mutant to reveal faults. 
To that end, we explore the capability of several features, which are designed to reflect specific code properties which may discriminate 
a useful mutant from another. Let us consider a mutant $M$ associated to a code statement $S_M$ on which the mutation was applied.
This mutant can be characterized from various perspectives with respect to (1) the complexity of the relevant mutated statement, (2) the position
of the mutated code in the control-flow graph, (3) dependencies with other mutants, (4) the nature of the code block where  $S_M$ is located.

\noindent 
{\bf \em ML features for characterizing mutants.}
Recently, Petrovic and Ivankovic \cite{Goran} used  arid nodes and the AST graph to infer mutant utility. Therefore, in addition to the mutant types, typically considered by selective mutation approaches \cite{OffuttLRUZ96,NaminAM08,PapadakisSurvey}, we also considered the information encoded on the program AST. We include three such features, the Data type at the mutant location, the parent AST node (of the mutant expression) and the child AST node (of the mutant expression), in our machine learning classification scheme. 

Let $B_M$ be the control-flow graph (CFG) basic block associated to a mutated statement $S_M$ containing the mutated expression $E_M$. Table~\ref{tab:features} provides the list of all 28 features that we extract from each mutant. The features named \emph{TypeAstParent}, \emph{TypeMutant}, \emph{TypeStmtBB}, \emph{AstParentMutantType}, \emph{OutDataDepMutantType}, \emph{InDataDepMutantType}, \emph{OutCtrlDepMutantType}, \emph{InCtrlDepMutantType}, \emph{DataTypesOfOperands} and \emph{DataTypesOfValue} are categorical. We represented them using one hot encoding. Besides the categorical features listed above, all other features are numerical. The values of numerical features are normalized between 0 and 1 using \emph{feature scaling}, more precisely \emph{min-max normalization/scaling}.

\begin{table}[!t]
%\vspace{-1.0em}
\scriptsize
	\begin{tabular}{p{0.34\linewidth}|p{0.59\linewidth}}\hline%\hline
  	\bf Complexity & Complexity of statement $S_M$ approximated by the total number of mutants on $S_M$ \\\hline
  	\bf CfgDepth &  Depth of $B_M$ according to CFG\\\hline
  	\bf CfgPredNum & Number of predecessor basic blocks, according to CFG, of $B_M$\\\hline
  	\bf CfgSuccNum& Number of successors basic blocks, according to CFG, of $B_M$ \\\hline
	\bf AstNumParents& Number of AST parents of $E_M$ \\\hline
	\bf NumOutDataDeps& Number of mutants on expressions data-dependents on $E_M$\\\hline
	\bf NumInDataDeps& Number of mutants on expressions on which $E_M$ is data-dependent\\\hline
	\bf NumOutCtrlDeps& Number of mutants on statements control-dependents on $E_M$ \\\hline
	\bf NumInCtrlDeps& Number of mutants on expressions on which $E_M$ is control-dependent \\\hline
  	\bf  NumTieDeps& Number of mutants on $E_M$ \\\hline
\bf  AstParentsNumOutDataDeps& Number of mutants on expressions data-dependent on$E_M$'s AST
parent statement\\\hline
\bf  AstParentsNumInDataDeps& Number of mutants on expressions on which $E_M$'s AST parent expression is data-dependent\\\hline
\bf  AstParentsNumOutCtrlDeps& Number of mutants on statements control-dependent on $E_M$'s AST parent expression\\\hline
\bf  AstParentsNumInCtrlDeps& Number of mutants on expressions on which $E_M$'s AST parent expression is control-dependent\\\hline
\bf  AstParentsNumTieDeps& Number of mutants on $E_M$'s AST parent expression\\\hline
\bf TypeAstParent& Expression type of AST parent expressions of $E_M$ \\\hline
 \bf  TypeMutant& Mutant type of M as matched code pattern and replacement. Ex: $a + b \rightarrow a - b$ \\\hline 
\bf  TypeStmtBB& CFG basic block type of $B_M$. Ex: $if-then,\; if-else$ \\\hline 
\bf  AstParentMutantType& Mutant types of M's AST parents\\\hline
%\bf AstParentStmtBBType& CFG basic block type of M's AST parent's basic block\\\hline
 
\bf  OutDataDepMutantType& Mutant types of mutants on expressions data-dependents on $E_M$ \\\hline
%\bf  OutDataDepStmtBBType& CFG basic block types of basic blocks containing statements data- dependent on $S_M$ \\\hline
\bf  InDataDepMutantType&Mutant types of mutants on expressions on which $E_M$ is data-dependent\\\hline
%\bf  InDataDepStmtBBType& CFG basic block types of basic blocks containing statements on which
%$S_M$ is data-dependent\\\hline
\bf  OutCtrlDepMutantType& Mutant types of mutants on statements control-dependents on $E_M$ \\\hline
%\bf  OutCtrlDepStmtBBType& CFG basic block types of basic blocks containing statements control- dependent on $S_M$ \\\hline
\bf  InCtrlDepMutantType&Mutant types of mutants on expressions on which $E_M$ is control-dependent\\\hline
%\bf  InCtrlDepStmtBBType& CFG basic block types of basic blocks containing statements on which
%$S_M$ is control-dependent\\\hline %\hline
\bf AstChildHasIdentifier& AST child of expression $E_M$ has an identifier\\\hline
\bf AstChildHasLiteral& AST child of expression $E_M$ has a literal\\\hline
\bf AstChildHasOperator& AST child of expression $E_M$ has an operator\\\hline
\bf DataTypesOfOperands& Data types of operands of $E_M$\\\hline
\bf DataTypeOfValue& Data type of the returned value of $E_M$\\\hline
 	\end{tabular}
\caption{Description of the static code features}
\label{tab:features}
\vspace{-2.2em}
\end{table}

A demonstrating example on how mutant features are computed is given in the following subsection (section \ref{Example}). After extracting feature values, we feed them to a machine learning classification algorithm along with the killable and fault revealing information for each mutant 
for a set of faults. The training process then produces two classifiers (one for the equivalent and one for the fault revealing mutants) which, given the feature values of a given mutant, they are capable of computing the utility probabilities for this mutant, i.e., its probability to be killable and its probability to be fault revealing. 

By using these two classifiers we form three approaches, two of them using each one of the classifiers alone and one of them by combining them. The first two, named \toolname and \toolnameC, respectively classify mutants according to their probability to be fault revealing and killable. The third one, named \toolnameB, divides the mutant set in two subsets, likely killable and likely equivalent, ranks them according to their fault revealing probability and concatenates them by putting the likely killable subset first.  

We implement a prioritization scheme by considering the ranking of all mutants 
in accordance to the values of the developed probability measure.  This forms our mutant prioritization approaches. Our mutant selection strategy sets a threshold probability value (e.g., $0.5$) or a cut-off point according to the number of the top ranked mutants  to keep only mutants with higher utility probability scores in the selected set.  This forms our mutant selection approach. For the combined approach (\toolnameB) we divide  the mutant set in the killable and equivalent subsets by using a cut-off point of $0.5$.  

\subsection{Implementation}
We implemented \toolname as a collection of tools in C++.
We leverage stochastic gradient boosting~\cite{friedman2002stochastic} (decision trees) to perform supervised learning. % on program features to predict mutant utility in revealing faults. 
Gradient boosting is a powerful ensemble learning technique which combines several trained weak models to perform classification. Unlike common ensemble techniques, such as random forests~\cite{breiman2001random}, that simply average models in the ensemble, boosting methods follow a constructive strategy of ensemble formation where models are added to the ensemble sequentially. At each particular iteration, a new weak, base-learner model is trained with respect to the error of the whole ensemble learnt so far~\cite{natekin2013gradient}. 
We use the FastBDT~\cite{keck2016fastbdt} implementation by setting the number of trees to 1,000 and the trees depth to 5.

%For data dependence analysis, we rely on the LLVM Dependence Graph (DG)~\cite{chalupaslicing} whose nodes
%are constituted by statements and vertices represent data and control dependences.

%\subsubsection{Killable Mutant Selection}

%\subsubsection{Killable Mutant Prioritization}

\subsection{Demonstrating Example}
\label{Example}

\begin{figure*}[!t]
	\centering
	\vspace{-1.0em}
	\includegraphics[width=0.85\linewidth]{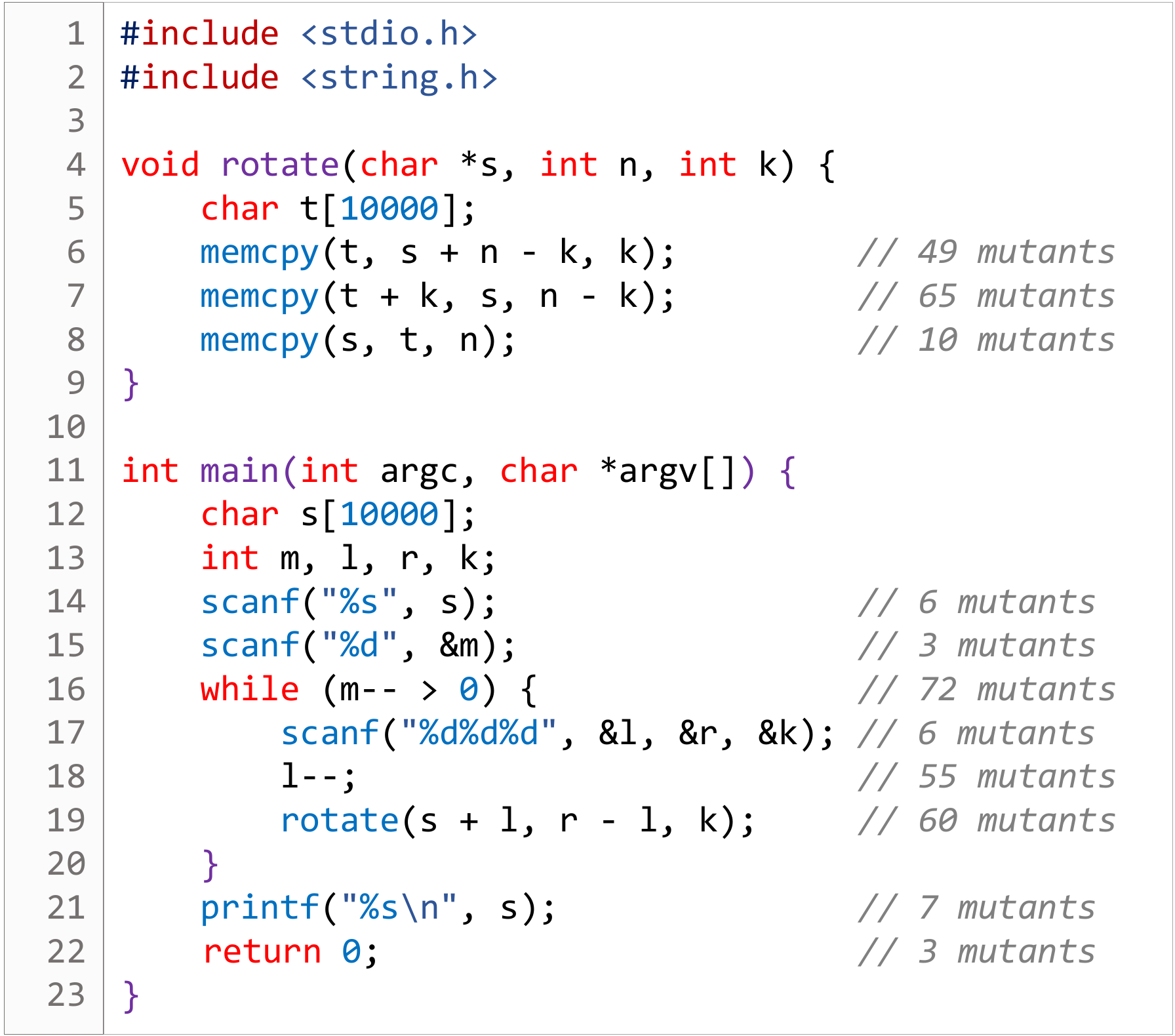}
	\caption{Example program where mutation is applied. The C language comments on each line show the number of mutants generated on the line.}
		\label{fig:example-program}
\end{figure*}

\begin{figure*}[!t]
	\centering
	\vspace{-1.0em}
	\includegraphics[width=1.0\linewidth]{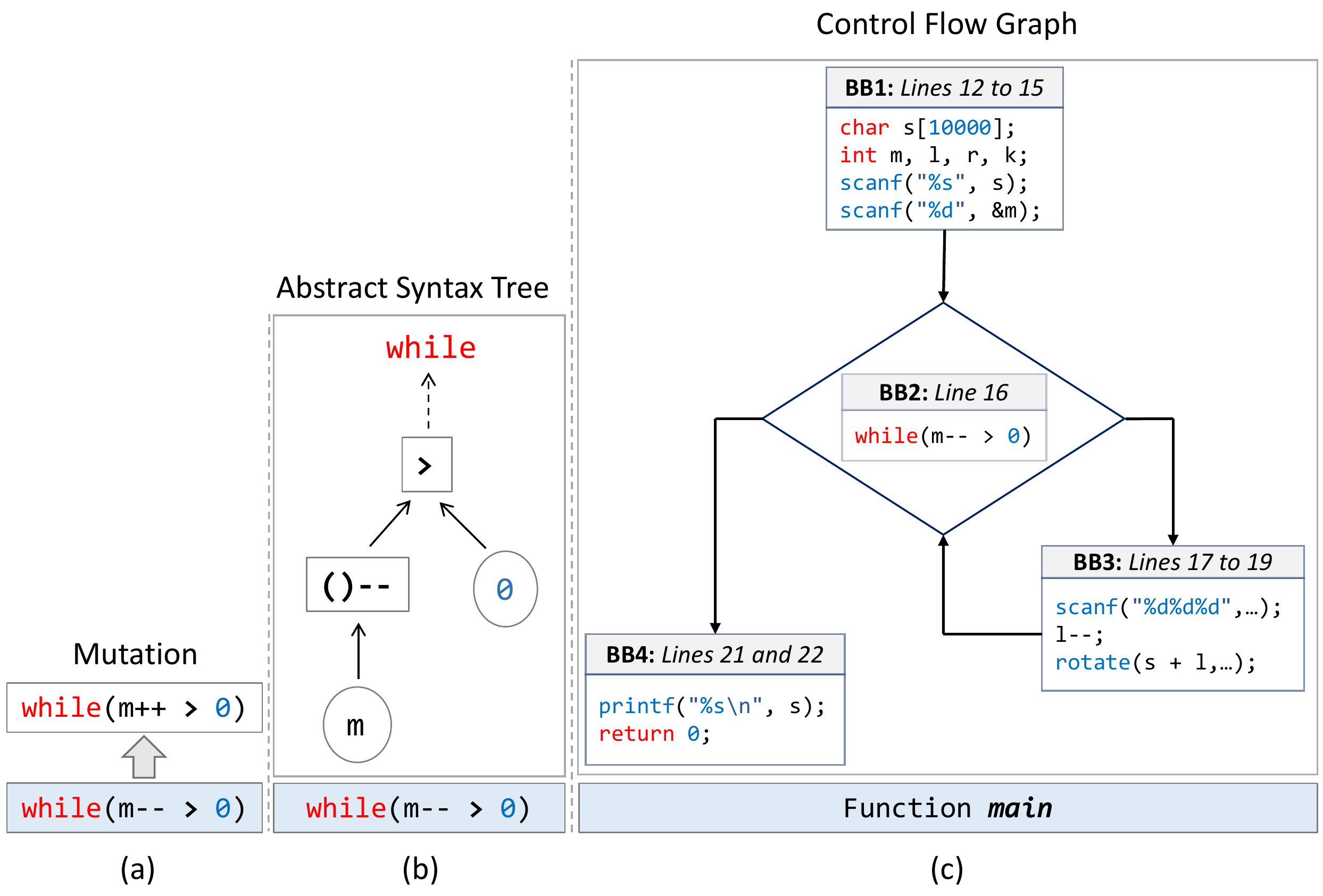}
	\caption{(a) An example of mutant $M$ from the example program from Figure~\ref{fig:example-program}, (b) the abstract syntax tree of the mutated statement and (c) the control flow graph of the function containing the mutated statement.}
		\label{fig:example-program-AST-CFG}
\end{figure*}

Here we provide an example on how the features of Table \ref{tab:features} are computed. We consider the program in figure~\ref{fig:example-program} (extracted from the Codeflaws benchmark, ID: 598-B-bug-17392756-17392766), on which mutation is applied. We present the feature extraction for a mutant $M$, which is created by replacing the right side decrement operator by the right side increment operator on line 16 ($m--$ becomes $m++$). %Thus the mutated expression is $m--$. 
We also present in figure~\ref{fig:example-program-AST-CFG}-a the mutant, the the abstract syntax tree (AST) of the mutated statement (\emph{while} condition) in figure \ref{fig:example-program-AST-CFG}-b and  in figure \ref{fig:example-program-AST-CFG}-c the control flow graph (CFG) of the function containing the mutated statement. 

The features, for mutant $M$, are computed as following: \\
- The \emph{complexity} feature value is the number of mutants generated on the statement containing the mutant $M$ (Line 16). In this case 72 mutants. Thus, the \emph{complexity} is 72. \\
- The \emph{CfgDepth} feature value is the minimum number of basic blocks to follow, along the CFG, from \emph{main} function’s entry point to the basic block containing $M$ (\textit{BB2}). In this case 1 basic block as shown in Figure~\ref{fig:example-program-AST-CFG}-c. Thus, the \emph{CfgDepth}  is 1. \\
- The \emph{CfgPredNum} feature value is the number of basic blocks directly preceding the basic block containing $M$ (\textit{BB2}) on the control flow graph. In Figure~\ref{fig:example-program-AST-CFG}-c there are 2 basic blocks (\textit{BB1} and \textit{BB3}). Thus, the \emph{CfgPredNum}  is 2. \\
- The \emph{CfgSuccNum} feature value is the number of basic blocks directly following the basic block containing $M$ (\textit{BB2}) on the control flow graph. In Figure~\ref{fig:example-program-AST-CFG}-c there are 2 basic blocks (\textit{BB3} and \textit{BB4}). Thus, the \emph{CfgSuccNum}  is 2. \\
- The \emph{AstNumParents} feature value is the number of AST parents of the mutated expression. In this case, the only AST parent is the relational expression, in Figure~\ref{fig:example-program-AST-CFG}-b, whose sub-tree is rooted on the greater than sign ($>$). Thus the feature value is 1. \\
- The \emph{NumOutDataDeps} feature value is the number of mutants on expressions data dependent on the mutated expression. In this case, looking at Figure~\ref{fig:example-program}, the value of variable $m$ written in the mutated expression $m--$ is only used in the same expression. Thus the feature value is the number of mutants on the mutated expression $m--$. \\
- The \emph{NumInDataDeps} feature value is the number of mutants on expressions on which the mutated expression is data dependent. In this case, looking at Figure~\ref{fig:example-program}, the value of variable $m$ used on the mutated expression $m--$ is either written on the \emph{scanf} statement at line 15, or in the same expression. Thus the feature value is the sum of the number of mutants on the statement at line 15 and the number of mutants on the mutated expression $m--$. \\
- The \emph{NumOutCtrlDeps} feature value is the number of mutants on statements control dependent on the mutated expression. In this case, looking at Figure~\ref{fig:example-program}, no statement is control dependent on the mutated expression $m--$. Thus the feature value is 0. \\
- The \emph{NumInCtrlDeps} feature value is the number of mutants on expressions on which the mutated statement is control dependent. In this case, looking at Figure~\ref{fig:example-program}, no expression controls the mutated expression. Thus the feature value is 0.  \\
- The \emph{NumTieDeps} feature value is the number of mutants on the right decrement expression (mutated expression). \\
- The \emph{AstParentsNumOutDataDeps} feature value is the number of mutants on expressions data dependent on the AST parent of the mutated expression. In this case, looking at Figures~\ref{fig:example-program} and \ref{fig:example-program-AST-CFG}-b, the value of the relational expression (AST parent of $m--$) is not used in other expressions. Thus the feature value is 0. \\
- The \emph{AstParentsNumInDataDeps} feature value is the number of mutants on expressions on which the AST parent of the mutated expression is data dependent. In this case, looking at Figures~\ref{fig:example-program} and \ref{fig:example-program-AST-CFG}-b, the value of the relational expression (AST parent of $m--$) only depends on the value of expression $m--$. Thus the feature value is the number of mutants on expression $m--$. \\
- The \emph{AstParentsNumOutCtrlDeps} feature value is the number of mutants on statements control dependent on the AST parent of the mutated expression. In this case, looking at Figures~\ref{fig:example-program} and \ref{fig:example-program-AST-CFG}-b, all the statements in basic block $BB3$ are control dependent on the relational expression (AST parent of $m--$). Thus the feature value is the sum of the number of mutants in lines 17, 18 and 19 of the code in Figure~\ref{fig:example-program}. \\
- The \emph{AstParentsNumInCtrlDeps} feature value is the number of mutants on expressions on which the AST parent of the mutated expression is control dependent. In this case, looking at Figures~\ref{fig:example-program} and \ref{fig:example-program-AST-CFG}-b, no expression controls the relational expression (AST parent of the mutated expression $m--$). Thus the feature value is 0.  \\
- The \emph{AstParentsNumTieDeps} feature value is the number of mutants on the relational expression, AST parent of the mutated right decrement expression. The feature value here is the number of mutants of the relational expression of operator greater than. \\
- The \emph{TypeAstParents} feature value is AST type of the AST parent expression of the mutated expression. Here, that is the AST type of the relational expression with operator greater than.  \\
- The \emph{TypeMutant} feature value is the type of the mutant as a string representing the matched and replaced pattern. The feature value is “$()-- \rightarrow ()++$”.  \\
- The \emph{TypeStmtBB} feature value is the type of the basic block containing the mutated statement. The feature value here is the type of \emph{BB2} (see Figure~\ref{fig:example-program-AST-CFG}-c), which is “While Condition”. \\
- The \emph{AstParentMutantType} feature value is the aggregation of types of the mutants on the AST parents of the mutated expression. That is the aggregation of the mutants types of the relational expression whose sub-tree is rooted on the greater than sign ($>$) as shown in Figure~\ref{fig:example-program-AST-CFG}(b). The aggregation of a set of mutant types is performed by summing up the one encoding vectors of the mutants types, allowing each mutant type to be represented in the encoding. \\
- The \emph{OutDataDepMutantType} feature value is the aggregation (as computed for \emph{AstParentMutantType}) of the mutant types of the mutants counted to compute \emph{NumOutDataDeps}. \\
- The \emph{InDataDepMutantType} feature value is the aggregation (as computed for \emph{AstParentMutantType}) of the mutant types of the mutants counted to compute \emph{NumInDataDeps}. \\
- The \emph{OutCtrlDepMutantType} feature value is the aggregation (as computed for \emph{AstParentMutantType}) of the mutant types of the mutants counted to compute \emph{NumOutCtrlDeps}. \\
- The \emph{InCtrlDepMutantType} feature value is the aggregation (as computed for \emph{AstParentMutantType}) of the mutant types of the mutants counted to compute \emph{NumInCtrlDeps}. \\
- The \emph{AstChildHasIdentifier} feature value is the Boolean value representing whether the mutated expression has an identifier as operand. In this case, the mutated expression has the identifier \emph{m} as operand. Thus, the value of the feature is 1 (True). \\
- The \emph{AstChildHasLiteral} feature value is the Boolean value representing whether the mutated expression has a literal as operand. In this case, the mutated expression does not have the literal as operand. Thus, the value of the feature is 0 (False). \\
- The \emph{AstChildHasOperator} feature value is the Boolean value representing whether the mutated expression has an operator. In this case, the mutated expression has the operator right decrement operator $--$. Thus, the value of the feature is 1 (True). \\
- The \emph{DataTypesOfOperands} feature value is the datatype of the operand of the right decrement operation $--$. That is the datatype of \emph{m} which is \emph{“int”}. \\
- The \emph{DataTypeOfValue} feature value is the datatype of the value of the mutated expression, Which is \emph{“int”} as the data type of \emph{m}. \\

%% file: RQs.tex
\section{Research Questions}
\label{sec:rqs}

When building prediction methods, the first thing to investigate is their prediction ability. %This step ensures that the trained classifiers predict the actual elements of interest. 
Thus, our first question can be stated as: 

\begin{description}
\item[\textbf{RQ1}:] 
  \emph{ How well does our machine learning method predicts the killable mutants? }
\end{description}

Similarly, our second question can be stated as: 

\begin{description}
\item[\textbf{RQ2}:] 
  \emph{ How well does our machine learning method predicts the fault revealing mutants? }
\end{description}

After demonstrating that our classification method predicts satisfactorily the fault revealing mutants, we continue by investigating its ability to practically support mutant selection with respect to the actual measure of interest, the revealed faults, and with respect to the random baseline techniques. Therefore, we investigate:

\begin{description}
\item[\textbf{RQ3}:] 
  \emph{ How do our methods compare against the random strategies with respect to the fault revealing mutant selection problem? }
\end{description}

In addition to the random strategies, we also compare with the current state-of-the-art mutant selection methods. Thus, we ask: 

\begin{description}
\item[\textbf{RQ4}:] 
  \emph{ How do our methods compare against the E-Selection and SDL with respect to the fault revealing mutant selection problem? }
\end{description}

As we already discussed an alternative mutant cost reduction technique is mutant prioritization. Hence, we ask:

\begin{description}
\item[\textbf{RQ5}:] 
  \emph{ How do our methods compare against the random strategies with respect to the fault revealing mutant prioritization problem? }
\end{description}

In addition to the random strategies, we also compare with the defect prediction mutant prioritization baseline. Therefore, we ask: 

\begin{description}
\item[\textbf{RQ6}:] 
  \emph{ How do our methods compare against the defect prediction mutant prioritization method? }
\end{description}

Finally, by demonstrating the benefits of our approach, we turn to investigate the generalization ability of our approach on larger and complex programs. Therefore we conclude by asking:

\begin{description}
\item[\textbf{RQ7}:] 
  \emph{ How well do our method generalise its findings on independently selected programs that are much larger and complex? }
\end{description}
%
%The answer to this questions will remove any doubts on the applicability and generalisability of our approach. It will also provide empirical evidence that the use of numerous small program units can provide reliable results that project to real-world and mature programs.

%% file: Experiment.tex
\section{Experimental Setup}
\label{sec:setup}

%We now discuss the benchmarks, mutation tools and experimental protocol for evaluating the \toolname approach.
\subsection{Benchmarks: Programs and Fault(s)}

For the purposes of our study we need a large number of programs that are not trivial and are accompanied with real faults. The fault set has to be large and of diverse types. Unfortunately, mutation testing is costly and its experimentation requires generating strong test suites \cite{ChekamPTH17}. Therefore, there are two necessary tradeoffs, between the number of faults to be considered, the strengths of the test suites to be used and the size of the used programs. 

To account for these requirements, we used the Codeflaws benchmark \cite{TanYYMR17}. This benchmark consists of 7,436 programs (among wich 3,902 are faulty) selected from the Codeforces\footnote{\url{http://codeforces.com/}} online database of programming contests. These contests consist of three to five problems, of varied difficulty levels. Every user submits its programs that resolve the posed problems. In total, the benchmark involves programs from 1,653 users ``with diverse level of expertise''~\cite{TanYYMR17}. 

Every fault in this benchmark has two program instances: the rejected {\em `faulty'} submission and the accepted {\em `correct'} submission. Overall, the benchmark contains 3,902 faulty program versions of 40 different defect classes. It is noted that every faulty program instance in our dataset is unique, meaning that every program we use is different from the others (in terms of implementation). To the best of our knowledge, this is the largest number of faults used in any of the mutation testing studies. The size of the programs varies from 1 to 322 with an average of 36 lines of code. Applying mutation testing on Codeflaws yielded 3,213,543 mutants and required a total of 8,009 CPU days for all computations. 

To strengthen our results and demonstrate the ability of our approach to handle faults made by actual developers, we also used the CoREBench~\cite{BohmeR14} benchmark. CoREBench includes real-world complex faults that have been systematically isolated from the history of C open source projects. These programs are of 9-83 KLoC and are accompanied by developer test suites. It is noted that every CoREBench fault forms a single fault instance (it differs from the other faults). 

We used the available test suites augmented by KLEE~\cite{CadarDE08}.  Although, these test suites greatly increased the cost of our experiment, we considered their use of vital importance as otherwise our results could be subject to ``noise effects''~\cite{ChekamPTH17}. %Also, these test suites represent the limits that are experimentally achievable by the current tools. 

%Unfortunately, the current version of KLEE does not handle calls to the file system directories, making it inapplicable to many cases of the Findutils and Make programs. We therefore, had to exclude these two programs from our analysis. %Two more faults were exclude from our analysis as their respective versions did not compile in our execution environment. 
%We also excluded the faults that required more than three weeks of execution time. We therefore, restricted our analysis on 25 faults (18/22 in Coreutils and 7/15 in Grep). Test generation resulted in a test pool composed of 122,261 and 17,921 test cases related to Codeflaws, CoREBench.
%Unfortunately we had technical difficulties at collecting data for many of the faulty versions, i.e., some did not compile, others (make program) was not working well with KLEE and others took exceedingly large amount of time to run.   
Due to the very high cost of the experiments and technical difficulties to reproduce some faults, we conducted our analysis on 45 faults (22 in Coreutils, 12 in Find and 11 in Grep). Applying mutation testing on these 45 versions yielded 1,564,614 mutants and required a total of 454 CPU days of computation (without considering the test generation and machine learning computations and evaluations). Test generation resulted in a test pool composed of 122,261 and 22,477 test cases related to Codeflaws, CoREBench.
 
The goal of our study is to evaluate the fault revealing ability of the mutants we select. However, approximately half of our faults are trivial ones (triggered by most of the test cases), and their inclusion in our analysis would artificially inflate our results. Thus, we restrict our analysis on the faults that are revealed by less than 25\% of the test cases involved in our test suites. Taking such a threshold is usual in fault injection studies \cite{SIR}, but it ensures that the targeted faults and our focus is on faults that are hard enough to find. Practically, taking a lower threshold will significantly reduce the number of faults to be considered hindering our ability to train, while taking a higher threshold will make all the approaches perform similarly, as the faults will be easy to reveal. Overall, from the Codeflaws benchmark we consider 1,692 out of the 3,902 ones (1,692 are the non-trivial faults) and  from the CoreBench benchmark 45 faults.

\begin{figure}[!t]
	\centering
	\vspace{-.6em}
	\includegraphics[width=0.9\linewidth]{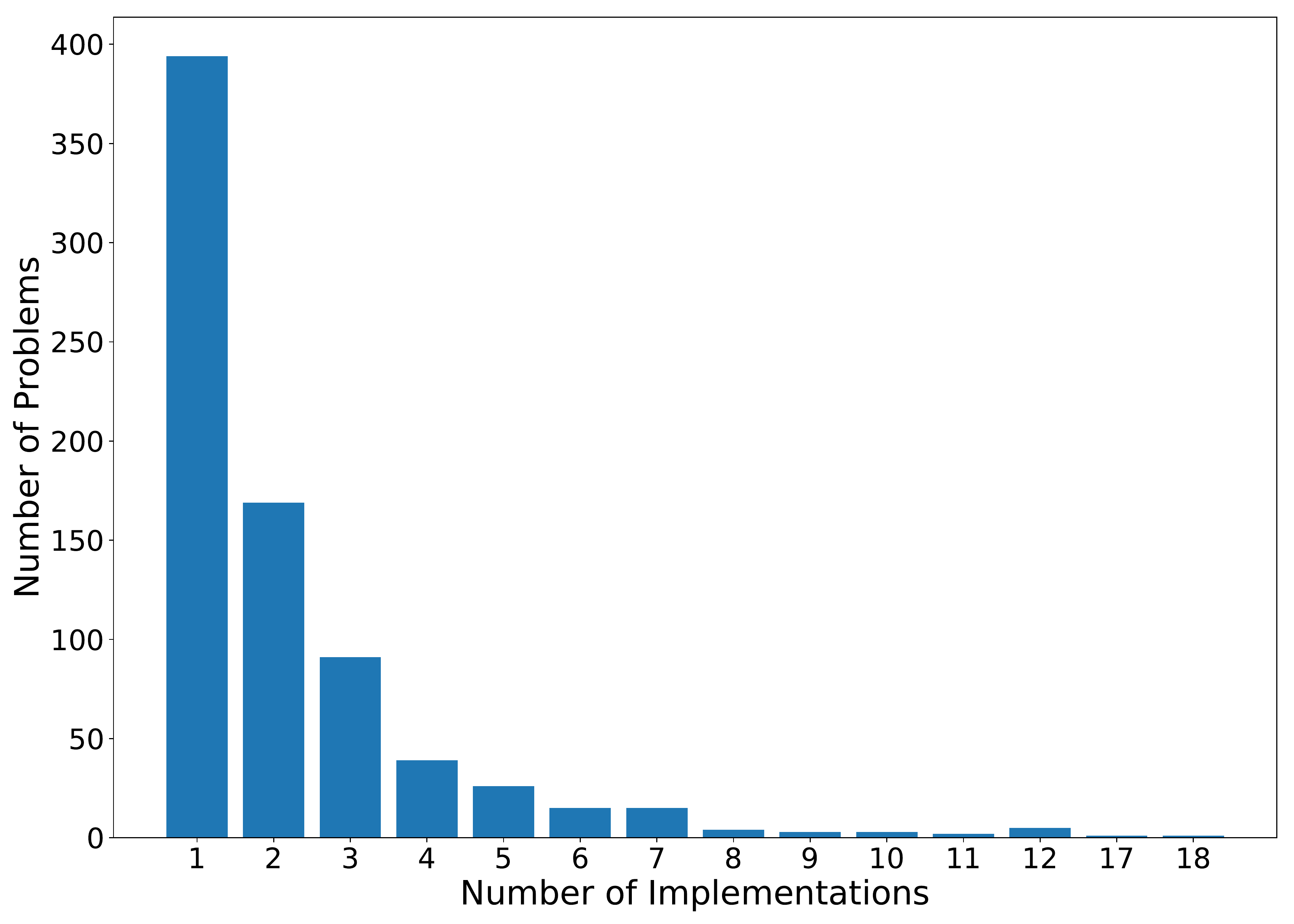}
	\caption{Distribution of Codeflaws Benchmark problems by number of implementations.}
%	\vspace{-1.0em}
		\label{fig:problemsbyimplemetations}
\end{figure}

Figure \ref{fig:problemsbyimplemetations} shows the distribution of number of problems by number of implementations for the considered faulty programs from Codeflaws. We observe that 85\% of the problem are having at most 3 implementations. 

\begin{figure}[!t]
	\centering
	\vspace{-.6em}
	\includegraphics[width=0.9\linewidth]{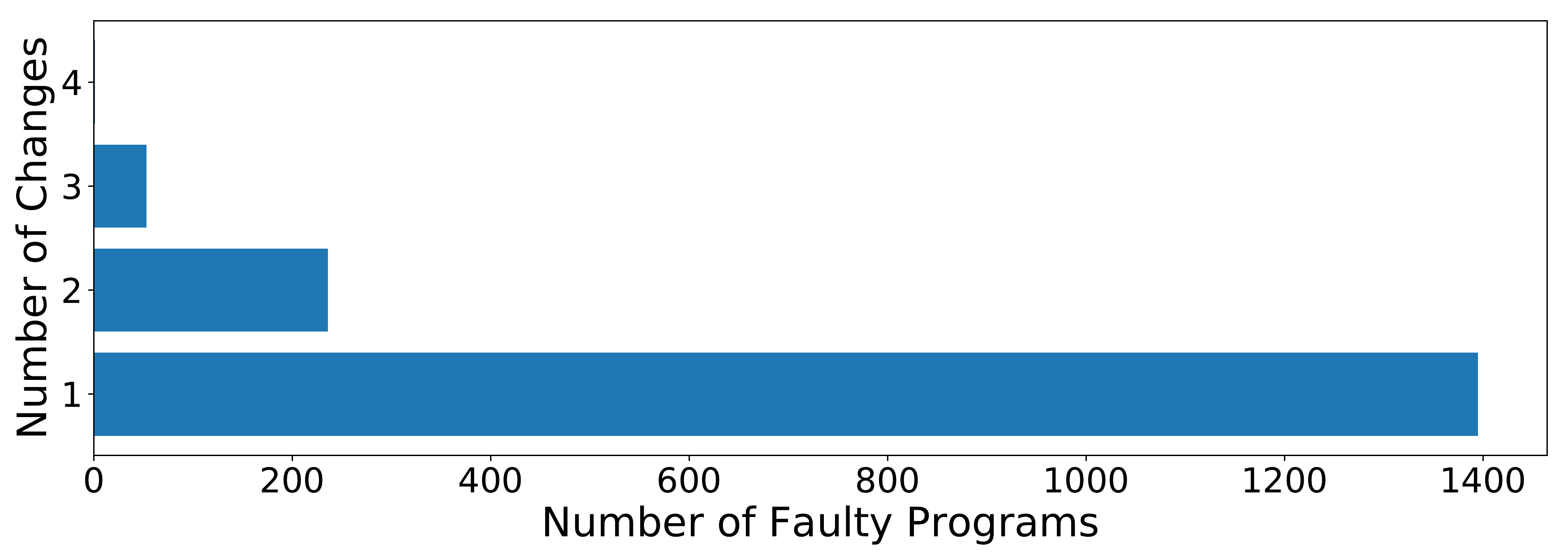}
	\caption{Distribution of Codeflaws Benchmark faulty programs by number of lines of code changed to fix the fault.}
%	\vspace{-1.0em}
		\label{fig:numChanges}
\end{figure}

Despite that Codeflaws benchmark faults were mined from a programming contest, the faults nevertheless are relatively small syntactical mistakes. We observe on figure \ref{fig:numChanges} that 82\% of the faults are fixed by modifying a single line of source code. This ensures that we are compatible with the competent programmer hypothesis\footnote{The competent programmer hypothesis states that programs have a small syntactic distance from the correct version so that we need a few keystrokes to correct the program}, which is one of the basic assumptions of mutation testing \cite{HintTestDataSelection1978}. 

\input{mutation}

\subsection{Experimental Procedure}

To answer our research questions we performed an experiment composed of three parts. The first part regards the prediction ability of our classification method, answer RQ1 and RQ2, the second regards the fault revealing ability of the approaches, answer RQ3-RQ6, and the third regards the fault revealing ability of our approach on large independently selected programs, answer RQ7. To account for our use case scenario, in our experiments we always train and evaluate our approach on a different sets of programs (CodeFlaws) or program versions (CoREBench). 

As a first step we used KLEE to generate test cases for all the programs we study and formed a pool of test cases by joining the generated and the available test cases. 
We then constructed a mutation-fault matrix, which records for every test case the mutants that it kills and whether it reveals the fault or not (we construct a matrix for every single fault we study). We also record the execution time needed to execute every mutant-test pair so that we can simulate the execution cost of the approaches. We make the data available\footnote{https://mutationtesting.uni.lu/farm}.

To measure fault revelation we mutated the faulty program versions. This is important in order to avoid making any assumption related to the interaction of mutants and faults, aka Clean Program Assumption \cite{ChekamPTH17}. Based on this matrix we compute the fault revealing ratio for each mutant. The \textit{fault revealing ratio} is the ratio of tests that kill the mutant and reveal the fault to the total number of tests that kill the mutant. 

\emph{First experimental part:}
The first task of prediction modeling is to evaluate the contribution of the used features. We computed the information gain values for each one of the used features. Higher information gain values represent more informative features for decision trees. Demonstrating the importance of our features helps us understand what is the most important factors affecting the utility of mutants. Having measured information gain, we then measure the prediction ability of our classification method by evaluating its ability to predict killable and fault revealing mutants. For this part of the experiment we considered as fault revealing the mutants that have fault revealing ratio equal to 1. We relax this constraint in the second part of the experiment. 

We evaluate the trained classifiers using four typically adopted metrics such as the precision, recall, F-measure and Area Under Curve (AUC). The \emph{precision} of a classifier is defined as the number of items that are truly relevant among the items that the classifier predicted to be relevant. The \emph{recall} of a classifier is defined as the number of items that are predicted to be relevant by the classifier among all the truly relevant items. The F-measure of a classifier is defined as the weighted harmonic mean of the precision and recall, it is also named \emph{F1 score}. The Area Under Curve (AUC) of a classifier is the area under the Receiver Operating Characteristic (ROC) curve (The ROC curve shows how many true positive classifications can be gained as more and more false positives are allowed) \cite{AliceZhengEvalML}. Precision represents the ratio of the identified killable and fault revealing mutants out of those classified as such. Recall represents the ratio of the identified killable and fault revealing mutants out of all existing ones. In classification usually recall and precision are competitive metrics in the sense that higher values of one imply lower values for the other. To better compare classifiers researchers use the F-measure and AUC metrics. These measure the general classification accuracy of the classifier. Higher values denote better classification.

To reduce the risk of overfitting, we applied a 10-fold cross validation by partitioning our program set into 10 parts and iteratively train on 9 parts and evaluation on one. We report the results for all the partitions.

This experiment part was performed on the Codeflaws programs.

%ToDo Describe the following
%Threshold for mutant selection  
%100 repetitions for each bug 
%compare with Random and dummy classifiers  

\emph{Second experimental part:}
Our analysis requires comparing mutation-based strategies with respect to the actual value of interest, the number of faults revealed. Given that killing a mutant does not always result in revealing a fault, we train the classifier 
in accordance with the actual fault revealing ratios (i.e., the ratio of tests that kill a mutant and also reveal faults).

%ToDo Name the studied approaches and explain their use in the experimetn

We then select and prioritise our mutants. To evaluate and compare the studied approaches with respect to fault revelation, we follow a typical procedures~\cite{ChekamPTH17,KurtzAODKG16,NaminAM08} by randomly selecting test cases, from the formed test pools, that kill the selected mutants. In case none of the available test cases on our test pool kills the mutant we treat it as equivalent. We repeat this process for each one of the studied approaches. As done in the first part of the experiment we report results using a 10-fold cross validation. 

For the mutant selection problem we randomly pick a mutant and then randomly pick a test case that kills it. Then we remove all the killed mutants and pick another one. 
If the mutant is not killed by any of the test cases on our test pool we treat it as equivalent. We repeat this process 100 times and compute the probability of revealing each one of the faults. 

For the mutant prioritisation case we follow the mutant order by picking test cases that kills each mutant. We do not attempt to kill a mutant twice. Again, we repeat this process 100 times and compute the Average Percentage of Faults Detected (APFD) values, which is typical metric used test case prioritization studies \cite{HenardPHJT16}. Again we align the compared approaches with respect to their cost (number of mutants need manual analysis) and compare their effectiveness. 
%detail further the evaluation metrics...

To account for coincidental results and the stochastic selection of test cases and mutants we used the Wilcoxon test, which is a non-parametric test, to determine whether the Null Hypothesis (that there is no difference between the studied methods) can be rejected. In case the Null Hypothesis is rejected, then we have evidence that our approach outperforms the others. Even when the null hypothesis does not hold, the size of the differences might be small. To account for this effect we also measured the Vargha Delaney effect size $\hat{\text{A}}_{12}$ \cite{Vargha00}, which quantifies the size of the differences (aka statistical effect size). $\hat{\text{A}}_{12} = 0.5$ suggests that the data of the two samples tend to be the same. $\hat{\text{A}}_{12} > 0.5$ values indicate that the first dataset has higher values, while $\hat{\text{A}}_{12} < 0.5$ indicate the opposite. 

This experiment part was performed on the Codeflaws programs.

\emph{Third experimental part:}
To further evaluate the fault revealing ability of our approach, we applied it on the CoreBench programs. We also adopted the 10-fold cross validation as for the experiments on Codeflaws. We report results related to both fault revelation and APFD values. The CoreBench corpus is small in size and hence \toolname might not be particularly important. However, in case the signal of our features is strong, we will be able to experience the benefits of our method even with those few data.   
%To further evaluate the fault revealing ability of our approach, we applied it on the CoreBench programs. Since CoreBench faults are very few, we adopted a leave-one-out strategy, where we train on all but one faults and evaluate on the `left-out' fault. We report results related to both fault revelation and APFD values. 
%To eliminate any concern related to the overfitting or to the generalizability of our approach, we simply evaluate the trained models from the second part of the experiment to the larger programs. In other words we trained on the CodeFlaws and evaluate on the CoreBench fault sets. We report results related to both fault revelation and APFD values. 

%detail exactly what we need to evaluate....

\subsection{Mutant Selection and Effort Metrics}

When comparing methods, a comparison basis is required. In our case we measure fault revelation and effort. While measuring fault revelation based on the fault set we use is direct, measuring effort/cost is hard. Effort/cost depends on a large number of uncontrolled parameters, such as the followed procedure, level of automation, skills, underlying infrastructure and the learning curve. Therefore, we have to account for different scenarios. and we adopt three frequently used metrics; the number of selected mutants, the number of test cases generated and the number of mutants requiring analysis. 

The first metric (selected mutants) represents the number of mutants that one should use when applying mutation testing. This is a direct and intuitive metric as it suggest that developers should select a particular set of mutants to generate (form an actual executable codes), execute and analyse. Although, such a metric conforms to our working scenario, it does not focus on the required test generation effort involved. Generating test cases is mostly a manual task (due to the test oracle problem) and so, we also consider a second metric, the number of test cases that can be generated based on a selected set of mutants.

We also adopt a third metric, the number of mutants that need to be analysed (equivalent mutants and those we pick, i.e., analysed in order to generate test cases). This metric somehow reflects the effort a tester needs to put in order to kill or identify as equivalent the selected mutants (under the assumption that equivalent mutants require the same effort as the test generation). 

 To fairly compare the random selection methods, we select mutants until we analyse the same number of mutants as analysed by our selection method. This establishes a fixed cost point for all the approaches and compare their effectiveness.

 There are other cost factors, such as the  mutant-test execution cost and the analysis of equivalent mutants (for the first two metrics) that we investigate separately. The reason for that is that we would like to see if our approaches are also faster to execute and require reasonably less equivalent mutants.

%% file: mutation.tex
\subsection{Automated Tools Used}
\label{subsec:mutation}

We used KLEE \cite{CadarDE08} to support the test generation. We used  KLEE with a relatively large timeout limit, equal to two hours per program, the Random Path search strategy,  with Randomize Fork Enabled, Max Memory 2048 MB, Symbolic Array Size 4096 elements, Symbolic Standard input size 20 Bytes and Max Instruction Time of 30 seconds. 
%This resulted in 26,229 and 2,727 test cases for CodeFlaws and CoREBench. 
This resulted in 26,229 and 1,942 test cases for CodeFlaws and CoREBench. 
Since the automatically generated test cases do not include any test oracle, we used the programs' fixed version as oracle. We considered as failing, every test case that resulted in different observable program output when executed in the `faulty' from that in the `correct'-fixed one. Similarly, we used the program output to identify the killed mutants. We deemed a mutant as killed if it resulted in a different output than in the original program. 

We built a mutation testing tool that operates on LLVM bitcode. Actually all our metrics and analysis were performed on the LLVM bitcode. Our tool implements 18 operators, composed of 816 transformation rules.  
These include all those that are supported by modern mutation testing tools~\cite{OffuttLRUZ96,PapadakisSurvey,ColesLHPV16} and are detailed in Table \ref{MutTypes}.  %We thus, used the following 19 operators:  Statement Deletion, Trap statements insertion, binary operation operand swaping, Left/Right operand, Logical connector Replacement,  Absolute Value Insertion, Non-Pointer Unary Operator Insertion,  Non-Pointer Unary Operator Stripping, Constant Value Replacement,   Non-Pointer Binary Operator Replacement, Function Call's arguments Shuffling, Switch's Cases shuffling, Pointer Binary Operation Replacement,   Pointer Unary Operation Replacement, Pointer Dereference,  Binary Operation to Unary operator. 
%We  used the 18 operators recorded on Table \ref{MutTypes}. 

Each mutation operation consists of matching an instruction type (original instruction type) and replacing with another instruction type (mutated instruction type). Thus, a mutation operator is defined as pair of original instruction type and mutated instruction type. The instruction types are defined as following ($p$ refers to pointer values and $s$ refers to scalar values):
\begin{itemize}
  \item \textbf{ANY STMT}
  	refers to matching any type of statement (only original instruction type).
  \item \textbf{TRAPSTMT}
  	refers to a \emph{trap}, which cause the program to abort its execution (only mutated instruction type).
  \item \textbf{DELSTMT}
  refers to statement deletion, i.e., replacing by the empty statement which is equivalent to deleting the original statement (applies only on the mutated instruction type).
  \item \textbf{CALL STATEMENT}
  refers to a function call.
  \item \textbf{SWITCH STATEMENT}
  refers to a C language like \emph{switch} statement.
  \item \textbf{SHUFFLEARGS}
  can only be a mutated instruction type and, when the orignal instruction type is a function call. It refers to the same function call as the original but with arguments of, same type, swapped. (e.g. $f(a,b)\rightarrow f(b,a)$)
  \item \textbf{SHUFFLECASESDEST}
  can only be used as mutated instruction type and, when the orignal instruction type is a \emph{switch} statement. It refers to the same \emph{switch} statement as the original but with the basic blocks of the \emph{cases} swapped. (e.g. \{$case\ a:B_1;\ case\ b:B_2;\ default:B_3;\} \rightarrow \{case\ a:B_2;\ case\ b:B_1;\ default:B_3;\}$)
  \item \textbf{REMOVECASES}
  only be used as mutated instruction type and, when the orignal instruction type is a \emph{switch} statement. It refers to the same \emph{switch} statement as the original but with some \emph{cases} deleted (the corresponding values will lead to execute the \emph{default} basic block). (e.g. \{$case\ a:B_1;\ case\ b:B_2;\ default:B_3;\} \rightarrow\ \{case\ a:B_2;\ default:B_3;\}$)
  \item \textbf{SCALAR.ATOM}
  refers to any non pointer type variable or constant (only original instruction type).
  \item \textbf{POINTER.ATOM}
  refers to any pointer type variable or constant (only original instruction type).
  \item \textbf{SCALAR.UNARY}
  refers to any non pointer unary arithmetique or logical operation (e.g. $abs(s)$, $-s$, $!s$, $s++$ ...).
  \item \textbf{POINTER.UNARY}
  refers to any pointer unary arithmetique operation (e.g. $p++$, $--p$ ...).
  \item \textbf{SCALAR.BINARY}
  refers to any non pointer binary arithmetique, relational or logical operation (e.g. $s_1+s_2$, $s_1\&\&s_2$, $s_1>>s_2$, $s_1<=s_2$ ...).
  \item \textbf{POINTER.BINARY}
  refers to any pointer binary arithmetique or relational operation (e.g. $p+s$, $p_1>p_2$ ...).
  \item \textbf{DEREFERENCE.UNARY}
  refers to any combination of pointer dereference and scalar unary arithmetic operation, or combnation of pointer unary operation and pointer dereference (e.g. $(*p)--$, $*(p--)$ ...).
  \item \textbf{DEREFERENCE.BINARY}
  refers to any combination of pointer dereference and scalar binary arithmetic operation, or combnation of pointer binary operation and pointer dereference (e.g. $(*p)+s$, $*(p+s)$ ...).
\end{itemize}

 %%%%%
Applying mutation testing on CodeFlaws and CoREBench yielded 3,213,543 and 1,564,614 mutants. %Their execution resulted in killing the 87\% and 54\% of the mutants and required a total of 8,009 and 454 CPU days of computations for CodeFlaws and CoREBench.  
 
To reduce the influence of redundant and equivalent mutants, we applied TCE~\cite{PapadakisJHT15, HaririSCKM16, KintisPJMTH18}. Since we operate on LLVM bitcode we compared the mutated optimized LLVM codes using the llvm-diff utility.  llvm-diff is a tool like the known Unix diff utility but for LLVM bitcode. TCE Detected 1,457,512 and  715,996 mutant equivalences on CodeFlaws and CoREBench. Note that the equivalent and redundant mutants detected by TCE are removed from the mutants set and neither executed nor considered in the experiments.

The execution of the mutants post TCE resulted in killing the 87\% and 54\% of the mutants for CodeFlaws and CoREBench. It is important to note that our tool applies mutant test execution optimizations by recording the coverage and program state at the mutation points avoiding the execution of mutants that do not infect the program state \cite{PapadakisM10b}. This optimization enables huge test execution reductions and forms the current state of the art at the test execution optimizations \cite{PapadakisSurvey}. Despite these optimization our tool required a total of 8,009 and 454 CPU days of computations for CodeFlaws and CoREBench indicating the large amount of computation resources required to perform such an experiment. 

\begin{table}[!t]
\vspace{-1.0em}
\caption{Mutant Types}
\label{MutTypes}
\begin{tabularx}{4.5in}{l l l}
\toprule
\textbf{\scriptsize{Mutated Instruction}} & \textbf{\scriptsize{Original Instruction Type}} & \textbf{\scriptsize{Mutated Instruction Type}}\tabularnewline
\midrule
\multirow{5}{*}{STATEMENT}  & \scriptsize{ANY STMT} & \scriptsize{TRAPSTMT}\tabularnewline
  & \scriptsize{ANY STMT} & \scriptsize{DELSTMT}\tabularnewline
 & \scriptsize{CALL STATEMENT} & \scriptsize{SHUFFLEARGS}\tabularnewline
 & \scriptsize{SWITCH STATEMENT} & \scriptsize{SHUFFLECASESDESTS}\tabularnewline
  & \scriptsize{SWITCH STATEMENT} & \scriptsize{REMOVECASES}\tabularnewline
\midrule
\multirow{13}{*}{EXPRESSION}  & \scriptsize{SCALAR.ATOM} & \scriptsize{SCALAR.UNARY}\tabularnewline
  & \scriptsize{SCALAR.BINARY} & \scriptsize{SCALAR.BINARY}\tabularnewline
 & \scriptsize{SCALAR.BINARY} & \scriptsize{SCALAR.UNARY}\tabularnewline
 & \scriptsize{SCALAR.ATOM} & \scriptsize{SCALAR.BINARY}\tabularnewline
  & \scriptsize{SCALAR.BINARY} & \scriptsize{TRAPSTMT}\tabularnewline
  & \scriptsize{POINTER.BINARY} & \scriptsize{POINTER.BINARY}\tabularnewline
 & \scriptsize{SCALAR.BINARY} & \scriptsize{DELSTMT}\tabularnewline
 & \scriptsize{DEREFERENCE.BINARY} & \scriptsize{DEREFERENCE.BINARY}\tabularnewline
  & \scriptsize{SCALAR.UNARY} & \scriptsize{SCALAR.UNARY}\tabularnewline
  & \scriptsize{POINTER.BINARY} & \scriptsize{POINTER.UNARY}\tabularnewline
 & \scriptsize{DEREFERENCE.BINARY} & \scriptsize{DEREFERENCE.UNARY}\tabularnewline
 & \scriptsize{POINTER.ATOM} & \scriptsize{POINTER.UNARY}\tabularnewline
  & \scriptsize{POINTER.UNARY} & \scriptsize{POINTER.UNARY}\tabularnewline
\bottomrule
\end{tabularx}
\vspace{-1.0em}
\end{table}

%% file: predictions.tex
%We now report on the experiments for answering the research questions previously outlined.
\subsection{Assessment of killable mutant prediction (RQ1 and RQ2)}
\label{subsec:Killprediction}

To check the prediction performance of our classifier we performed a 10-Fold cross-validation for three different selected sets. These were the results of applying  \toolnameC to predict killable mutants and selecting the 5\%, 10\% and 20\% of the top ranked mutants. The \toolnameC classifier achieves  98.8\% 5.7\%, 10.7\% precision, recall and F-measure when selecting the 5\% of the mutants. With respect to 10\% and 20\% sets of mutants, it achieves 98.8\% and 98.7\% (precision), 11.4\% and 22.8\% (recall), 20.4\% and 37.0\%	F-measures. These values are higher than those that one can get by randomly sampling the same number of mutants. In particular the \toolnameC has 12.3\%, 12.2\% and 12.1\% higher precision, and 0.7\%, 1.4\% and 2.8\% higher recall values for the 5\%, 10\% and 20\% sets of mutants.   % 1.3\%, 2.5\% and 4.5\% higher F-Measure

When using \toolnameC to predict non killable mutant, the classifier achieves  95.1\% 35.0\%, 51.2\% precision, recall and F-measure when selects the 5\% of the mutants. With respect to 10\% and 20\% sets of mutants, it achieves 79.1\% and 49.3\% (precision), 58.6\% and 73.2\% (recall), 67.3\% and 58.9\%	F-measures. These values are higher than those that one can get by randomly sampling the same number of mutants. In particular the \toolnameC has 81.6\%, 65.7\% and 35.8\% higher precision, and 30.1\%, 48.7\% and 53.3\% higher recall values for the 5\%, 10\% and 20\% sets of mutants.

To train our models, approximately 48 CPU hours were required, while to perform the evaluation (perform mutant selection) it required less than a second. Since, training should only happen  once in awhile, the training time is considered acceptable. Of course the cost of selecting and prioritizing mutants is practically negligible.

\begin{figure}[!t]
	\centering
	\includegraphics[width=\linewidth]{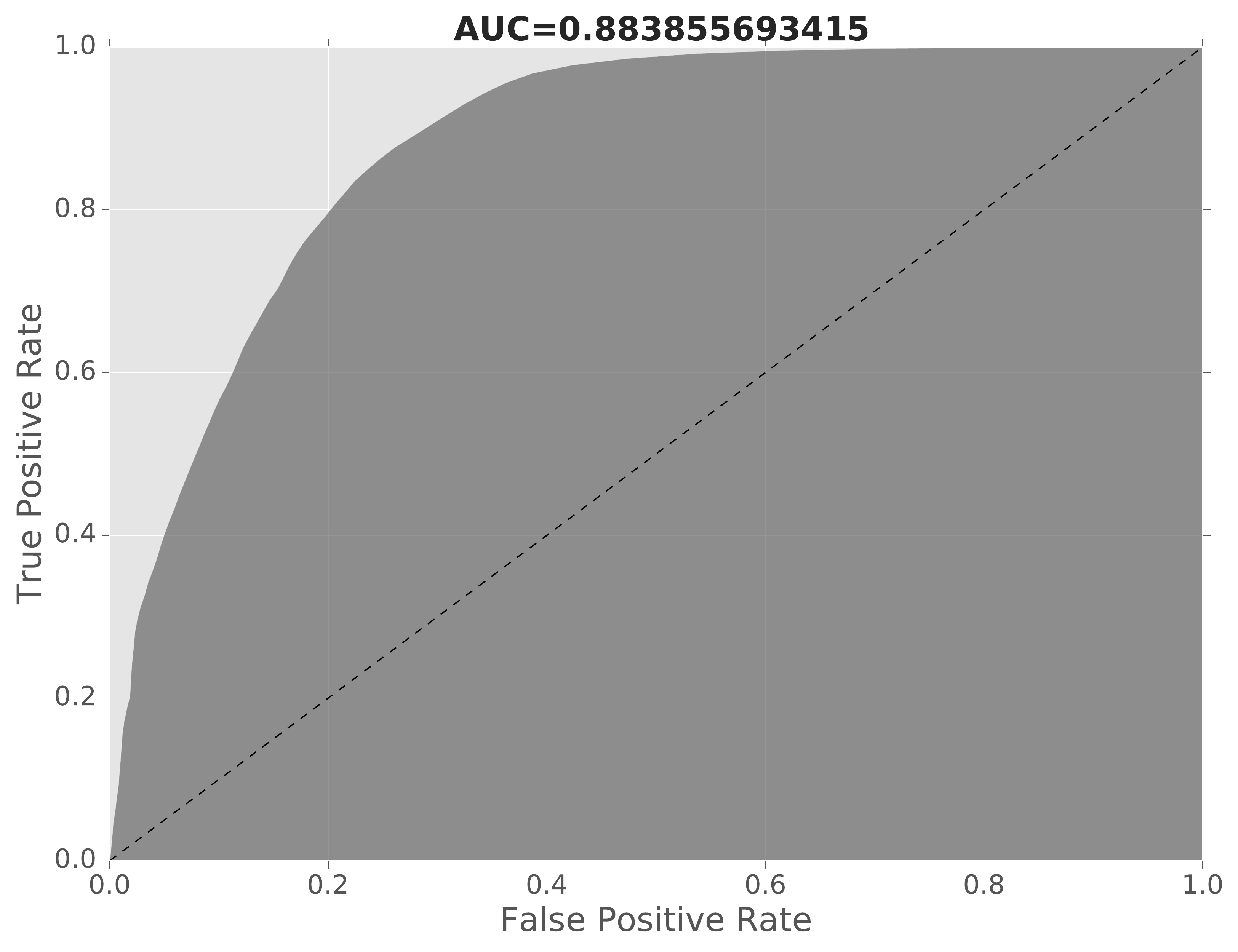}
	\caption{Receiver Operating Characteristic For Killable Mutants Prediction on Codeflaws}
	\label{fig:roc-killable}
	\vspace{-1em}
\end{figure}

The Receiver operating characteristic (ROC) shown in Figure~\ref{fig:roc-killable} further
illustrates performance variations of the classifier in terms of true positive and false positive rates when the discrimination threshold changes: the higher the area under curve (AUC), the better the classifier. Our classifier
achieves an AUC of 88\%. 
These results establish that the code properties that were leveraged as features for characterizing mutants provide, together, a good discriminative power for assessing the fault revealing potential of mutants.  

\subsection{Assessment of fault revelation prediction}
\label{subsec:prediction}
\paragraph{ML prediction performance}

Similarly to subsection \ref{subsec:Killprediction} we performed a 10-Fold cross-validation for three different selected sets in order to check the prediction performance of our classifier. These were the results of applying  \toolname and selecting the 5\%, 10\% and 20\% of the top ranked mutants. The \toolname classifier achieves  5.7\% 12.8\%, 7.8\% precision, recall and F-measure when selects the 5\% of the mutants. With respect to 10\% and 20\% sets of mutants, it achieves 4.9\% and 3.9\% (precision), 22.0\% and 35.1\% (recall), 8.0\% and 7.0\%	F-measures. These values are higher than those that one can get by randomly sampling the same number of mutants. In particular  \toolname has 3.5\%, 2.7\% and 1.7\% higher precision, and 7.8\%, 12.1\% and 15.1\% higher recall values for the 5\%, 10\% and 20\% sets of mutants.

The cost of training and evaluation are same as those reported in section~\ref{subsec:Killprediction}.

\begin{figure}[!t]
	\centering
	\includegraphics[width=\linewidth]{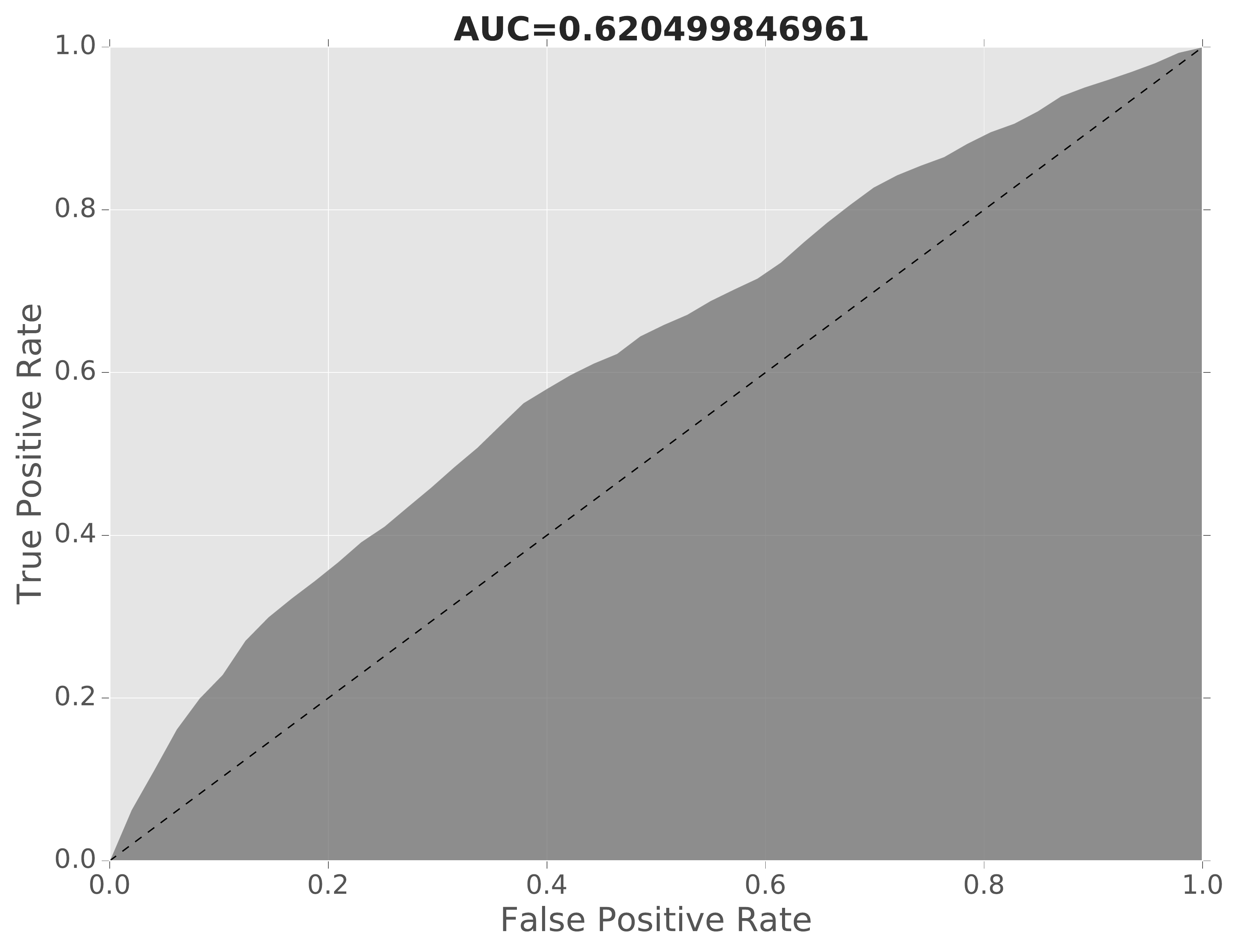}
	\caption{Receiver Operating Characteristic For Fault Revealing Mutants Prediction on Codeflaws}
	\label{fig:roc}
	\vspace{-1em}
\end{figure}

The Receiver operating characteristic (ROC) shown in Figure~\ref{fig:roc} further
illustrates performance variations of the classifier in terms of true positive and false positive rates when the discrimination threshold changes: the higher the area under curve (AUC), the better the classifier. Our classifier
achieves an AUC of 62\%. % for all three sets. 

We believe that  such a result is encouraging due to the nature of the developer mistakes. As developers make mistakes in an non-systematic way, for the same problem, some may make mistakes while some others may not, the only thing we can hope for is to form good heuristics, i.e., identify mutants that maximize the chances to reveal faults. Therefore, it is hard to get much higher AUC values. Nevertheless, we expect future research to built on and improve our results by forming better predictors. 

Overall, the above results demonstrate that the code properties that were leveraged as features for characterizing mutants provide, together, a discriminative power to assess the fault revealing potential of mutants. 

\begin{figure*}[!t]
	\centering
	\vspace{1em}
	\includegraphics[width=\linewidth]{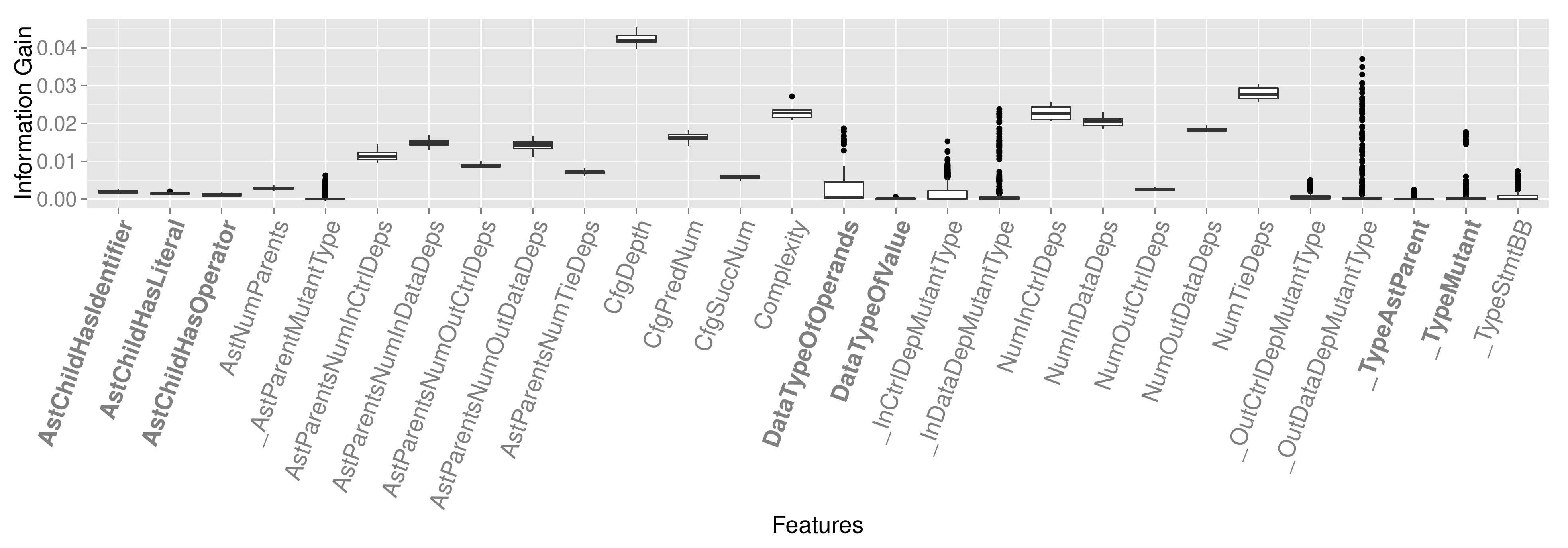}
	\caption{Information Gain distributions of ML features on Codeflaws}
	\label{fig:rq1-IG}
\end{figure*}

\paragraph{Considered features} We provide in Figure~\ref{fig:rq1-IG} the distribution of information Gain values for the various features considered in this work. 
Information gain (IG) measures how much ``information" a feature gives us about the class we want to predict. The IG values are computed by the supervised learning algorithm during the training process. 
These data enable the assessment of the potential contribution of every feature to a prediction model. 
Experimental training process provides evidence in Figure~\ref{fig:rq1-IG} that the suggested features (in bold) contribute significantly less than several other features that we have designed for \toolname. Interestingly, together with complexity, the features related to control and data dependencies are the most informative ones. Here we should note that IG values do not suggest which features to select and which not. Actually our results show that we need all the features.

%% file: comparison.tex
\subsection{Mutant selection}
\label{subsec:comparison}

\subsubsection{Comparison with Random (RQ3)}

\begin{figure}[!t]
\vspace{-1.0em}
\centering
\includegraphics[width=\linewidth]{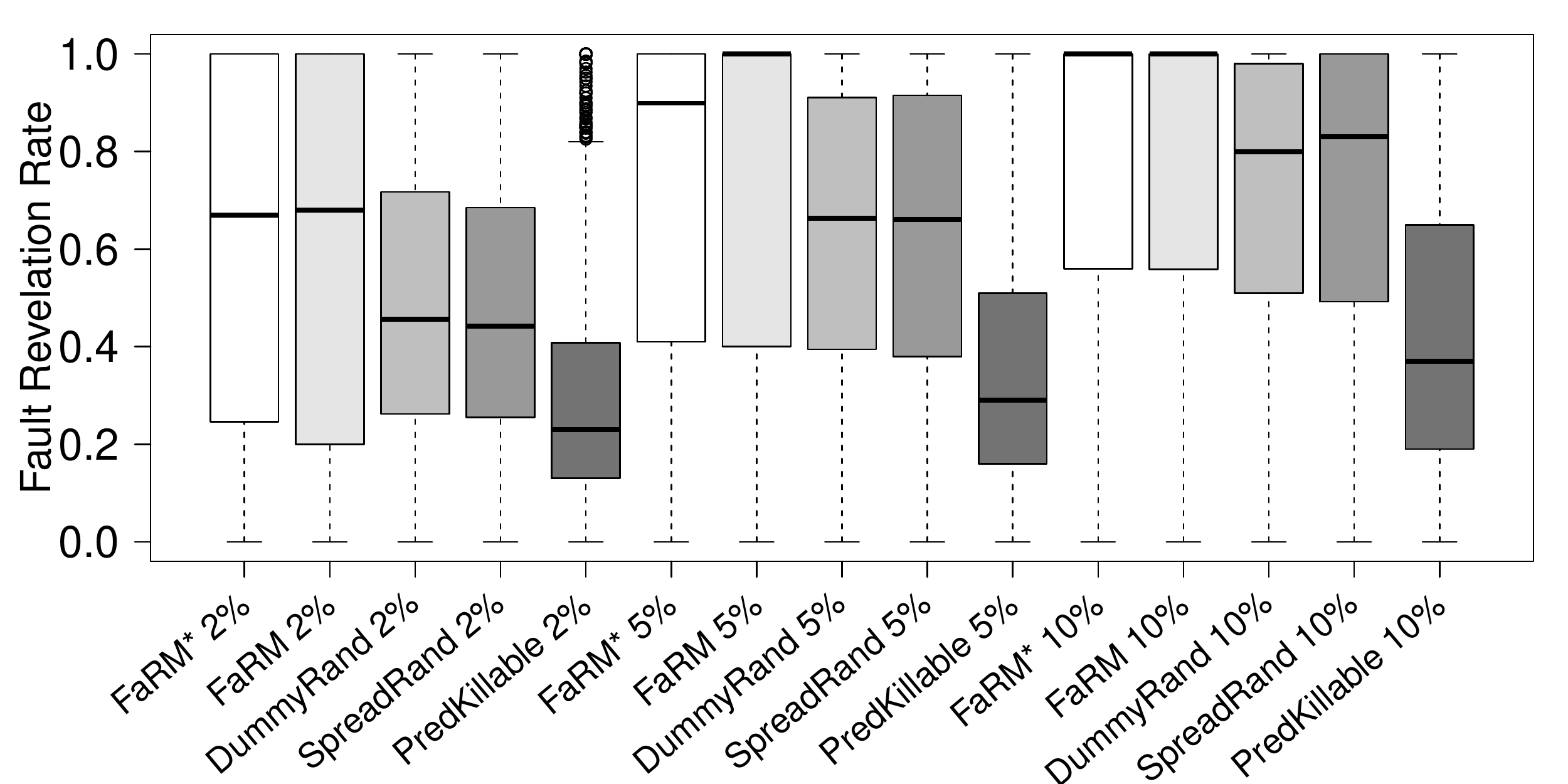}
\caption{Fault revelation of the mutant selection strategies on Codeflaws. All three  \toolname and \toolnameB sets  outperform the random baselines. }
\label{fig:rq2}
\vspace{-1.0em}
\end{figure}

Figure~\ref{fig:rq2} shows the distribution of the fault revelation of the mutant selection strategies when selecting the  2\%, 5\% and 10\% of the top ranked mutants. As can be seen from the plot, both \toolnameB and \toolname outperforms both DummyRandom and spreadRandom. Both DummyRandom and spreadRandom outperform \toolnameC. When selecting 2\% of the mutants the difference, for both \toolname and \toolnameB, of the median values is 22\% and 24\% for the DummyRandom and SpreadRandom respectively. This difference is increasing when  selecting the 5\% of the mutants and goes to 34\% and 34\% for \toolname and, 24\% and 24\% for \toolnameB. When selecting 10\% of the mutants the difference becomes 20\% and 17\% for both \toolname and \toolnameB. %, and for the 20\% threshold is 9\% and 4\%.
Regarding \toolnameC, the difference with DummyRandom and SpreadRandom at the 2\% mutant selection threshold is 23\% and 21\% respectively. This difference increase for the 5\% to 37\% and 37\%. For the 10\% threshold is 43\% and 46\%.

To check whether the differences are statistically significant we performed a Wilcoxon rank-sum test, which is a non-parametric test that measures whether the values of one sample are higher than those of the second sample. We adopt a statistically significant level $a<0.01$ below of which we consider the differences as statistically significant. We also computed the Vargha Delaney $\hat{A}_{12}$ effect size value between the approaches.

The statistical test showed that \toolname and \toolnameB outperforms both DummyRandom and SpreadRandom with statistically significant difference. both DummyRandom and SpreadRandom outperform \toolnameC with statistically significant difference. As expected the differences between DummyRandom and SpreadRandom are not significant. 
It is noted that all comparisons are aligned with respect to the number of mutants that need analysis, which as we already explained represents the manual effort involved. 
The Vargha Delaney $\hat{A}_{12}$ value between the approaches show that for the 2\% threshold, \toolname is better than DummyRandom and SpreadRandom in 60\% and 63\% of the cases respectively. These values are slightly higher for \toolnameB where it is better than DummyRandom and SpreadRandom in 62\% and 65\% of the cases respectively. DummyRandom and SpreadRandom are respectively better than \toolnameC in 84\% and 82\% of the cases. 
For the 5\% threshold, \toolname is better than DummyRandom and SpreadRandom in 66\% of the cases. \toolnameB is better than DummyRandom and SpreadRandom in 64\% and 65\% of the cases respectively. DummyRandom and SpreadRandom are respectively better than \toolnameC in 88\% and 84\% of the cases.
For the 10\% threshold, \toolname is better than DummyRandom and SpreadRandom in 65\% and 63\% of the cases respectively. \toolnameB is better than DummyRandom and SpreadRandom in 64\% and 61\% of the cases respectively. DummyRandom and SpreadRandom are respectively better than \toolnameC in 87\% and 85\% of the cases.

Regarding the test execution time of the involved methods, our approach has an advantage but this is minor. The median difference between \toolname and DummyRandom and SpreadRandom was 12 and 39 seconds per program respectively. This means that \toolname required 12 and 29 seconds less execution time than the random baselines. While these differences are considered as minor they demonstrate that \toolname has significantly higher fault revelation ability than the compared baselines without introducing any major overhead. 

Overall, our results suggest that \toolname and \toolnameB significantly outperforms the random baselines with practically significant differences, i.e., improvements on the ratios of revealed faults were between 4\% to 34\%.  \toolnameC is outperformed by all the approaches.

\subsubsection{Comparison with SDL \& E-Selective (RQ4)}
\label{subsubsec:farm_vs_sdl_vs_eselective}

\begin{figure}[!t]
\vspace{-1.0em}
\centering
\includegraphics[width=\linewidth]{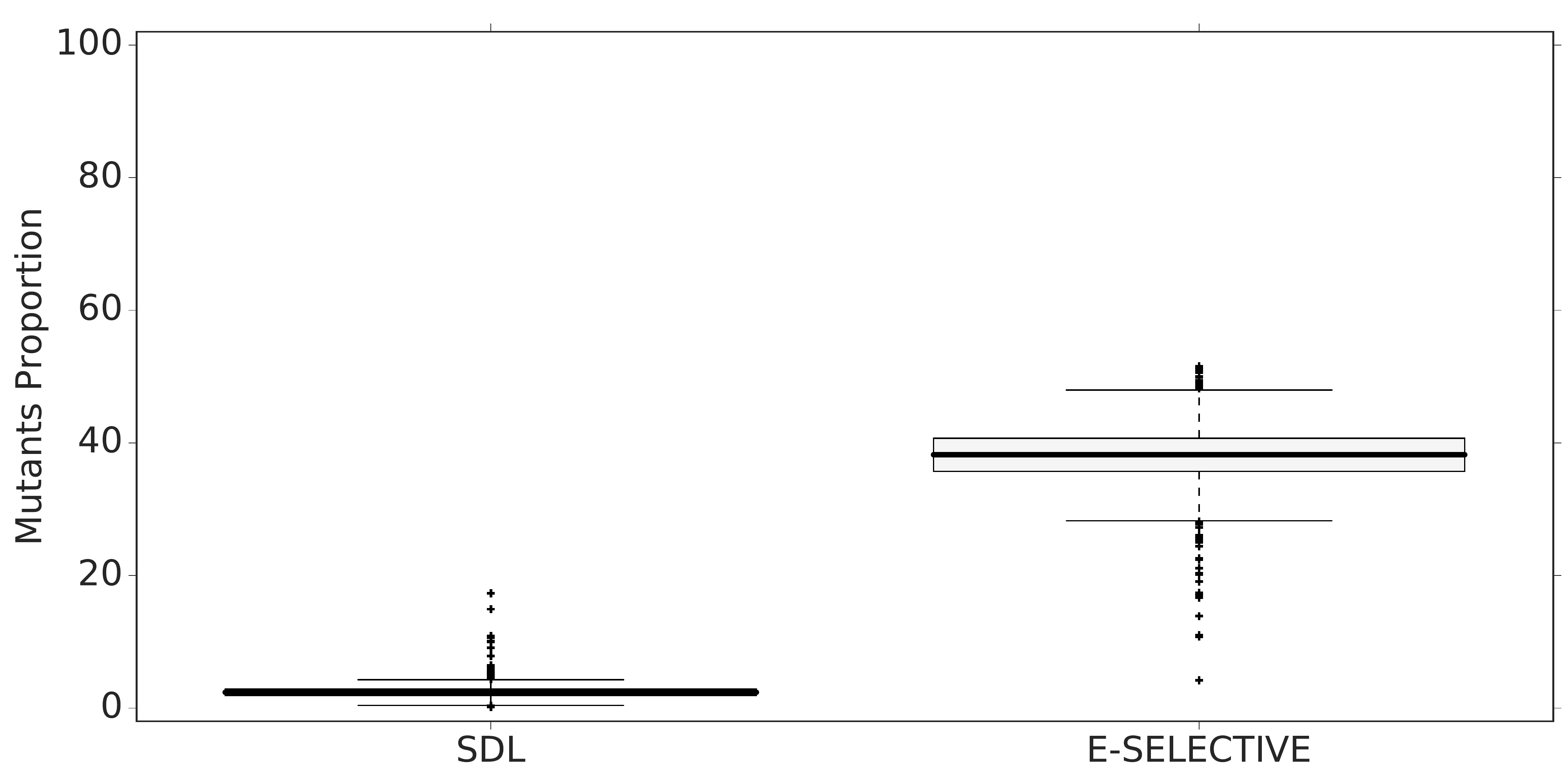}
\caption{Proportion of SDL and E-SELECTIVE mutants among all mutants for Codeflaws subjects.}
\label{fig:sdl_eselective_proportions}
\vspace{-1.0em}
\end{figure}

This section aims to compare the fault revelation of our approach with that of the SDL and the E-Selective mutants sets.

%\subsubsection{Comparison \toolname with SDL}
In order to compare our approach with SDL selection, the selection size is set to the number of SDL mutants. In the Codeflaws subjects, SDL and E-SELECTIVE mutants represent in median respectively 2\% and 38\% of all mutant as seen in Figure~\ref{fig:sdl_eselective_proportions}.

\begin{figure}[!t]
\vspace{-1.0em}
\centering
\includegraphics[width=\linewidth]{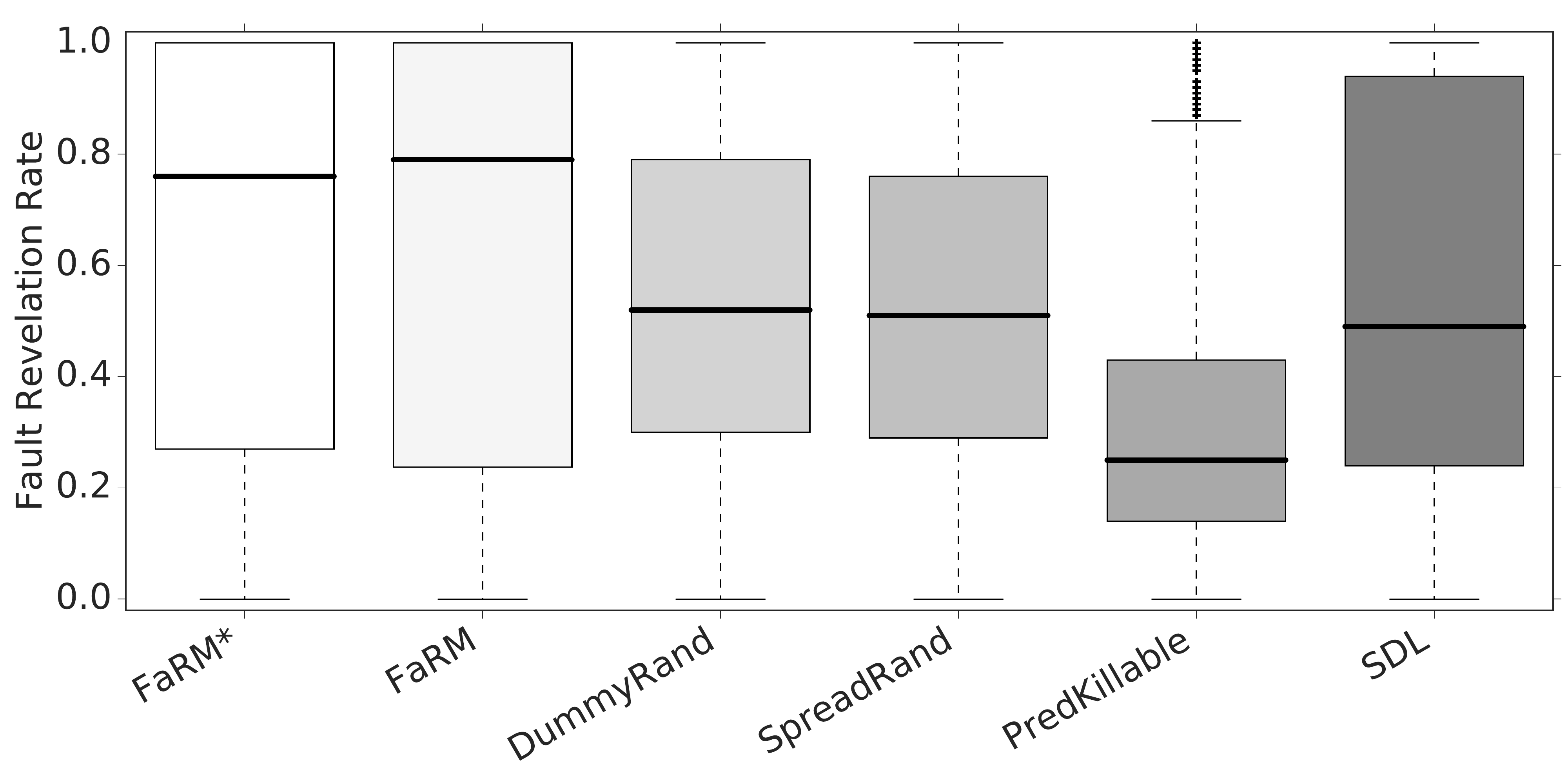}
\caption{Fault revelation of \toolname compared with SDL on Codeflaws. \toolname sets  outperform the SDL selection. Approximately 2\% (number of SDL mutants) of all the mutants are selected.}
\label{fig:farm_vs_sdl}
\vspace{-1.0em}
\end{figure}

Our analysis is designed as following. For each subject, the $|SDL|$ top ranked mutants of \toolname  are selected (where $|SDL|$ is the total number of SDL mutants). We also select the $|SDL|$ top ranked mutants with the random approaches. Then, the fault revelation of each approach's selected mutants set is computed for comparison and presented in Figure~\ref{fig:farm_vs_sdl}. We observe that \toolname and \toolnameB respectively have 30\% and 27\% higher median fault revelation than SDL. \toolnameC has 25\% lower median fault revelation than SDL. We also observe that SDL has similar fault revelation with the random selections (respectively 3\% and 2\% lower than DummyRandom and SpreadRandom).

We also performed the Wilcoxon rank-sum test as in section~\ref{subsec:comparison}. The statistical test showed that both \toolname and \toolnameB outperform SDL, and SDL outperforms \toolnameC. The difference between SDL and DummyRandom and SpreadRandom is not statistical significant. We also computed the Vargha Delaney $\hat{A}_{12}$ value between the approaches and found that \toolname and \toolnameB are respectively better than SDL in 54\% and 55\% of the cases.  %58\% of the cases.
SDL is better than \toolnameC in 79\% of the cases.

\begin{figure}[!t]
\vspace{-1.0em}
\centering
\includegraphics[width=\linewidth]{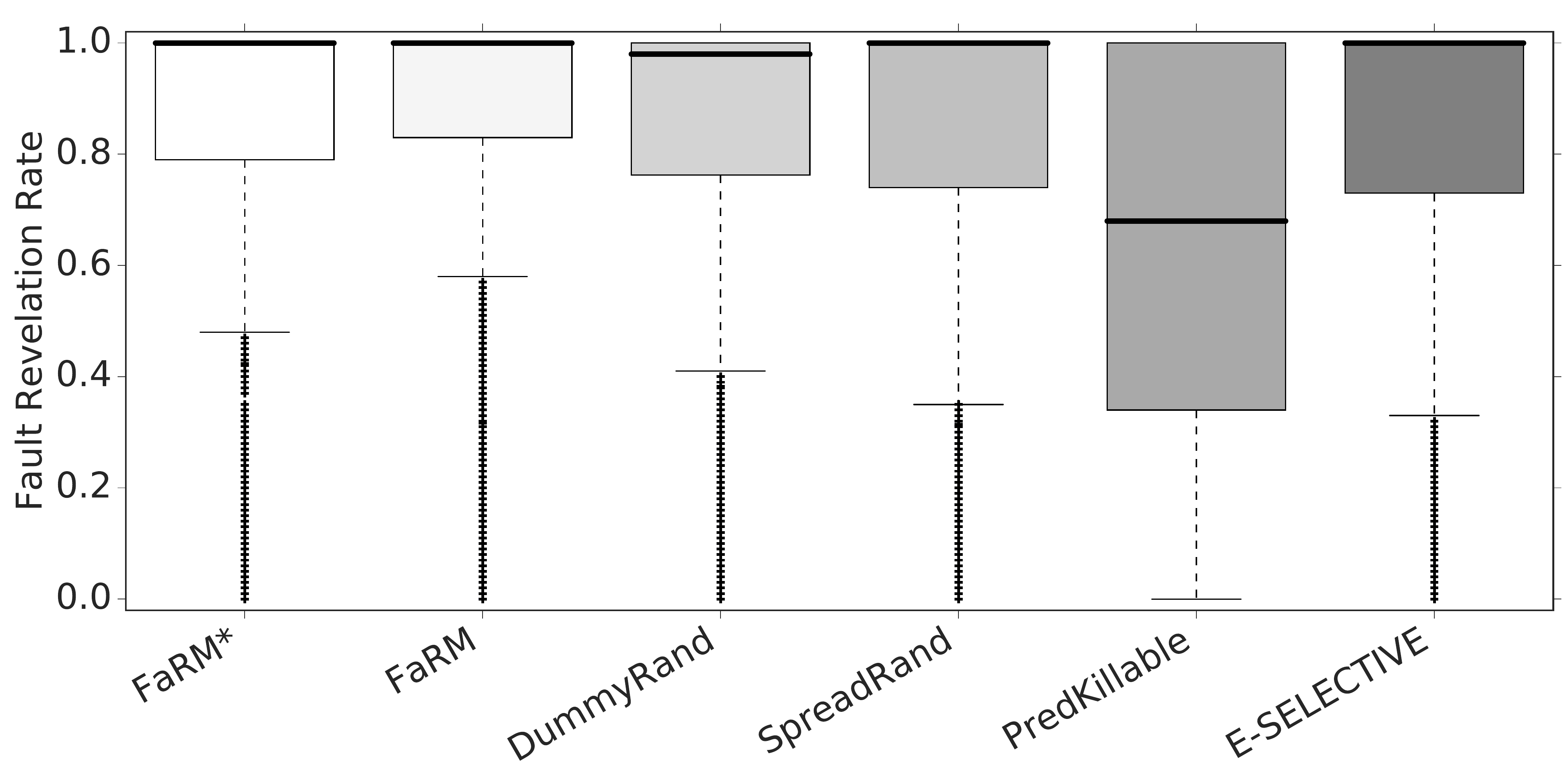}
\caption{Fault revelation of \toolname compared with E-Selective on Codeflaws. Approximatively 38\% (number of E-Selective mutants) of all the mutants are selected.}
\label{fig:farm_vs_eselective}
\vspace{-1.0em}
\end{figure}

\begin{figure}[!t]
\vspace{-1.0em}
\centering
\includegraphics[width=\linewidth]{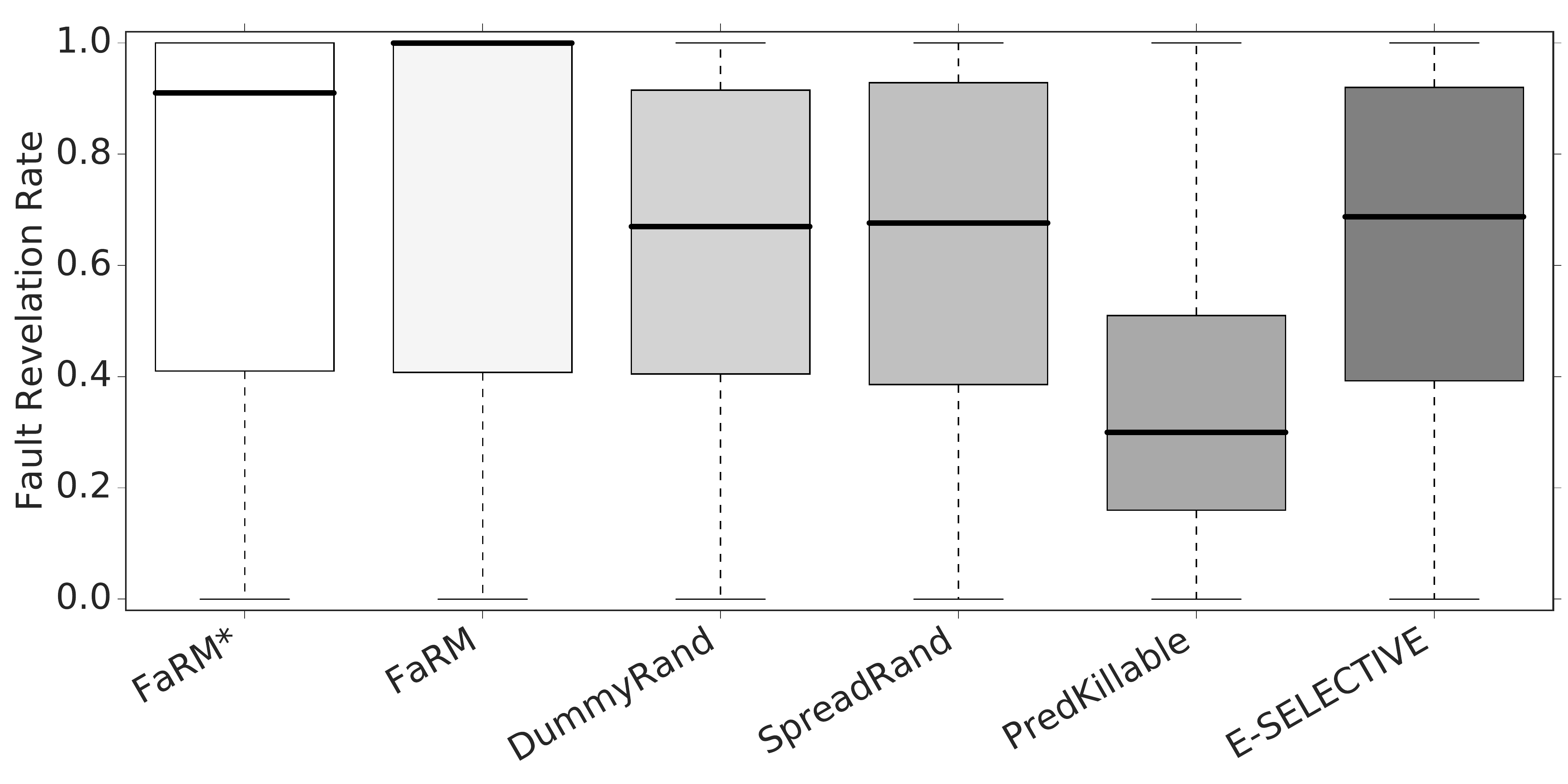}
\caption{Fault revelation of \toolname compared with E-Selective for selection size 5\% of all mutants. \toolname and \toolnameB sets  outperform E-Selective selection. }
\label{fig:farm_vs_eselective_5percent}
\vspace{-1.0em}
\end{figure}

\begin{figure}[!t]
\vspace{-1.0em}
\centering
\includegraphics[width=\linewidth]{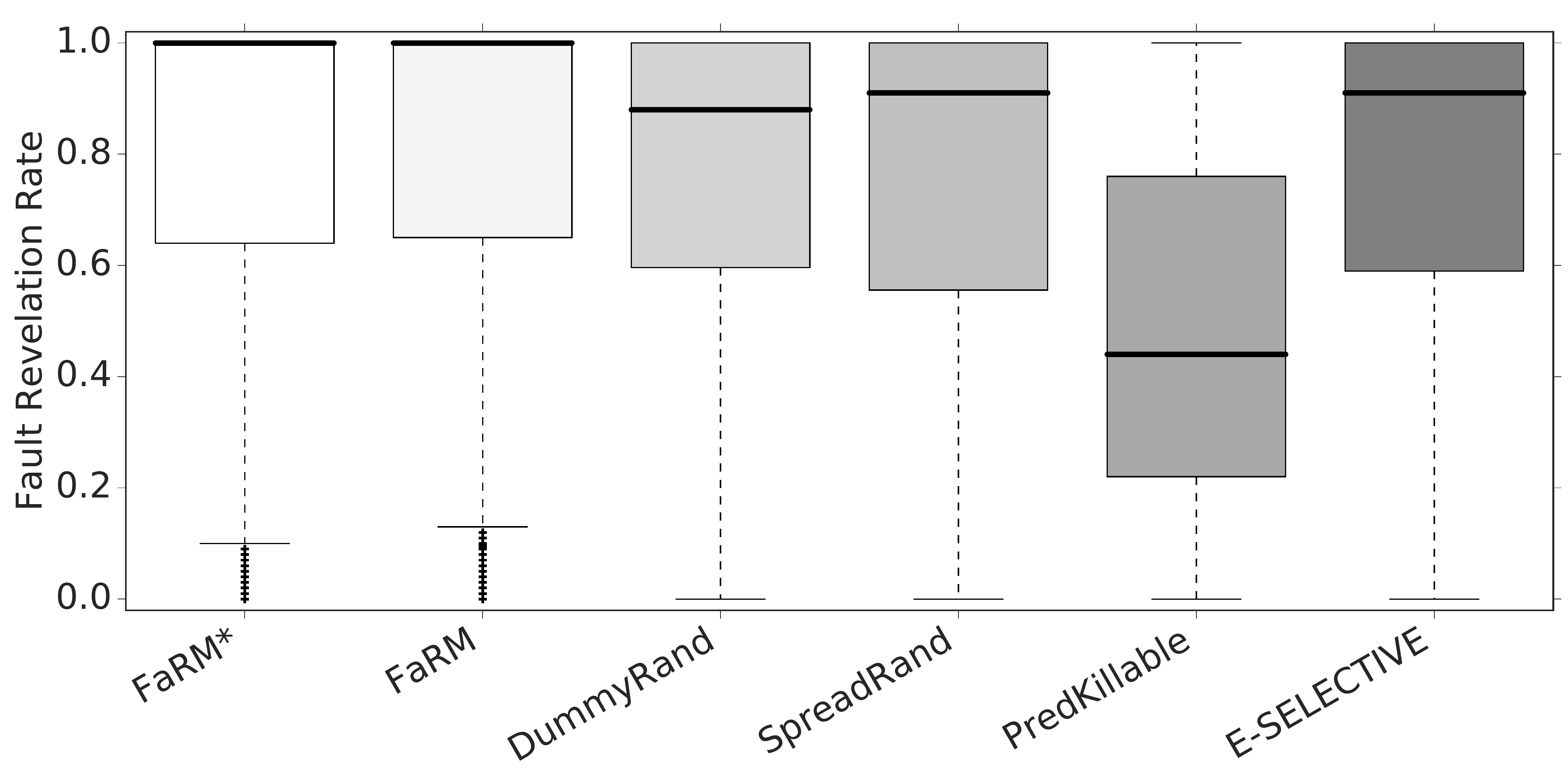}
\caption{Fault revelation of \toolname compared with E-Selective for selection size 15\% of all mutants. \toolname sets  outperform E-Selective selection. }
\label{fig:farm_vs_eselective_15percent}
\vspace{-1.0em}
\end{figure}

%\subsubsection{Comparison \toolname with E-Selective}
Similar to the experiment performed above to compare our approach with SDL, we perform another experiment to compare \toolname with E-Selective selection. The fault revelation results are presented in Figure~\ref{fig:farm_vs_eselective}. %We can observe that \toolname has 18.5\% higher median fault revelation than E-Selective. We also observe that E-Selective has similar fault revelation with the random selections (respectively 2\% and 0.4\% higher than DummyRandom and SpreadRandom).
We observe that for a selection size equal to the number of E-Selective mutants, all selection approaches except \toolnameC and DummyRandom achieve the highest median fault revelation. Given that E-Selective mutants are roughly 38\% of all the mutants, which is relative large set, we make the comparison with the E-Selective set for smaller selection size namely 5\% and 15\% thresholds of the top ranked mutants (w.r.t all mutants). The E-Selective mutants of the given sizes are randomly selected from the whole E-Selective mutant set.
The fault revelation results are presented in Figures~\ref{fig:farm_vs_eselective_5percent} and \ref{fig:farm_vs_eselective_15percent}. We can observe that \toolname and \toolnameB respectively have 31\% and 22\% higher median fault revelation than E-Selective for thresholds 5\%. For the 15\% threshold, both have 9\% higher median fault revelation. \toolnameC has 38\% and 47\% lower median fault revelation than E-Selective for thresholds 5\% and 15\% respectively. We also observe that E-Selective has similar fault revelation with the random selections (respectively 2\% and 1\% higher than DummyRandom and SpreadRandom for selection size 5\% and respectively 3\% and 0\% higher than DummyRandom and SpreadRandom for selection size 15\% ).
 
The Wilcoxon rank-sum statistical test shows that both \toolname and \toolnameB outperform E-Selective, and E-Selective outperforms the \toolnameC. The difference between E-Selective and the random approaches is not statistical significant. We also computed the Vargha Delaney $\hat{A}_{12}$ effect size value between the approaches and found that for the 5\% and 15\% thresholds, \toolname is better than E-Selective in 64\% and 63\% of the cases respectively. \toolnameB is better in 62\% and 61\% of the cases respectively, and \toolnameC is worse in 86\% and 82\% of the cases respectively. 

%We observe that the effect size value between \toolname and E-Selective (67\% with respect to $\hat{A}_{12}$) is larger than the value between \toolname and SDL (58\% with respect to $\hat{A}_{12}$). The difference is smaller with SDL because very few mutants (2\%) are considered. This makes the benefits of \toolname  smaller. 

\subsection{Mutant prioritization}

\subsubsection{Comparison with Random (RQ5)}

\noindent 
{\bf \em Selected Mutants Cost Metric.}

\begin{figure}[!t]
\vspace{-1.0em}
	\centering
	\includegraphics[width=\linewidth]{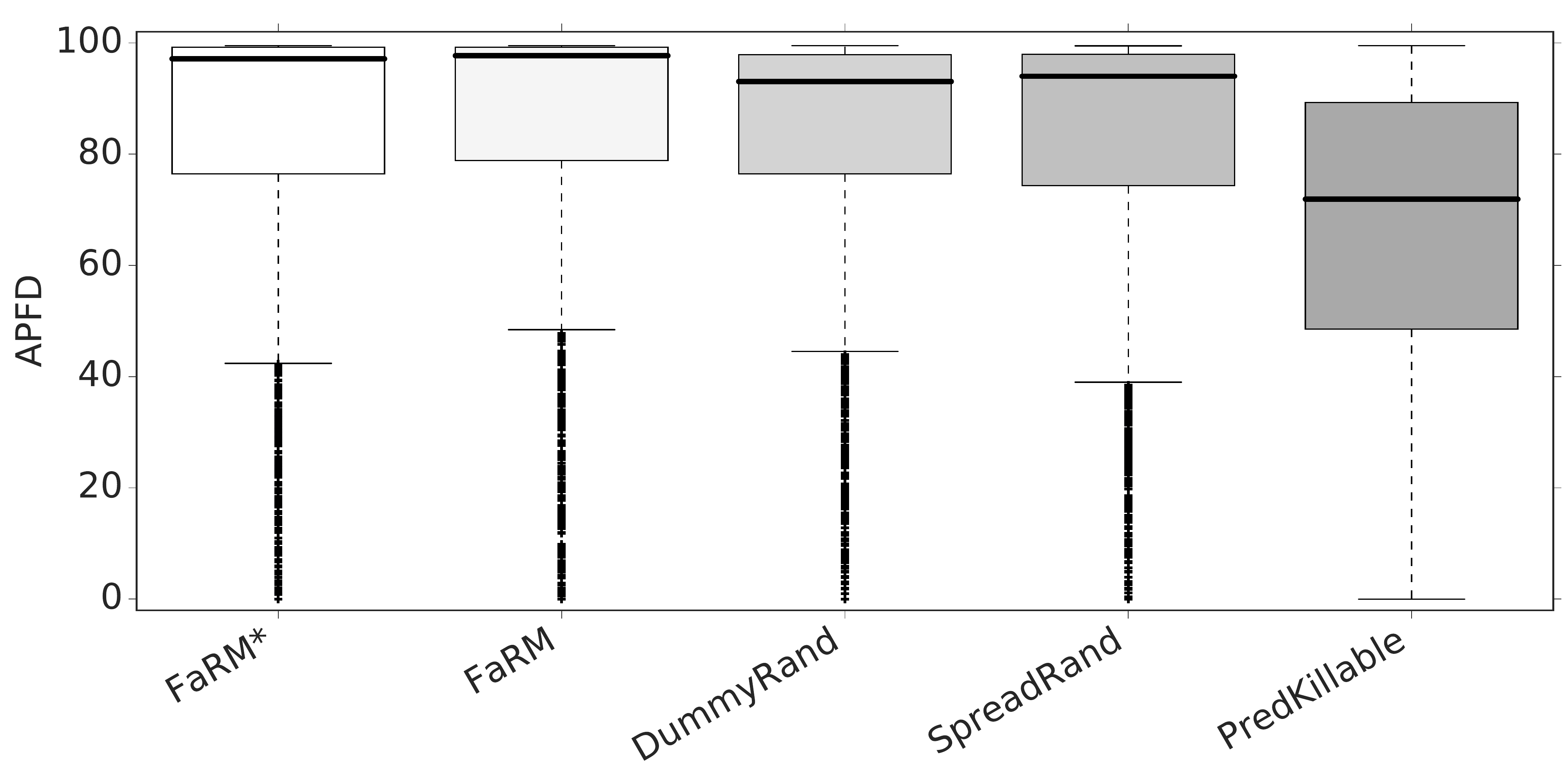}
	\caption{APFD measurements considering all mutants for the selected mutants cost metric for Codeflaws. The \toolname prioritization outperform the random baselines.}
	\label{fig:rq3-apfd-all}
\end{figure}

Figure~\ref{fig:rq3-apfd-all} shows the distributions of APFD (Average Percentage of Faults Detected) values 
for all faults, using the five approaches under evaluation. While \toolname and \toolnameB respectively yield an APFD median of 98\% and 97\%, and \toolnameC yields an APFD median of 72\%, DummyRandom and SpreadRandom reach median APFD values of 93\% and 94\% respectively. These results reveal that the general trend is in favour to our approach. As our approaches \toolname and \toolnameB are better than the random baseline, when the main cost factor (number of mutants that need analysis) is aligned, we can infer that it is generally better with practically important differences (of 4\%). Note that the highest possible improvement over the random baseline is 6\% (DummyRandom has a median APFD value of 94\%). Nonetheless, \toolnameC is worse than the random baseline.

To account for the stochastic nature of the compared methods and increase the confidence on our results, we further perform a statistical test. Wilcoxon test results yielded p-values much lower than our significance level for the compared data, i.e., samples of \toolname and DummyRandom, \toolname and SpreadRandom, \toolnameB and DummyRandom, \toolnameB and SpreadRandom, \toolnameC and DummyRandom, and  \toolnameC and SpreadRandom respectively.  Therefore, we can definitively conclude that \toolname and \toolnameB outperform random mutant selection with statistically significance while random mutant selection outperform \toolnameC. On the other hand, as expected,  the Wilcoxon test revealed that there is no statistical difference between the performance of DummyRandom and that of SpreadRandom.

When examining mutant selection strategies there are two main parameters that influence the application cost. These are the killable and equivalent mutants that testers need to analyse. When analysing a killable mutant our ability to select fault revealing ones is important, while increasing the chance to get a killable mutant is also important. Therefore, it could be that our \toolname is better simply because it selects killable mutants and not fault revealing ones. 
To account for this factor we removed all non-killable mutants from our sets and recompute our results. This helped eliminating the influence of non-killable mutants, from both approaches. 

\begin{figure}[!t]
\vspace{-1.0em}
	\centering
	\includegraphics[width=\linewidth]{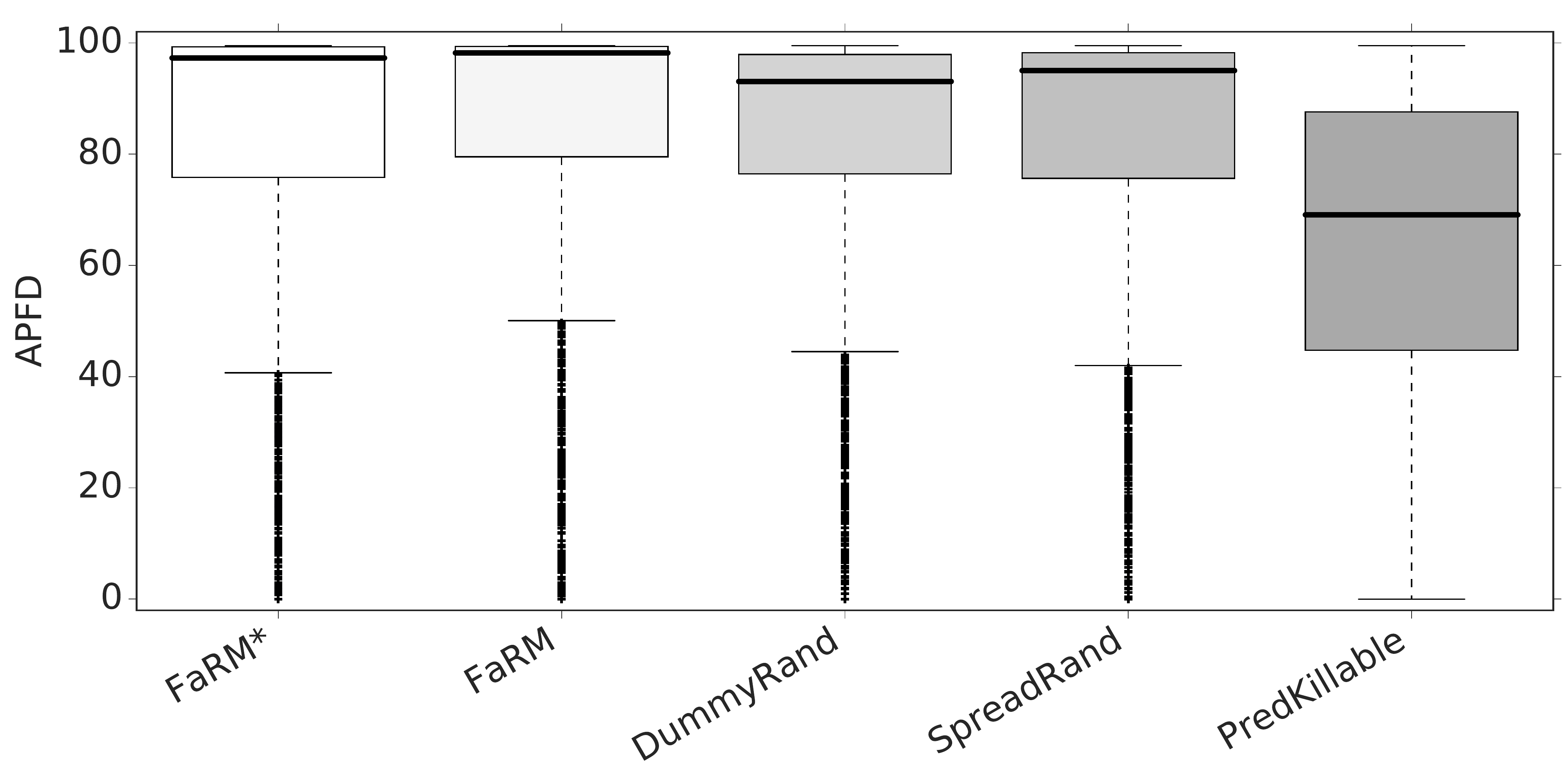}
	\caption{APFD measurements considering only killable mutants for the selected mutants cost metric on Codeflaws. The \toolname prioritization  outperform the random baselines, independent of non-killable mutants.} 
	\label{fig:rq3-apfd-killed}
\end{figure}

Our results show that the performance improvement of \toolname and \toolnameB over SpreadRandom and DummyRandom is also effective when considering only killable mutants (approximated by our test suites). Figure~\ref{fig:rq3-apfd-killed} 
shows the relevant distributions of APFD, which are visibly similar to the distributions for all mutants (all values are slightly higher when considering only killable mutants). This result suggest that \toolname and \toolname* are indeed capable of identifying fault revealing mutants, independent of the equivalent mutants involved. 

\begin{figure}[!t]
\vspace{-1.0em}
	\centering
	\includegraphics[width=0.9\linewidth]{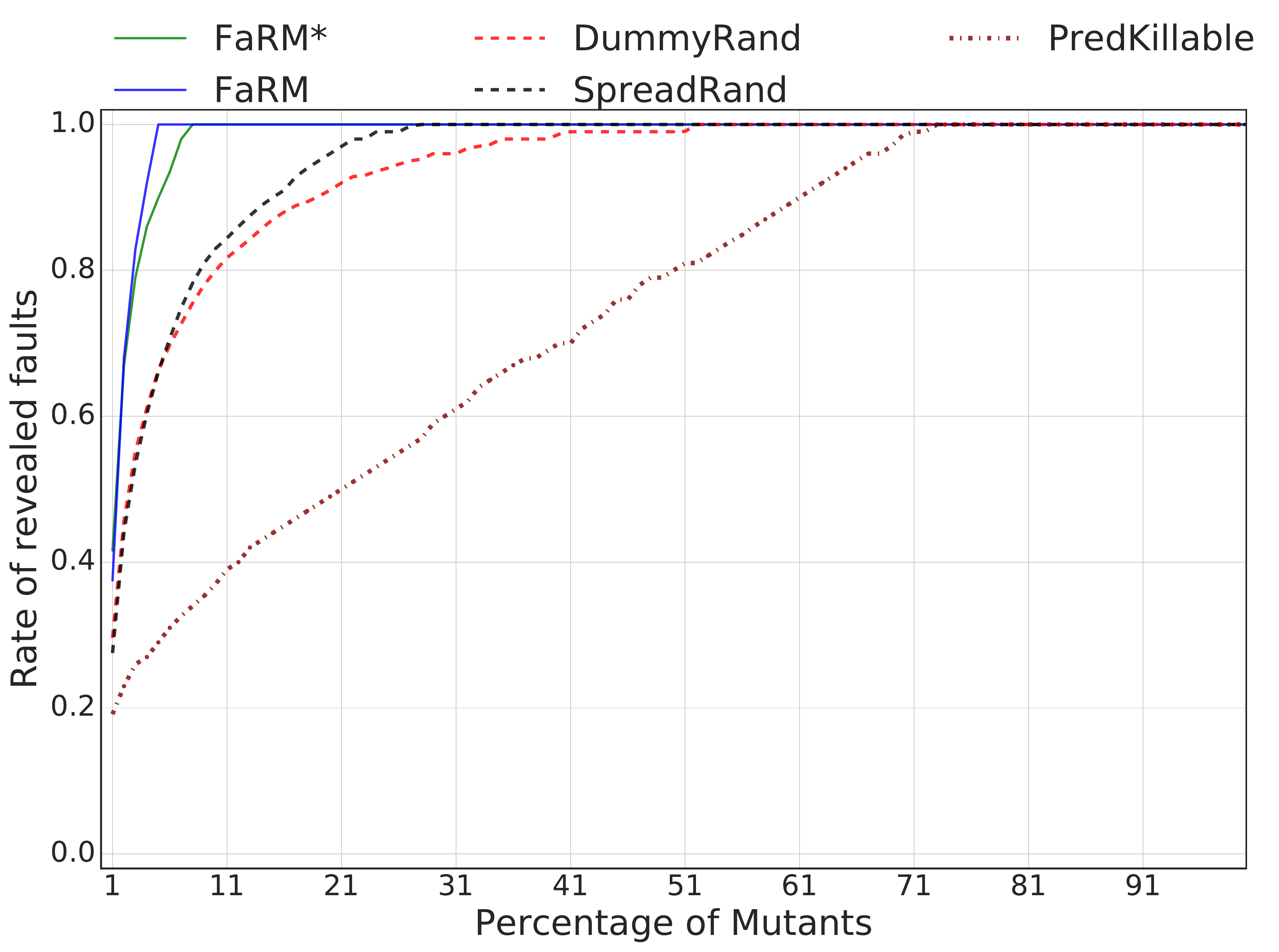}
	\caption{Mutant prioritization performance in terms of faults revealed (median case) for the selected mutants cost metric on CodeFlaws. The x-axis represent the number of considered mutants. The y-axis represent the ratio of the fault revealed by the strategies. }
	\label{fig:rq3-all}
\end{figure}

To provide a general view of the trends, Figure~\ref{fig:rq3-all} illustrates the overall (median) effectiveness of the mutant prioritization by \toolname, \toolnameB and \toolnameC in comparison with random strategies. We note that for
all percentages of mutants, \toolname and \toolnameB outperforms random-based prioritization while \toolnameC is outperformed by the random-based prioritization. Overall, we observe that the fault revelation benefit of \toolname over the random approaches is above 20\% (maximum difference is 34\%) when selecting 2\% to 8\% of mutants. \toolname reaches a plateau around 5\% of mutants, as the median fault revelation is maximal. This suggests that a hint for the mutant selection size for \toolname could be 5\% of the mutants.

%The performance improvement goes around 10\% to 15\% of more faults revealed   when 30\% until 50 \% of mutants are executed.

\begin{figure}[!t]
	\centering
	\begin{subfigure}[b]{0.8\linewidth}
		\includegraphics[width=\linewidth]{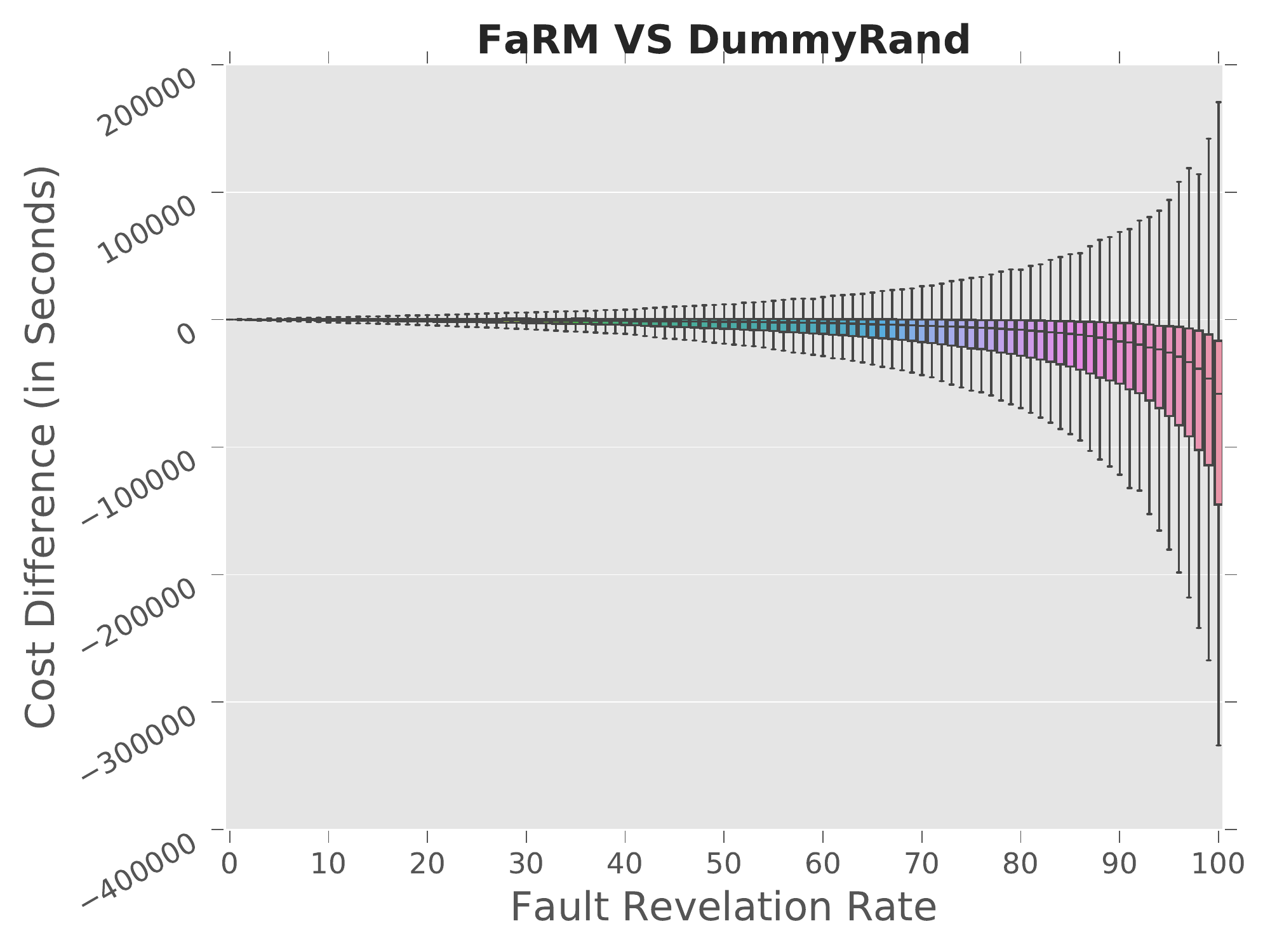}
		\caption{\normalfont Cost of \toolname - Cost of DummyRand.}
		\label{fig:costDummy}
	\end{subfigure}\\%
    \begin{subfigure}[b]{0.8\linewidth}
		\includegraphics[width=\linewidth]{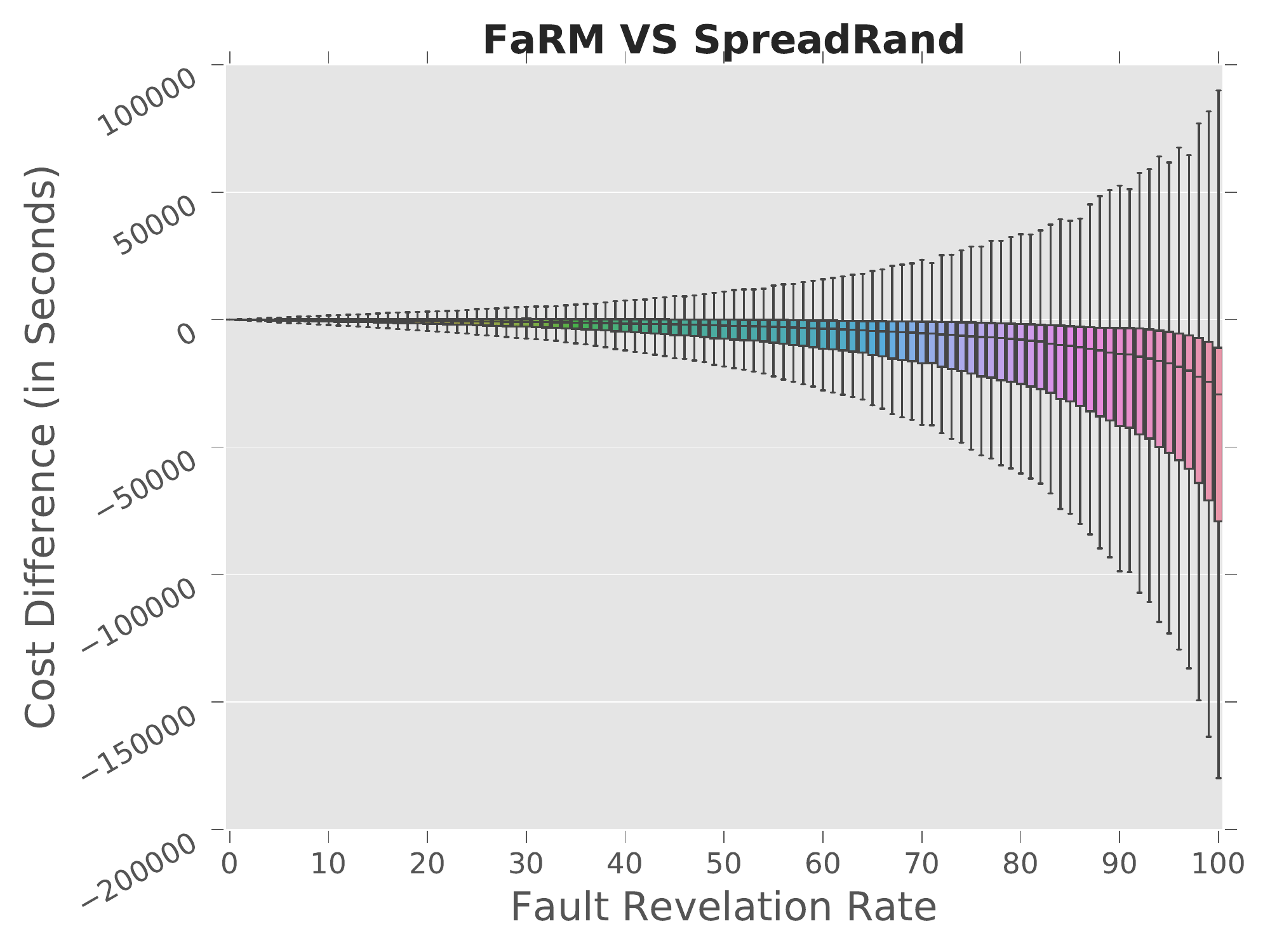}
		\caption{\normalfont Cost of \toolname - Cost of SpreadRand}
		\label{fig:costSpread}
	\end{subfigure}
	\caption{Execution cost of prioritization schemes}
 	\label{fig:rq3-cost}
\end{figure}

Finally, we examined the differences between the approaches in terms of execution time. Although, we do not explicitly aim at reducing the test execution cost, 
we expect some benefits due to our methods' ability to prioritise the mutants, which results in a reduced execution time \cite{ZhangMK13}. 
Figure~\ref{fig:rq3-cost} illustrate, in a box-plot form, the overall execution time differences between the \toolname and the random baselines with respect to the attained fault revelation, measured in seconds. 
Although the differences can be significant in some (rare) cases, the expected (median values) ones are -58,167 and -29,373 seconds (-16 and -8 hours) for DummyRandom and SpreadRandom.  %This is the cost difference when both approaches attain 100% fault revelation.
This result indicates that our approach has also an advantage with respect to test execution, which sometimes becomes significant.  

Conclusively, our results  demonstrate that \toolname is indeed effective as it is statistically superior to random baselines, independent of the equivalent mutants involved. It provides 4\% higher APFD values, which means that when testers analyse mutants (to strengthen their test suites) they get a 4\% improvement on their fault revelation ability. Note that the highest possible improvement over the random baseline is 6\% (DummyRandom has a median APFD value of 94\%).

%%%%%%%%%

\noindent 
{\bf \em Required Tests Cost Metric.}

\begin{figure}[!t]
\vspace{-1.0em}
	\centering
	\includegraphics[width=\linewidth]{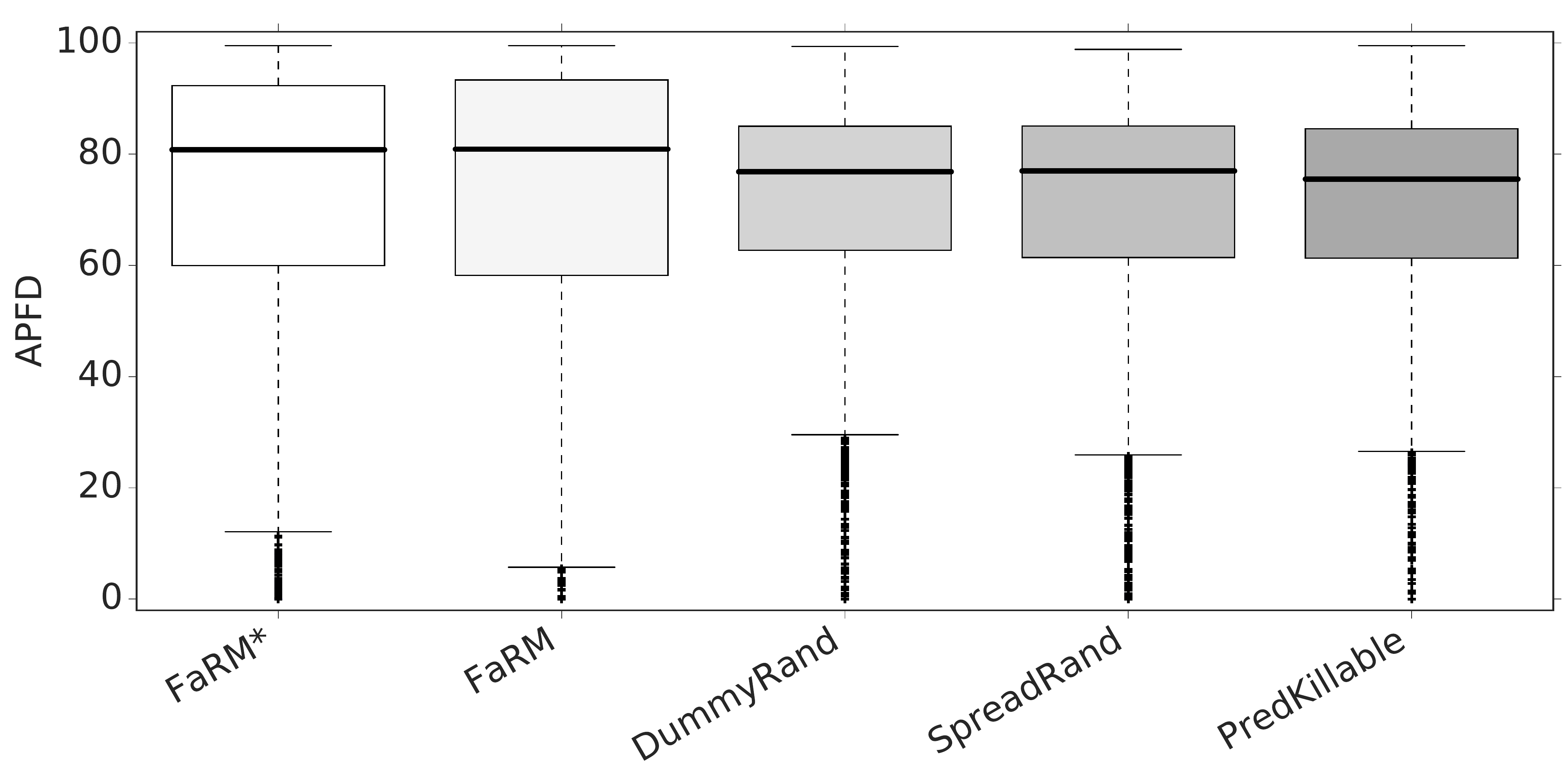}
	\caption{APFD measurements for the required tests cost metric on Codeflaws. The \toolname prioritization outperform the random baselines.}
	\label{fig:requiredtests-apfd}
\end{figure}

Figure~\ref{fig:requiredtests-apfd} shows the distributions of APFD (Average Percentage of Faults Detected) values 
for all faults, using the five approaches under evaluation. While both \toolname and \toolnameB yield an APFD median of 81\%, and \toolnameC yields an APFD median of 76\%, DummyRandom and SpreadRandom reach median APFD values of 77\%. These results reveal that the general trend is in favour to our approach. As our approaches \toolname and \toolnameB are better than the random baseline, when the main cost factor (number of test that need to be designed and executed) is aligned, we can infer that it is generally better with practically important differences (of 4\%). The \toolnameC performs quite similarly to the random baseline.

To account for the stochastic nature of the compared methods and increase the confidence on our results, we further perform a statistical test. Wilcoxon test results yielded p-values much lower than our significance level for the compared data, i.e., each of \toolname and \toolnameB compared with each of \toolnameC, DummyRandom and SpreadRandom. Therefore, we can definitively conclude that \toolname and \toolnameB outperform random baseline with statistically significance. On the other hand, the Wilcoxon test revealed that there is no statistical difference between the performance of \toolnameC, DummyRandom and SpreadRandom.

The results of the Vargha Delaney effect size show that \toolname is better than DummyRandom, SpreadRandom and \toolnameC in 58\%, 61\% and 60\% of the cases respectively. \toolnameB is better than DummyRandom, SpreadRandom and \toolnameC in 58\%, 61\% and 59\% of the cases respectively.

\begin{figure}[!t]
\vspace{-1.0em}
	\centering
	\includegraphics[width=0.9\linewidth]{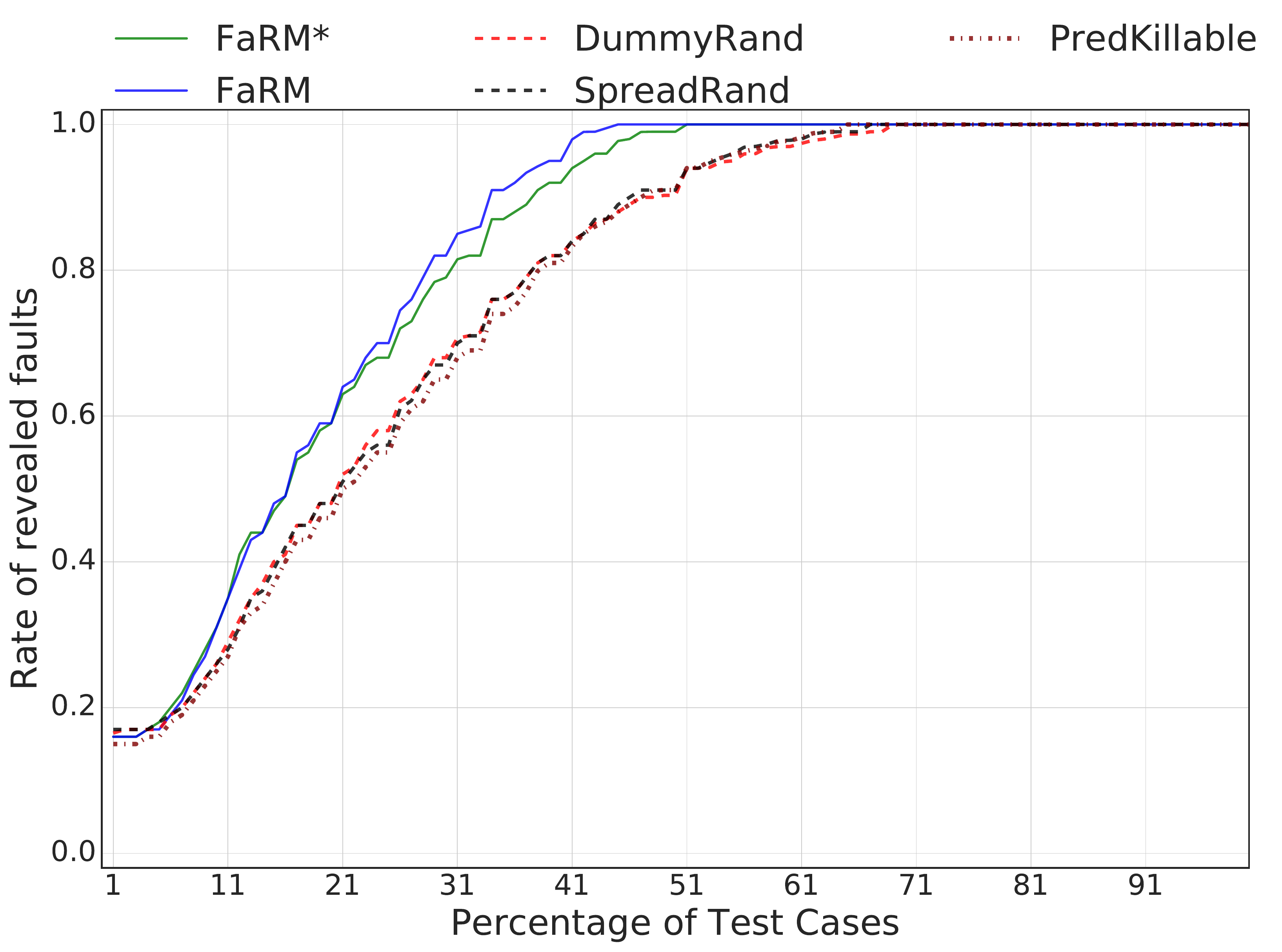}
	\caption{Required tests prioritization performance in terms of faults revealed (median case) on CodeFlaws. The x-axis represent the number of considered tests. The y-axis represent the ratio of the fault revealed by the strategies. }
	\label{fig:requiredtests-median}
\end{figure}

To provide a general view of the trends, Figure~\ref{fig:requiredtests-median} illustrates the overall (median) effectiveness of the required test prioritization by \toolname, \toolnameB and \toolnameC in comparison with random strategies. We note that for all percentages of tests, \toolname and \toolnameB outperforms random-based prioritization while \toolnameC is outperformed by the random-based prioritization. Overall, we observe that the fault revelation benefit of \toolname over the random approaches is above 10\% (maximum difference is 15\%) for the 20\% to 45\% top ranked tests.

%%%%%%%%%

\noindent 
{\bf \em Analysed Mutants Cost Metric.}

The analysed mutants cost metric measures the minimum number of mutants that need to be analysed, including equivalent mutants, following a mutant prioritization approach, before the fault is revealed.
A good mutant prioritization approach will minimized the analysed mutants cost. Following, we compare the analysed mutants cost metric between our approaches and the random baselines. The analysed mutants cost metric is calculated for each approach and for each bug of the benchmark. We compare the approaches statistically with Wilcoxon rank-sum test and the Vargha Delaney effect size.

The results show that \toolname, \toolnameB and \toolnameC are better than DummyRandom and SpreadRandom with statistical significance displayed by a p-value much lower than the significance level. \toolname is better than DummyRandom and SpreadRandom in 57\% and 61\% of the cases respectively. The performance difference is higher for \toolnameB where it is better than DummyRandom and SpreadRandom in 60\% and 64\% of the cases respectively. \toolnameC is better than DummyRandom and SpreadRandom in 60\% and 65\% of the cases respectively.

\toolnameB shows a larger improvement than \toolname over the random baseline, but there is no statistical significance difference between \toolname and \toolnameB. Furthermore, \toolnameB outperforms \toolnameC with statistical significant difference, and is better in 53\% of the cases. There is no statistical significant difference between \toolname and \toolnameC.

Conclusively, our results  demonstrate that \toolname and \toolnameB are indeed effective as they are statistically superior to random baselines.

%%%%%%%%%%%%%%%
\subsubsection{Comparison with Defect Prediction (RQ6)} 
\begin{figure}[!t]
\vspace{-1.0em}
	\centering
	\includegraphics[width=\linewidth]{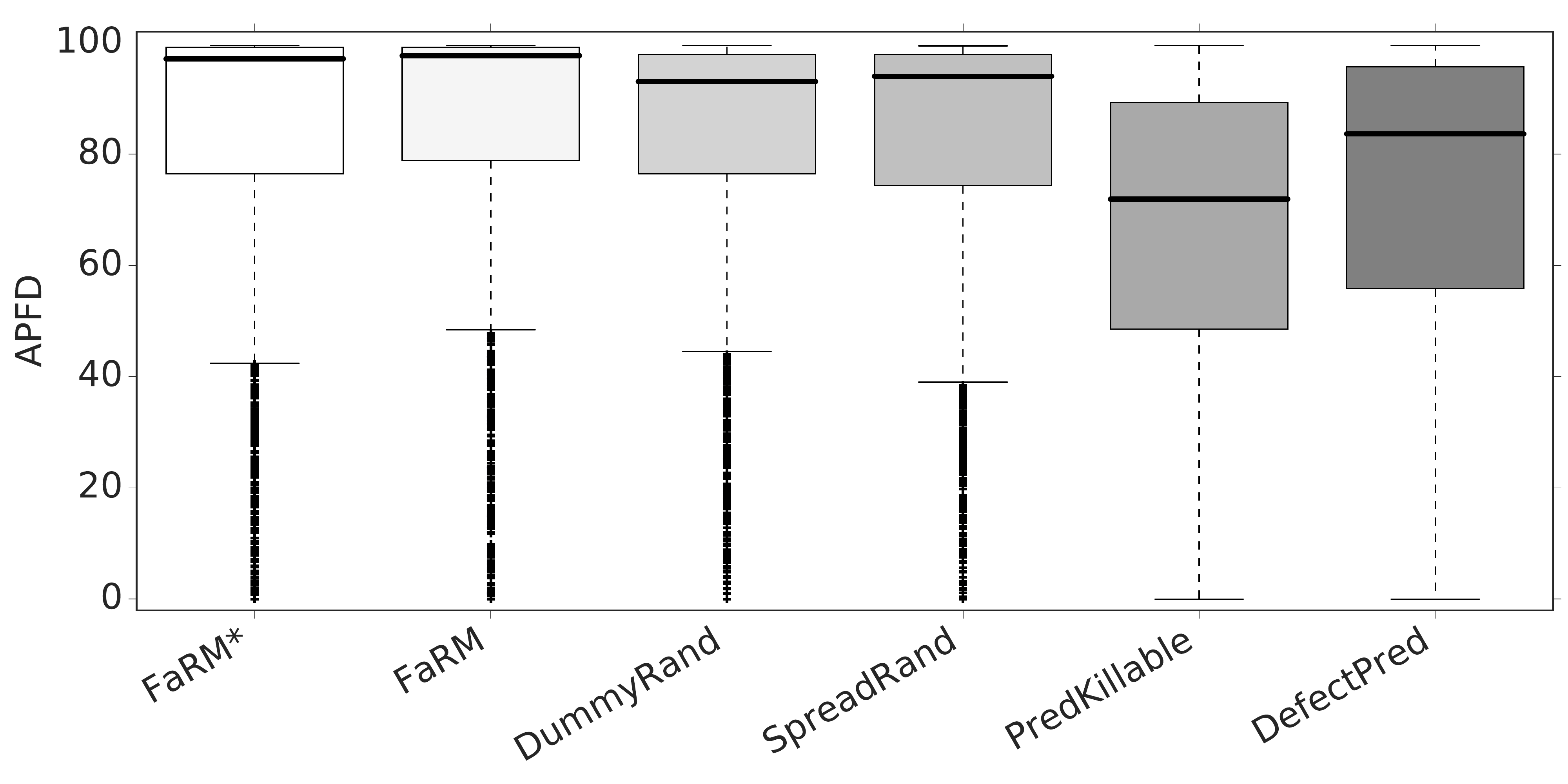}
	\caption{APFD measurements considering all mutants. The \toolname prioritization outperform the defect prediction.}
	\label{fig:farm_vs_defectpred_APFD}
\end{figure}

\noindent 
{\bf \em Selected Mutants Cost Metric.}
Figure~\ref{fig:farm_vs_defectpred_APFD} shows the distributions of APFD (Average Percentage of Faults Detected) values for all faults, using the \toolname, \toolnameB, \toolnameC and the random approaches. While \toolname yields an APFD median of 98.0\%, defect prediction (DefectPred) reach median APFD value of 83.7\%. These results reveal that the general trend is in favour to our approach. As our approach is much better than the defect prediction approach, when the main cost factor (number of mutants that need analysis) is aligned, we can infer that it is generally better with practically important differences (of 14\%). Even the random approaches  are better than the defect prediction approach. Nevertheless, \toolnameC is worse than defect prediction.

The Wilcoxon test results yielded p-values much lower than our significance level for the samples of  \toolname and DefectPred, and \toolnameB and DefectPred. Therefore, we can definitively conclude that \toolname and \toolnameB outperforms defect prediction with statistical significance. On the other hand, the Wilcoxon test also revealed that there is statistical significant difference between the performance of DefectPred and dummyRandom, and DefectPred and that of spreadRandom respectively. Nonetheless, DefectPred outperforms \toolnameC with statistical significance. 
The Vargha Delaney $\hat{A}_{12}$ effect size value shows that \toolname and \toolnameB are better than DefectPred in 76\% of cases. While DummyRandom and SpreadRandom are better than DefectPred in 71\% and 70\% of the cases respectively.

\begin{figure}[!t]
\vspace{-1.0em}
	\centering
	\includegraphics[width=0.9\linewidth]{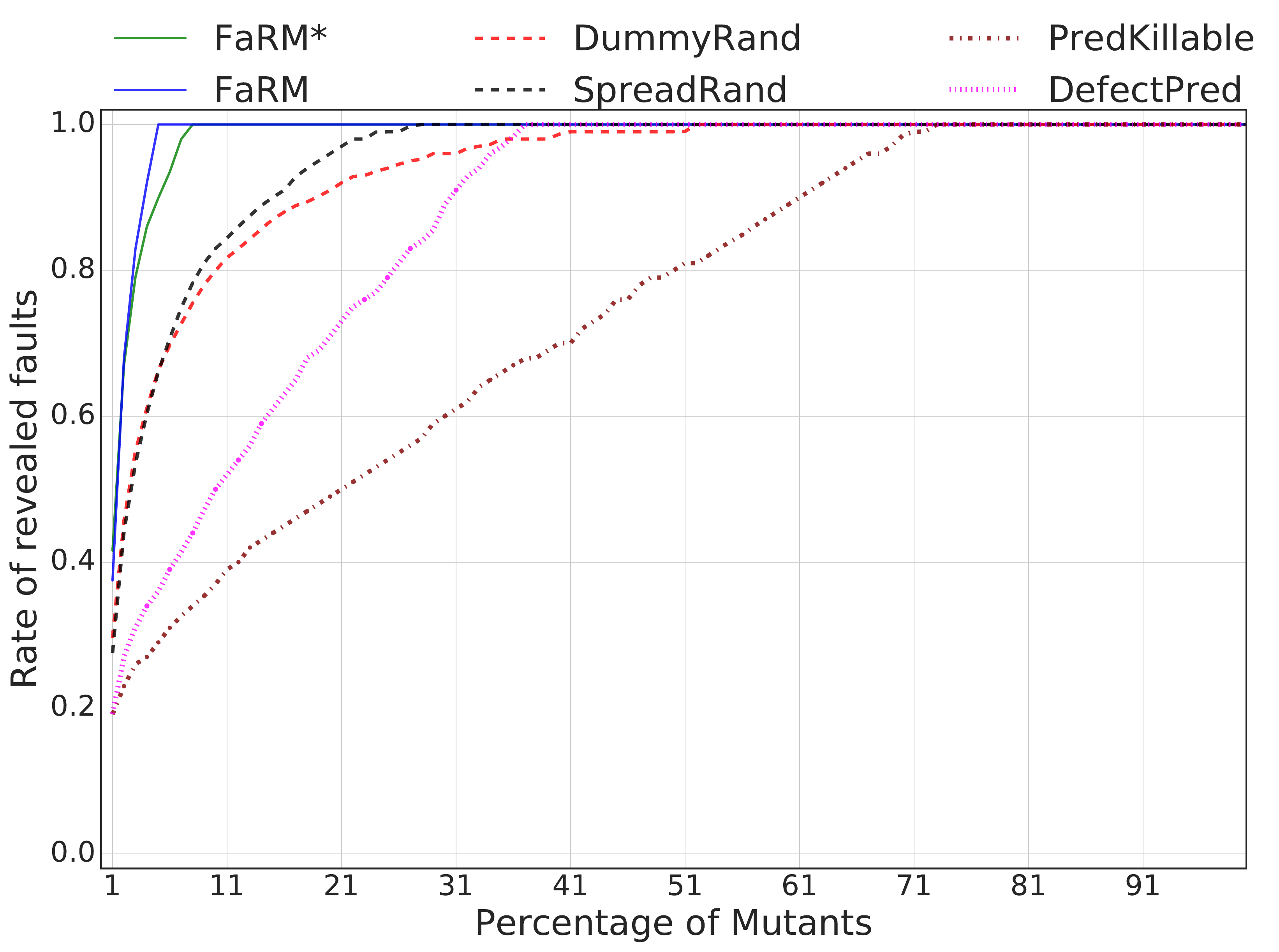}
	\caption{Mutant prioritization performance in terms of faults revealed (median case) on CodeFlaws. The x-axis represent the number of considered mutants. The y-axis represent the ratio of the fault revealed by the strategies. }
	\label{fig:farm_vs_defectpred_all}
\end{figure}

To provide a general view of the trends, Figure~\ref{fig:farm_vs_defectpred_all} illustrates the overall (median) effectiveness of the mutant prioritization by \toolname in comparison with the defect prediction approach. We note that for all percentage of mutants, \toolname outperforms the defect prediction approach. The performance improvement goes around 40\% to 66\% of more faults revealed when 2\% until 8\% of mutants are executed.

%
%

%%%%%%%%%%%%%%%%%%%%%%%%%%%%Needs change!!!

%
%
%\begin{figure}[!h]
%	\centering
%	\includegraphics[width=\linewidth]{graphs/rq3/KILLED-ONLY.png}
%	\caption{Prioritization effectiveness / only killed mutants}
%	\label{fig:rq3-all}
%\end{figure}
%
%
%\subsubsection{Execution cost}
%
%\begin{figure*}[!t]
%	\centering
%	\includegraphics[width=\linewidth]{graphs/rq3/COSTDIFF-ALL.png}
%	\caption{Difference of execution cost with all mutants}
%	\label{fig:rq3-costALL}
%\end{figure*}
%
%
%\begin{figure*}[!t]
%	\centering
%	\includegraphics[width=\linewidth]{graphs/rq3/COSTDIFF-KILLED-ONLY.png}
%	\caption{Difference of execution cost with killed mutants only}
%	\label{fig:rq3-costALL}
%\end{figure*}
%

%\begin{figure}[!h]
%	\centering
%	\begin{subfigure}[b]{0.5\linewidth}
%	\includegraphics[width=\linewidth]{graphs/rq1/IG.png}
%	\caption{\normalfont fig1.}
%	\label{fig:s1}
%	\end{subfigure}%
%	\begin{subfigure}[b]{0.5\linewidth}
%	\includegraphics[width=\linewidth]{graphs/rq1/IG.png}
%	\caption{\normalfont fig2.}
%	\label{fig:s2}
%	\end{subfigure}
%	\caption{with subfig}
%\end{figure}

\subsection{Experiments with large programs (RQ7)}
\noindent 
{\bf \em Selected Mutants Cost Metric.}

\begin{figure}[!t]
\vspace{1.0em}
	\centering
	\includegraphics[width=\linewidth]{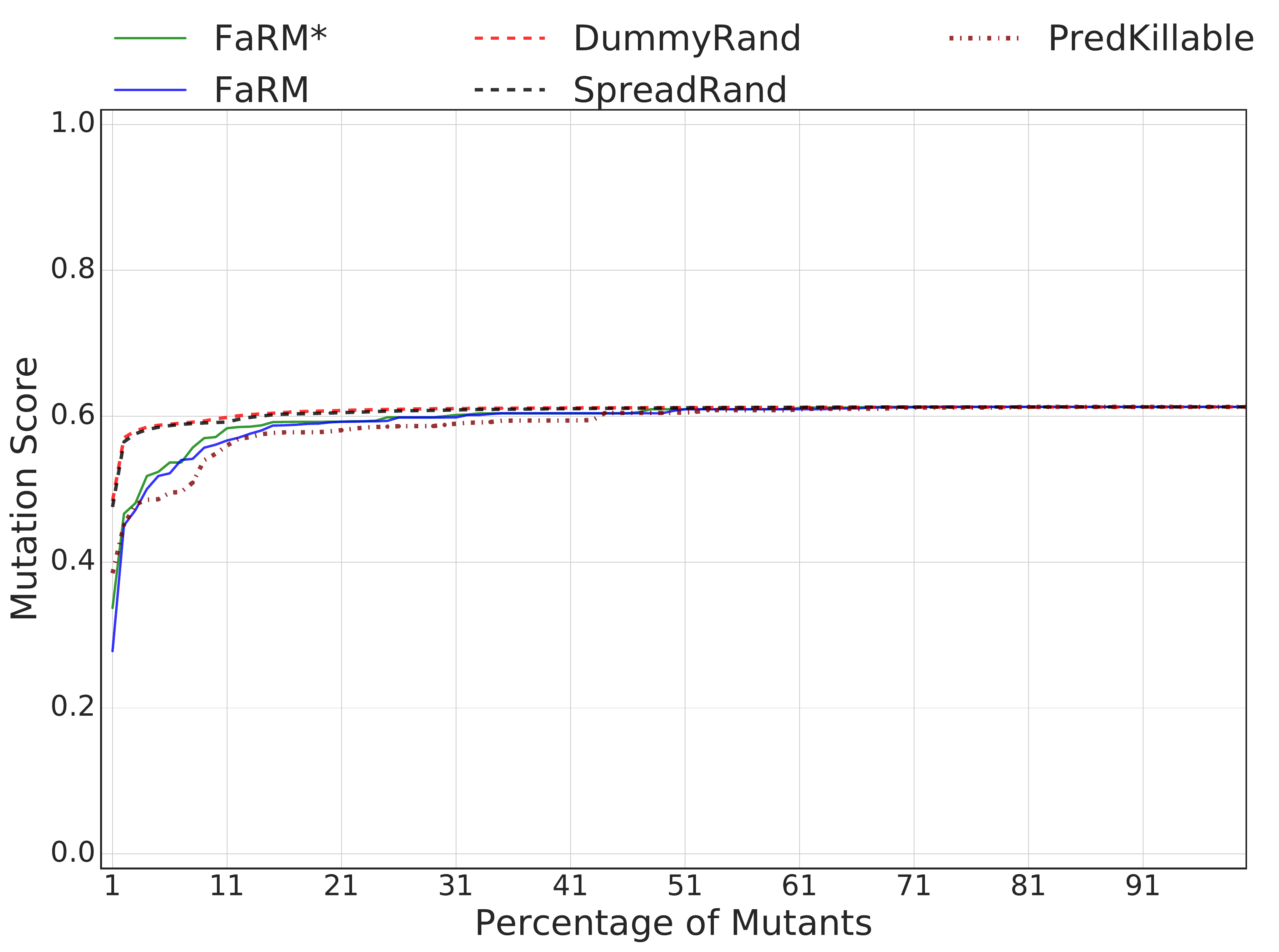}
	\caption{Mutation score (median case) on CoREBench. The x-axis represent the number of considered mutants, while the y-axis represent the mutation score attained by the strategies.}
	\label{fig:rq4-MedianMS}
\end{figure}

%Figure~\ref{fig:rq4APFD} depicts the distributions of the APFD values for the considered CoREBench faults. These data demonstrate %it is obvious that \toolname yields much higher APFD values than the two random baselines. 
%a trend favouring  our approach (\toolname has an APFD median of 93\%, dummyRandom and spreadRandom have median APFD values of 91\% and 88\%). 

%Interestingly, 
in CoREBench, all APFDs values are much higher than in CodeFlaws, with \toolname, \toolnameB, DummyRandom and SpreadRandom having median APFD value of 99\%, and \toolnameC a median APFD value of 94\%.  The maximum possible improvement is 1\% (given that the random baseline has a median of 99\%). This is caused by the large number of redundant mutants involved. To demonstrate this we check the relation between mutation score and percentage of considered mutants.  Figure~\ref{fig:rq4-MedianMS} illustrates the overall (median) mutation score achieved (y-axis) by the tests killing the percentage of mutants recorded in x-axis. From this graph we can see that all approaches reach their maximum median mutation score value when considering more than 30\% of the mutants. This implies that the benefits are reduced for every approach that consider more than 30\% of the involved mutants. 

\begin{figure}[!t]
\vspace{1.0em}
	\centering
	\includegraphics[width=\linewidth]{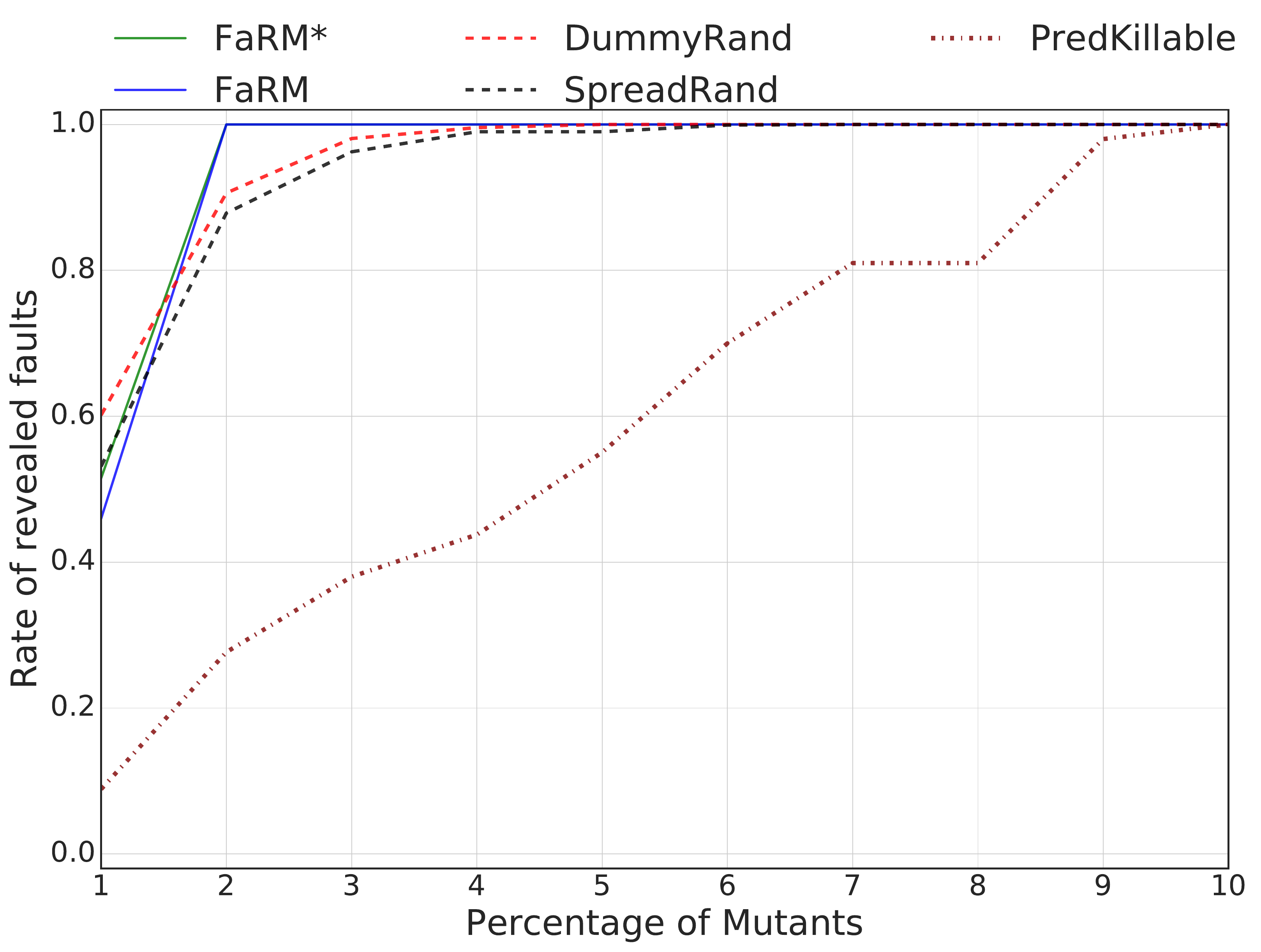}%{graphs/leaveOneOut/ALL-MEDIAN.pdf}
	\caption{\toolname performance in terms of faults revealed (median case) on CoREBench considering all mutants. The x-axis represent the number of considered mutants, while the y-axis represent the ratio of the fault revealed by the strategies.}
	\label{fig:rq4-Median}
\end{figure}

\begin{figure}[!t]
\vspace{1.0em}
	\centering
	\includegraphics[width=\linewidth]{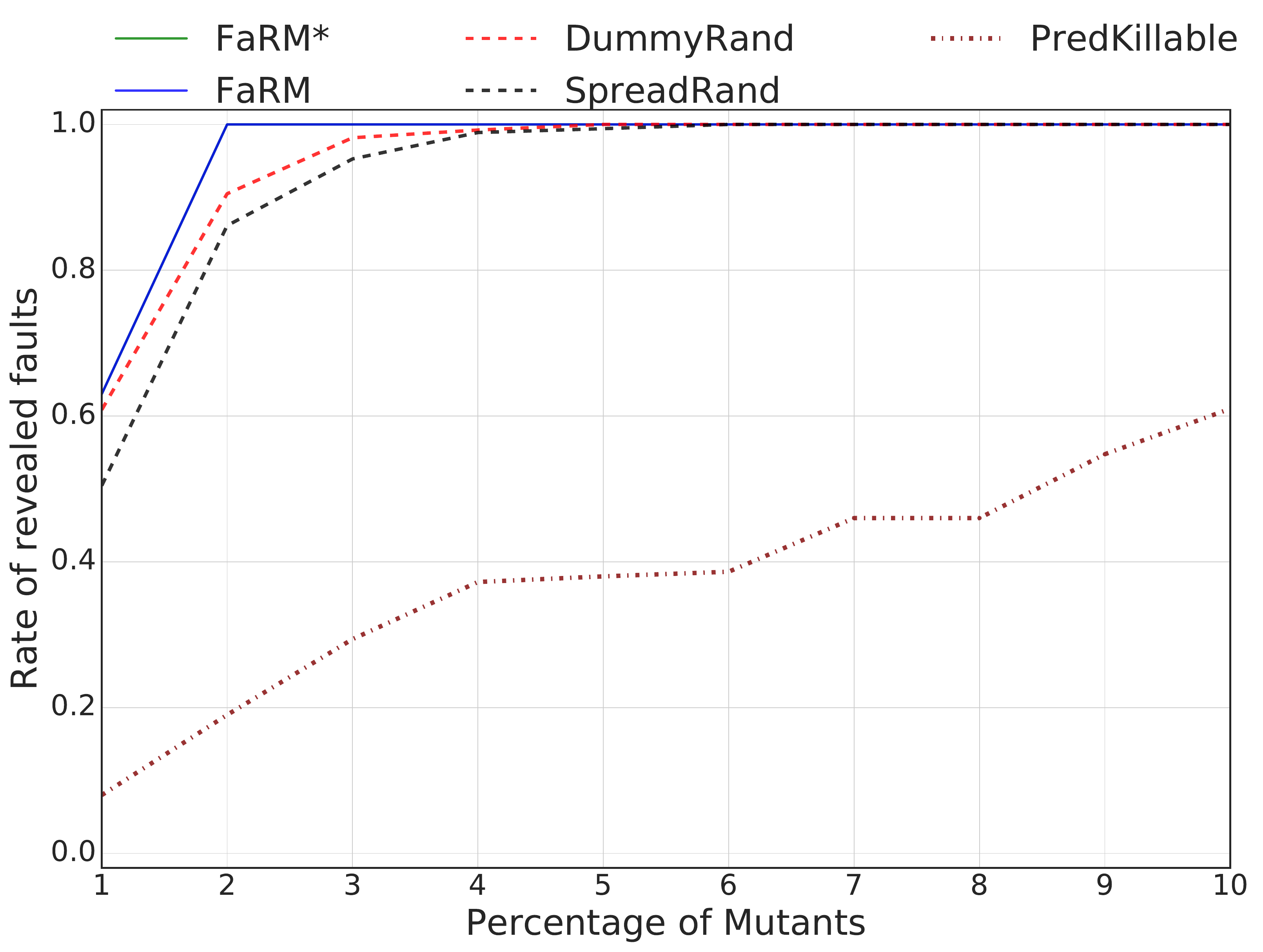}%{graphs/leaveOneOut/ALL-MEDIAN.pdf}
	\caption{\toolname performance in terms of faults revealed (median case) on CoREBench considering only killable mutants. The x-axis represent the number of considered mutants, while the y-axis represent the ratio of the fault revealed by the strategies.}
	\label{fig:rq4-Median-killed}
\end{figure}

\begin{figure}[!t]
\vspace{1.0em}
	\centering
	\includegraphics[width=\linewidth]{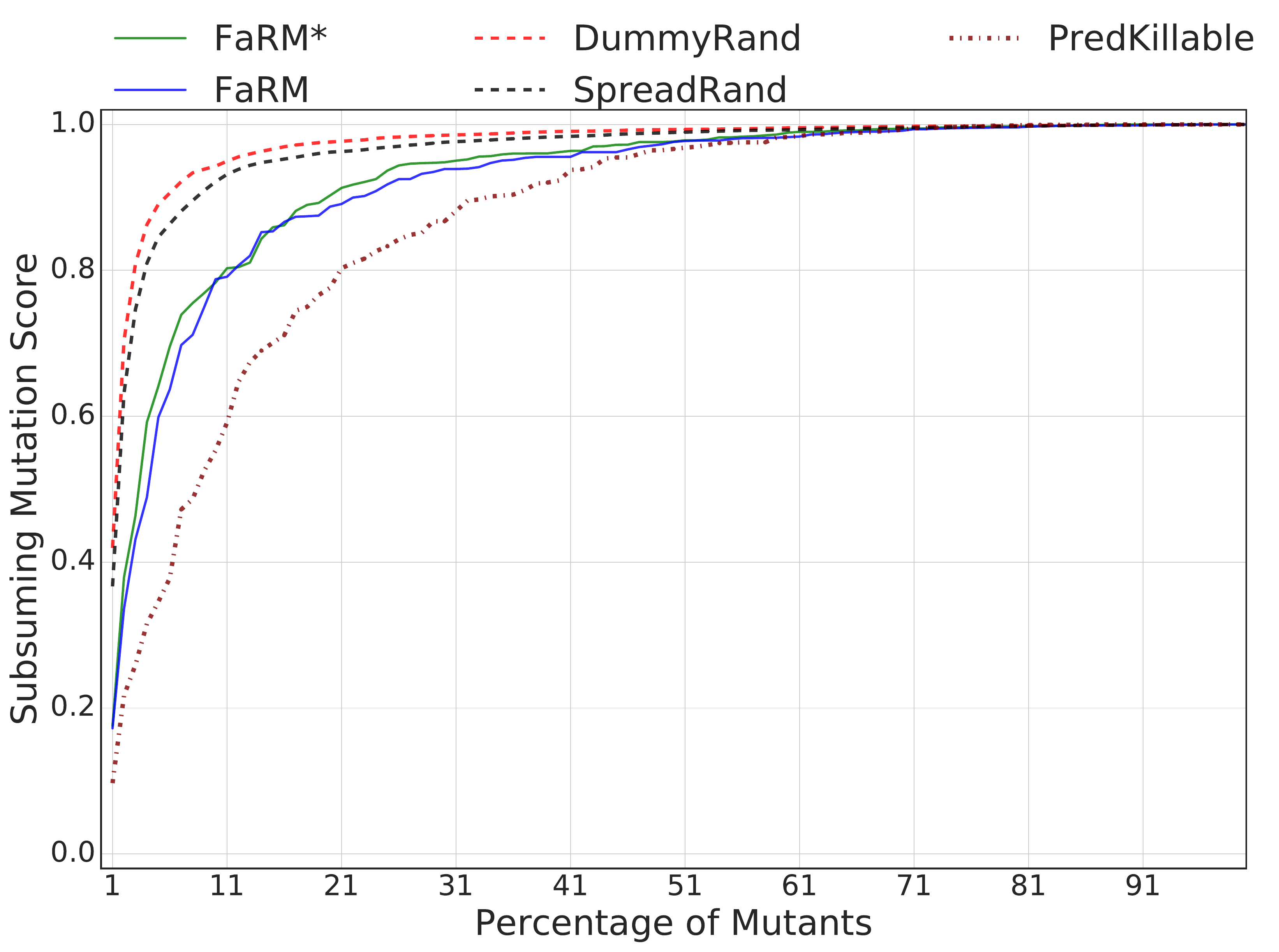}
	\caption{Subsuming Mutation score (median case) on CoREBench. The x-axis represent the number of considered mutants, while the y-axis represent the subsuming mutation score attained by the strategies.}
	\label{fig:rq4-MedianSubsumingMS}
\end{figure}

Interestingly, both Figure~\ref{fig:rq4-MedianMS} and Figure~\ref{fig:rq4-MedianSubsumingMS} demonstrate that \toolname guides the mutant selection towards mutants that do not maximize the mutation score nor the subsuming mutation score (random mutant selection achieves higher mutation and subsuming mutation scores than \toolname). Instead the selected mutant maximize fault revelation as demonstrated in Figures~\ref{fig:rq4-Median} and \ref{fig:rq4-Median-killed}.  

\begin{figure}[!t]
\vspace{1.0em}
	\centering
	\includegraphics[width=\linewidth]{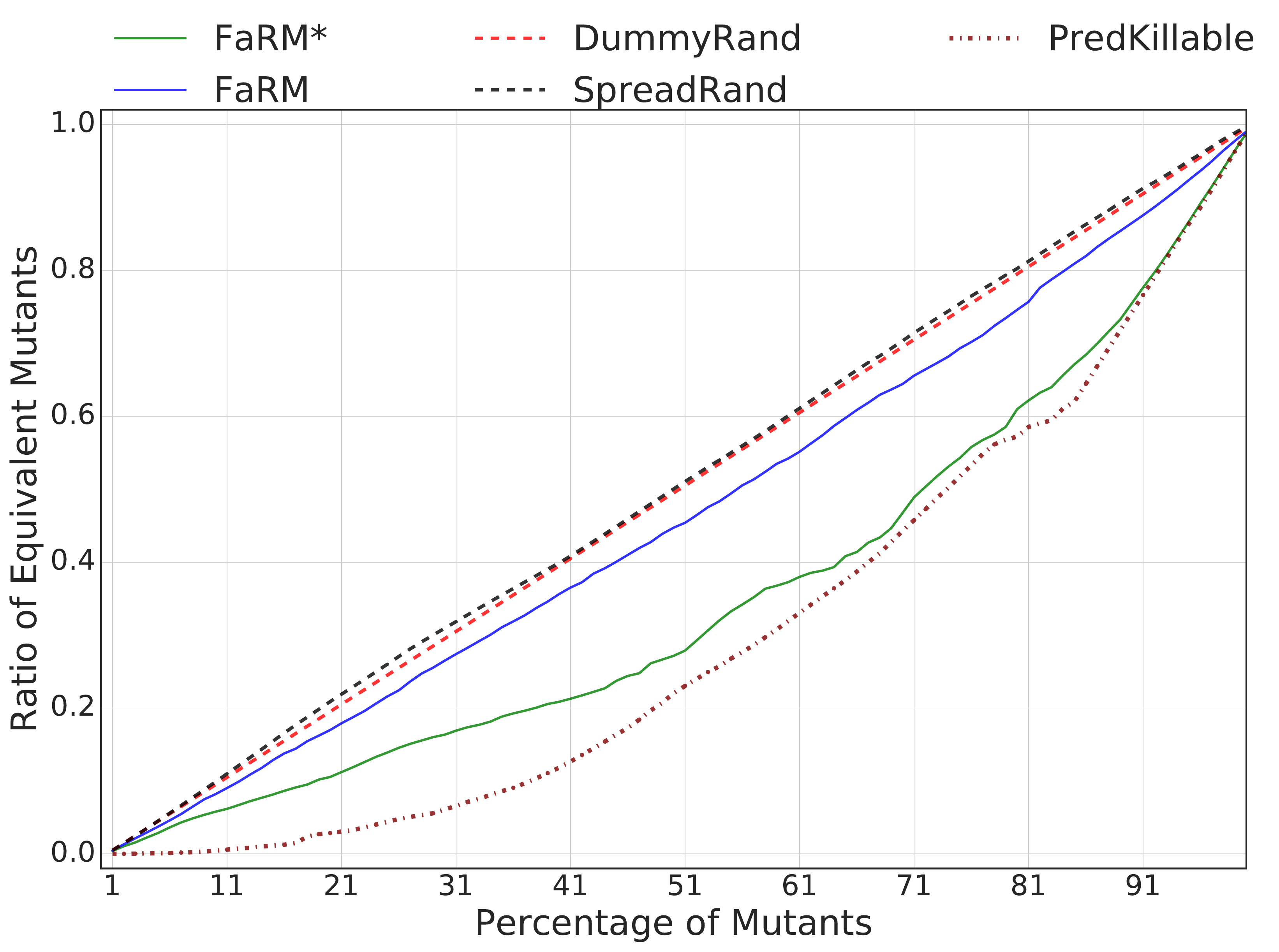}
	\caption{Ratio of equivalents (median case) on CoREBench. The x-axis represent the number of considered mutants, while the y-axis represent the proportion of equivalent mutants selected by the strategies.}
	\label{fig:numEquivalents}
\end{figure}

Given that a large proportion of the mutants are not killable (Figure~\ref{fig:rq4-MedianMS}), we present in Figure~\ref{fig:numEquivalents} the sensitivity of the approaches with regard to the equivalent mutants, to see how they are ranked.  We observe that \toolnameC does quite well at ranking the killable mutants first, and \toolnameB inherit of such characteristic from \toolnameB relatively well. We also observe that \toolname tend to keep equivalent mutants away from the top ranks.

To provide a general view of the fault revelation trend, Figures~\ref{fig:rq4-Median} and ~\ref{fig:rq4-Median-killed} illustrate the overall (median) effectiveness of the mutant prioritization by \toolname in comparison with random strategies for the ratios of selected mutants from 1\% to 10\%. We note that for all percentage of mutants, \toolname outperforms random-based prioritization. The performance improvement goes from 0\% to 10\% of more faults revealed when 5\% and 2\% of mutants are killed. These trends are  similar with those we observe on CodeFlaws, suggesting that \toolname effectively learns the properties of the important mutants. 

%We further confirm this by showing that the differences of APFD values are statistically significant, using the Wilcoxon test at the significant level $a<0.01$. The Vargha Delaney $\hat{A}_{12}$ value shows that \toolname is better than DummyRandom and SpreadRandom in 54\% and 61\% of the cases respectively. These results indicate that the signal of our features is strong enough and leads to significant benefits even when our training corpus involves few data.   

\begin{comment}
\begin{figure}[!t]
%\vspace{-1.0em}
	\centering
	\includegraphics[width=\linewidth]{graphs/NEWDATA-SUMMARY/merged-crossvalidation/corebench-merged/selected-mutants/{APFD-ALL.10}.pdf}%{graphs/leaveOneOut/APFD-ALL.pdf}
	%\includegraphics[width=\linewidth]{graphs/corebench_crossvalidation/3-ALL-MEDIAN.png}
	\caption{APFD measurements on CoREBench (up to top 10\% of the mutants). The \toolname prioritization outperform the random baselines.}
	\label{fig:rq4APFD}
\end{figure}
\end{comment}

%To provide a general view of the trends, Figure~\ref{fig:rq4-Median} illustrates the overall (median) effectiveness of the mutant prioritization by \toolname in comparison with random strategies. We note that for all percentage of mutants, \toolname outperforms random-based prioritization. The performance improvement goes from 15\% to 13\% of more faults revealed  when 5\% and 10\% of mutants are killed.
%when 30\% until 50\%of mutants are executed.

\
%%%%%%%%%

\noindent 
{\bf \em Required Tests Cost Metric.}

\begin{figure}[!t]
\vspace{-1.0em}
	\centering
	\includegraphics[width=\linewidth]{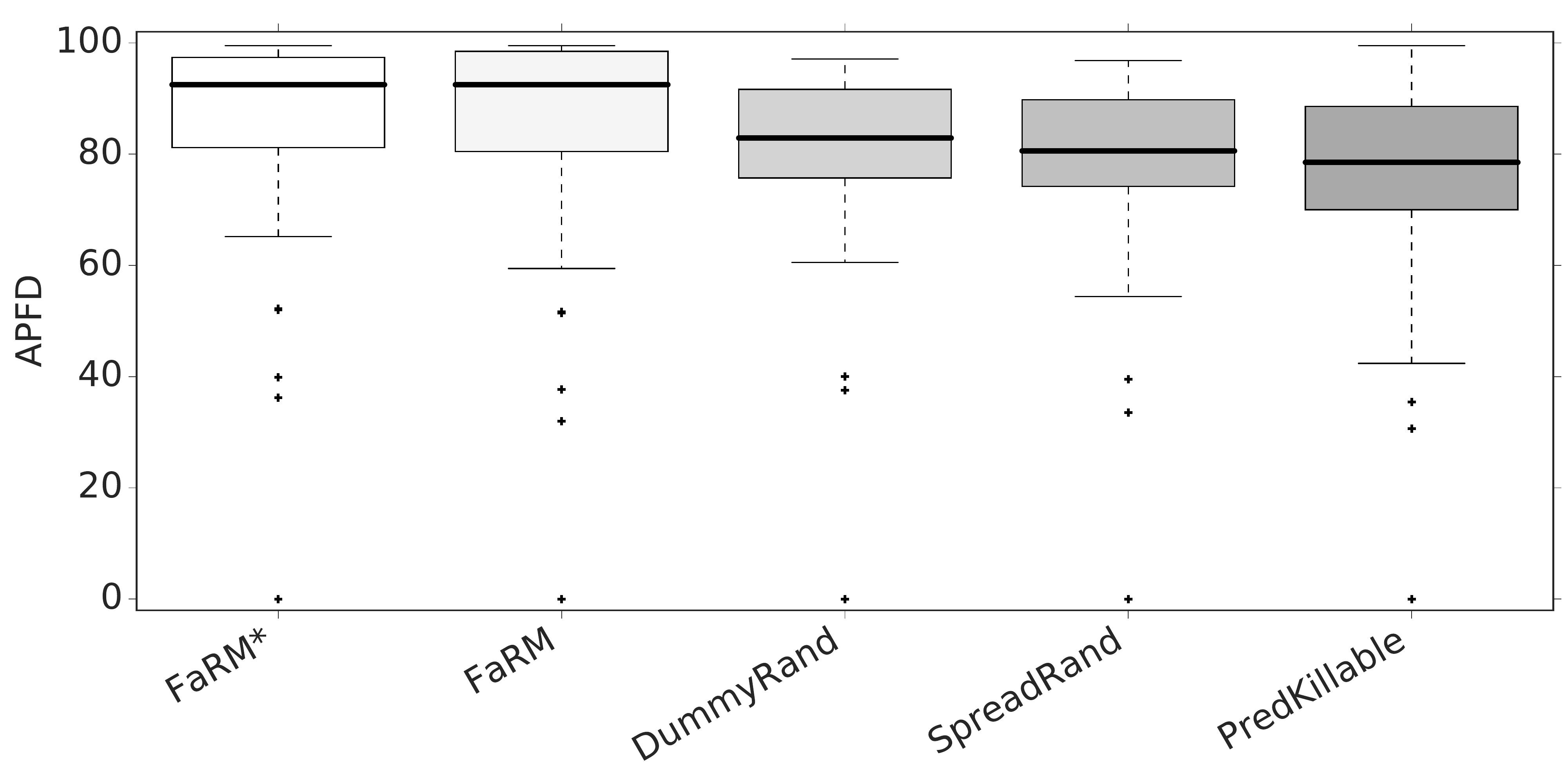}
	\caption{APFD measurements on CoREBench for the required tests cost metric. The \toolname prioritization outperform the random baselines.}
	\label{fig:bench-requiredtests-apfd}
\end{figure}

Figure~\ref{fig:bench-requiredtests-apfd} shows the distributions of APFD (Average Percentage of Faults Detected) values 
for all faults, using the five approaches under evaluation. While both \toolname and \toolnameB yield an APFD median of 92\%, and \toolnameC yields an APFD median of 79\%, DummyRandom and SpreadRandom reach median APFD values of 83\% and 81\% respectively. These results reveal that the general trend is in favour to our approach. As our approaches \toolname and \toolnameB are better than the random baseline, when the main cost factor (number of test that need to be designed and executed) is aligned, we can infer that it is generally better with practically important differences (of 9\%). The \toolnameC performs slightly worse than the random baseline.

%To account for the stochastic nature of the compared methods and increase the confidence on our results, we further perform a statistical test. Wilcoxon test results yielded p-values much lower than our significance level for the compared data, i.e., each of \toolname and \toolnameB compared with each of \toolnameC, DummyRandom and SpreadRandom. Therefore, we can definitively conclude that \toolname and \toolnameB outperform the random baselines with statistically significance. On the other hand, the Wilcoxon test revealed that there is no statistical difference between the performance of \toolnameC, DummyRandom and SpreadRandom.
The results of the Vargha Delaney $\hat{A}_{12}$ effect size show that \toolname is better than DummyRandom, SpreadRandom and \toolnameC in 74\%, 77\% and 86\% of the cases respectively. \toolnameB is better than DummyRandom, SpreadRandom and \toolnameC in 70\%, 74\% and 81\% of the cases respectively. \toolnameC is worse than the DummyRandom and SpreadRandom in 70\% and 66\% of the cases respectively.

\begin{figure}[!t]
\vspace{-1.0em}
	\centering
	\includegraphics[width=0.9\linewidth]{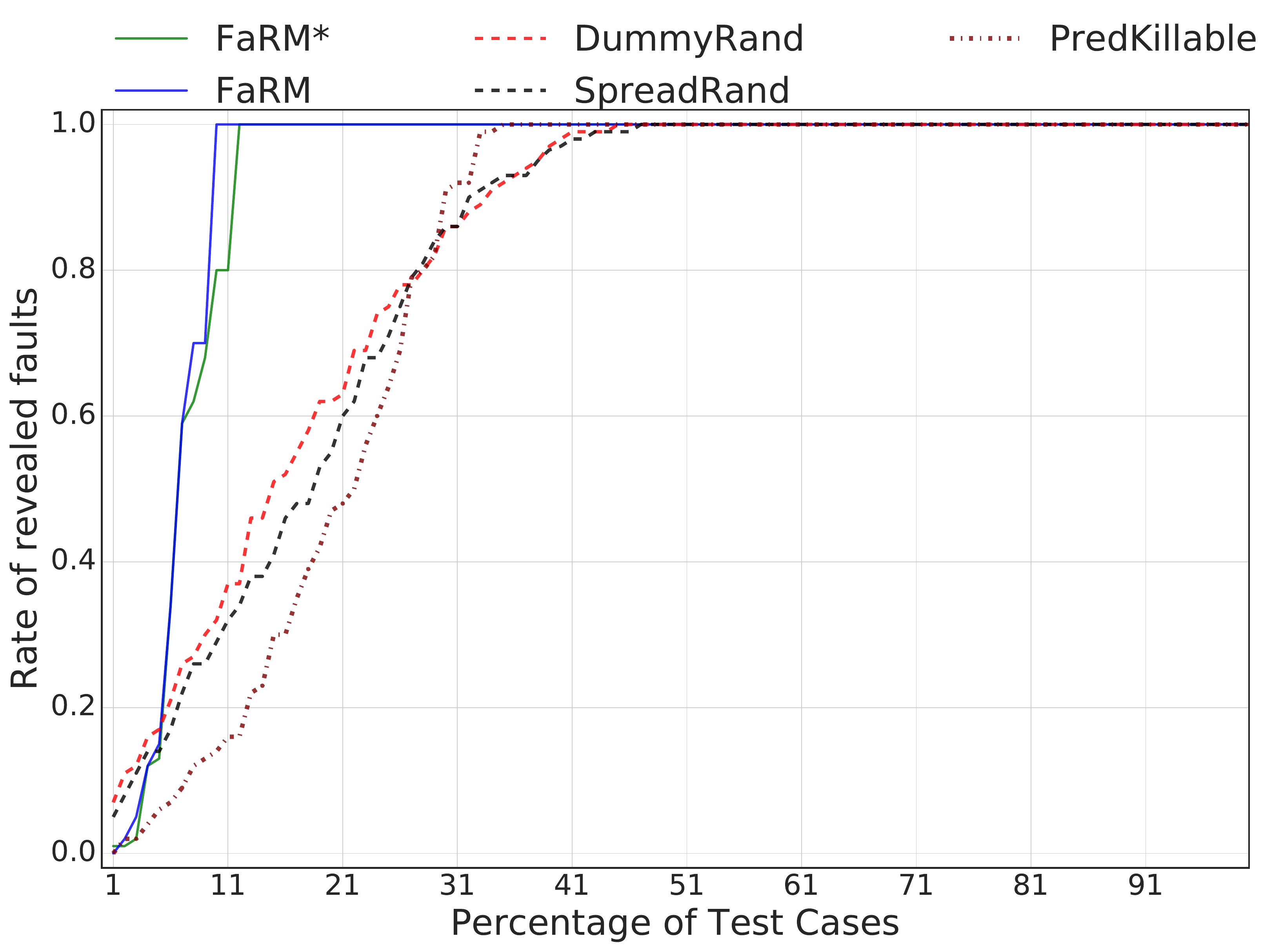}
	\caption{Required tests prioritization performance in terms of faults revealed (median case) on CoREBench. The x-axis represent the number of considered tests. The y-axis represent the ratio of the fault revealed by the strategies. }
	\label{fig:bench-requiredtests-median}
\end{figure}

To provide a general view of the trends, Figure~\ref{fig:bench-requiredtests-median} illustrates the overall (median) effectiveness of the required test prioritization by \toolname, \toolnameB and \toolnameC in comparison with random strategies. We note that \toolname and \toolnameB outperforms random-based prioritization while \toolnameC is outperformed by the random-based prioritization. Overall, we observe that the fault revelation benefit of \toolname over the random approaches is above 30\% (maximum difference is 70\%) for the 5\% to 20\% top ranked tests.

%%%%%%%%%

\noindent 
{\bf \em Analysed Mutants Cost Metric.}

The results of the Vargha Delaney $\hat{A}_{12}$ effect size values related to the analysed mutants cost metric show that \toolname and \toolnameB are better than DummyRandom and SpreadRandom. %$with statistical significance displayed by a p-value muck lover than the significance level. 
\toolname is better than DummyRandom and SpreadRandom in 58\% and 60\% of the cases respectively. The performance difference is higher for \toolnameB where it is better than DummyRandom and SpreadRandom in 61\% and 63\% of the cases respectively. \toolnameC is better than DummyRandom and SpreadRandom in 56\% and 58\% of the cases respectively.

\toolnameB shows a larger improvement than \toolname over the random baseline.

%%%%%%%%%%%%%%%

Taken together our results demonstrate that \toolname and \toolnameB achieves significant improvements over the random baselines on both CodeFlaws and CoREBench fault sets. 
Therefore, the improvements made by \toolname and \toolnameB can be considered as important.

%% file: threats.tex
\section{Discussion}
\label{Discussion}

\subsection{Working Assumptions}

Our approach uses machine learning to support mutation testing. As such it makes some assumptions that should hold in order to be applicable and effective. First, we assume that there are sufficient historical data from applications of the same context or previous software releases. This means that we need to have a diverse and comprehensive set of defects where mutation testing has been applied. Of course these defects need to belong to the targeted, by the testing procedure, class of defects. In the absence of sufficient defects, we can relax this requirement by training on hard-to-kill or subsuming mutants. This can be easily performed, the same way we train for equivalent mutants, as long as we have a large codebase that is sufficiently tested. 

Second, we assume that defect causes are repeated. This is an important assumption as in its absence machine learning cannot work. We believe that it holds given the evidence provided by the $n$-version programming studies \cite{0090600, KnightL86} and the empirical observations in the context of Linux kernel \cite{PalixTSCLM11}. 

Third, we assume that mutants are linked with targeted defects. This assumption comes with the use of mutation testing. We believe that it holds given the empirical evidence provided by recent studies \cite{ChekamPTH17, Goran, RamlerWK17, Papadakis2018hi}. Finally, we assume that fault revelation utility can be captured by static features such as the ones used in this study. We are confident that this assumption holds given the reports of Petrovic and Ivankovic \cite{Goran} on the utility of the AST features in mutant selection and the evidence we provide here. 

\subsection{Threats to Validity}
\label{threats}
We acknowledge the following threats that could have affected the validity of our results. One possible external validity threat lies in the nature of the test subjects we used. Individually, the majority of programs  in comparison experiments are small in size, and may not be representative of real-world programs. Our mitigation strategy is discussed in the following subsection (section \ref{Representativeness}). Moreover,  since the properties of the  fault revealing  mutants reside on the code parts that  are control and data dependent to and from the faults, the cumulative size of relevant code parts (based on which we get the feature values) should be small. Therefore, for such a study, the most important characteristics should be the faulty code area and its dependencies. Since we have a large and diverse set of real faults, we feel that this threat is limited. Future work should validate our findings and analysis to larger programs. 

Another potential threat relates to the mutation operators we used. Although we have considered a variety of operators, we cannot guarantee that they yield representative mutants. To diminish this threat we used a large number of operators (816 simple operators across 18 categorises) covering the most frequently used C features. We also included all the operators adopted by the modern mutation testing tools \cite{ChekamPTH17,PapadakisSurvey}. 

Threats to internal validity lie in the use of recent machine learning algorithms to the detriment of established and widely used techniques. Nevertheless, these threats are minimized as gradient boosting is gaining a momentum in the research literature as well as the practice of machine learning. 

Similarly, there might be some issues related to code redundancy, duplicated code, that may influence our results. We discuss our redundancy mitigation  strategy on section \ref{redundancy}.

 Another internal validity threat may be due to the features we use. These have not been optimized with any feature selection technique. This is not a big issue in our case as we use gradient boosting that automatically performs feature selection. To verify this  point we trained a Deep Learning model that also performs feature selection and checked its performance. The result showed insignificant differences from our method. Additionally, we retrained our classifiers using the features with information gain greater or equal to 0.02 and got results similar to random, suggesting that all our features are needed. Future research should seed light on this aspect by complementing and optimizing our feature set. 
 
Other internal validity threats are due to the way we treated mutants as equivalent. To deal with this issue, we used KLEE, a state of the art test generation tool and the accompanied test suites. As the programs we are using are small KLEE should not have a problem at generating effective test suites. Together these tools kill 87\% of all the mutants, demonstrating that our test suites are indeed strong. Since the 13\% of the mutants we treat as equivalent is in line with the results reported by the literature~\cite{PapadakisJHT15}, we believe that this threat is not important. Unfortunately, we cannot practically do much more than that, as the problem is undecidable \citep{Budd:1982:TNC:2697733.2697965}.

Finally, our assessment metrics may involve some threats to construct validity. Our cost measurement, number of selected, analysed mutants and number of test cases essentially captures the manual effort involved. Automated tools may reduce this cost and hence influence our measurements. Regarding equivalent mutants, we used a state-of-the-art equivalent mutant detection technique, TCE~\cite{PapadakisJHT15}, to remove all trivially equivalent mutants before conducting any experiment. Therefore, the remaining equivalent mutants are those that remain undetectable by the current standards. Regarding the test generation cost, we acknowledge that while automated tools manage to generate test inputs, they fail generating test oracles. Therefore, augmenting the test inputs with test oracles, remains a manual activity, which we approximate by measuring the number of tests. In our experiments we bypassed the oracle problem by using the `correct' program versions as oracles. An alternative scenario involves the use of automated oracles, but these are rare in practice and we did not considered them. Overall, we believe that with the current standards, our cost measurements approximate well the human cost involved.

All in all, we aimed at minimizing any potential threats by using various comparisons scenarios, clearly evaluating the benefit of the different steps in \toolname, and leveraging frequently used and established metrics. Additionally, to enable replication and future research we make our data publicly available\footnote{https://mutationtesting.uni.lu/farm}.

\subsection{Representativeness of test subjects}
\label{Representativeness}

\begin{figure}[!t]
	\centering
	\includegraphics[width=\linewidth]{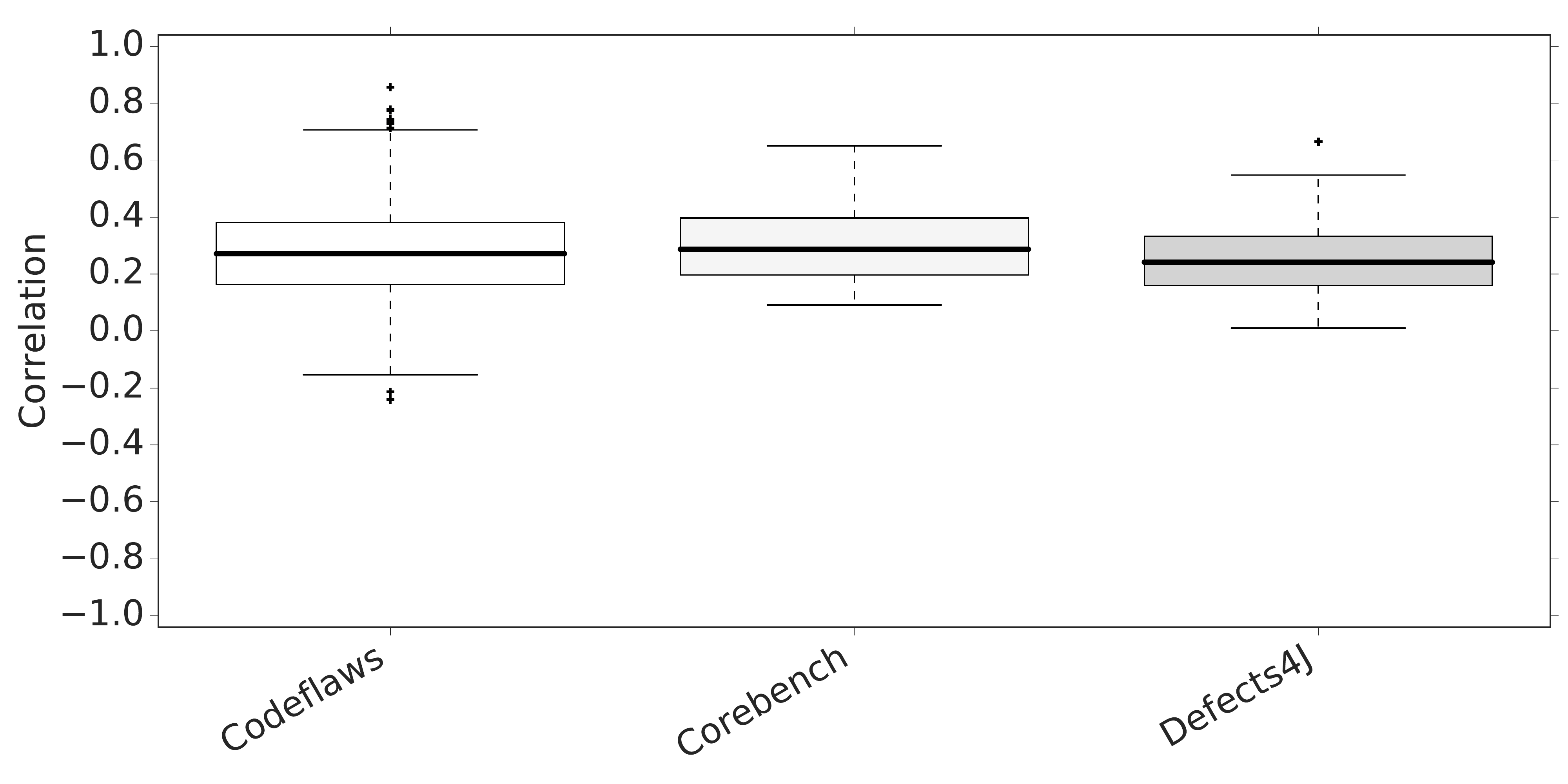}
	\caption{Correlations between mutants and faults in three defect datasets. Similar correlations are observed in all three cases suggesting that Codeflaws provides good indications on the fault revealing ability of the mutants. }
	\label{fig:Correlations}
	\vspace{-1em}
\end{figure}

Most of our results are based on Codeflaws. We used this benchmark because machine learning requires lots of data and Codeflaws is, currently, the largest benchmark of real faults on C programs. Also because of its manageable size, we can automatically generate a relatively large and thorough test pool and apply mutation testing. Still this required 8,009 CPU days of computations (only for the mutant executions), indicating that we reach the experimentally achievable limits. Similarly, applying mutation testing on the 45 faults on CoREBench required 454 CPU days of computations. 

The obvious differences between the size of the test subjects raise the question of whether our conclusions hold on other programs and faults. Fortunately, as already discussed our results on CoREBench have similar trends with those observed on Codeflaws. Training a classifier on CoREBench yields AUC values around 0.616, which is approximately the same (slightly lower) than the one we get from Codeflaws. 
This fact provides confidence that our features do capture the mutant properties we are seeking for. To further carter for this issue, we also selected the harder to reveal faults (faults revealed by less than 25\% of the tests). This is a quality control practice, used in fault injection studies, ensures that our faults are not trivial.

Additionally, we checked the syntactic distance of the Codeflaws faults and show that it is small (please refer to Figure \ref{fig:numChanges}), similarly to the one assumed by mutation testing. This property together with the subtle faults (faults revealed by less than 25\% of the tests) we select make our fault set compliant with the mutation testing assumptions, i.e., the Competent Programmer Hypothesis. 

Furthermore, we computed and contrasted the correlation between mutants and faults on three defect benchmarks; CoREBench, Codeflaws and Defect4J dataset\footnote{\url{https://github.com/rjust/defects4j}}. Our aim is to check whether there are major difference in the relation between the faults and mutants of the three benchmarks. 

Defect4J is a popular defect dataset for Java, with real faults from large open source programs. To compute the correlations on Defect4J we used the data from the study of Papadakis \etal \cite{Papadakis2018hi}, while for CoREBench and Codeflaws we used the data from this paper. We computed the Kendall correlations with uncontrolled test suite size between 1\% and 15\% of all tests. We make 10,000 random test sets each with size randomly chosen between 1\% and 15\% of all the tests. Then, we compute the mutation score and the fault revelation of each test set, and compute the Kendall correlation between the mutation score and fault revelation. Figure \ref{fig:Correlations} shows the correlations for Codeflaws, CoReBEnch and Defect4J. As can be seen, the correlations are similar in all three cases. Therefore, since the mutants and faults relations share similar properties on all cases, we believe that our defect set provide good indications on the fault revealing ability of our approach. %We further validate our 

\subsection{Redundancy between the considered faults}
\label{redundancy}

\begin{figure}[!t]
	\centering
	\includegraphics[width=\linewidth]{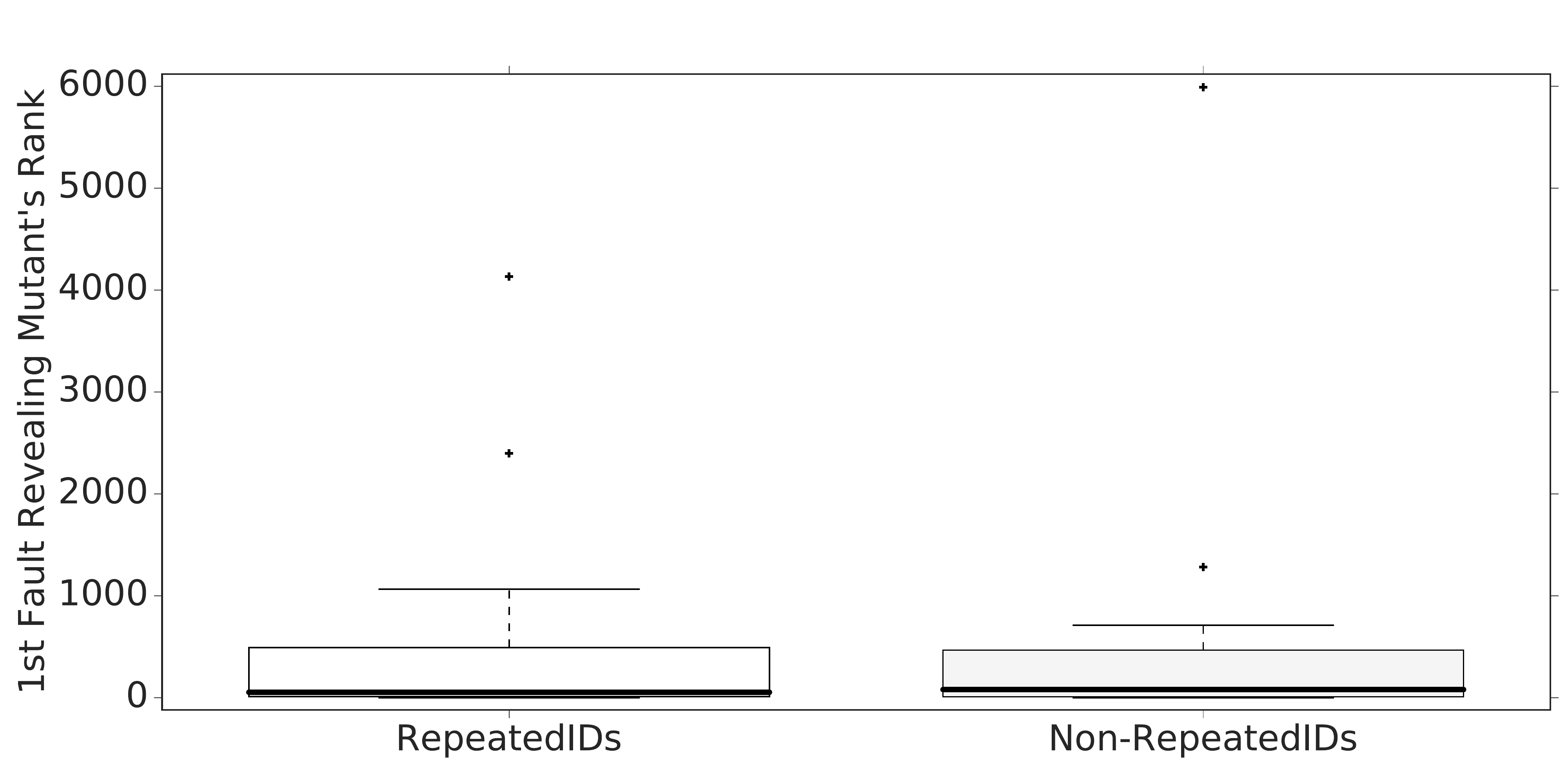}
	\caption{CoREBench results on similar (repeatedIDs) and dissimilar (Non-RepeatedIDs) implementation. We observe similar trend in both cases suggesting a minor or no influence of code similarity on \toolname performance. %XXX - needs explanation
	}
	\label{fig:Similarities}
	\vspace{-1em}
\end{figure}

Code redundancy may influence our results. As depicted in Figure \ref{fig:problemsbyimplemetations}, in  Codeflaws  the number of implementations for the same problem is usually higher than one. This introduces a risk that our evaluation, test defect set, may benefit from the knowledge gained during training, in case there is another implementation for the same problem in this set. Although, such a case is unlikely as all of our defects are different and form unique program versions, to remove any threat from such a factor we repeated our experiment by randomly splitting the Codeflaws subjects into training and test sets in such a way that all implementations of the same problem either appear in the training set or in the evaluation, but not both. We obtained almost identical results with the previous experiment, i.e., we get AUC values of 62\% when controlling for the implementations (having always different problem implementations on the training and test sets). 
%[32,35,27,  38,41, 47,48,  3,6,11,12,17,21,  4,5,  7,18,19, 9,20]
Another threat related to code redundancy may have affected our results in CoREBench. Among the 45 CoREBench faults we consider, only 20 of them are on the same components (13 in Coreutils, 3 in Find, and 4 in Grep). Note that the 13 instances of Coreutils form 3 separated set of 2, 5 and 6 bugs on the same component. We manually checked these defects and found that they all differ (they are located in different code parts and the code around the locations modified to fix the defects differs). Nevertheless, still there is a possibility that code similarities may impact positively or negatively our classifiers. Although such a case is compatible with our working scenario (we assume that we have similar historical data), so it is not a problem for our approach, it is still interesting to check the classifier performance on similar/dissimilar implementations. 

To deal with this case, we divided our fault set (CoREBench) into two sets, one with the faults having similar faulty functions and one with dissimilar ones. To do so, we used the Deckard~\cite{deckard_clonedetection2007} tool, which computes the similarity between code instances (at the AST level). For each faulty function, the tool compares the vector representation of the sub-trees of small code snippets and reports similarity scores. Two codes are considered as similar if they have code parts with high similarity scores on the utilized abstraction, i.e., above 95\% ~\cite{deckard_clonedetection2007}. 

Having divided the fault sets as similar and dissimilar we then contrast the results they provide. Overall, we found insignificant differences between the two sets. Figure \ref{fig:Similarities} compares the ranking position of the fault revealing mutants in the order provided by \toolname, when using the following tool parameters: similarity threshold 95, 4 stides and 50 minimum number of tokens. From these results we see that there are no significant differences between the two sets, suggesting that code redundancy does not affect our results.

\subsection{Other Attempts}

Our study demonstrates how simple machine learning approaches can help improving mutation testing. Since our goal was to demonstrate the benefits of using such an approach we did not attempted to manipulate our data in any way (apart from the exclusion of the trivial faults). We achieve this goal, but still there is room for improvement that future research can exploit. For instance it is likely that classification results can be improved by preprocessing training data, e.g., exclude fault types that are problematic \cite{Papadakis2018Mutation}, excluding fault types with few instances, excluding versions with low strength test suites, as well as by removing many other sources of noise.  

Data manipulation strategies we attempted during our study were oversampling, the exclusive use of features with high information gain, the use of a Deep Learning classifier and targeting irrelevant mutants (the mutants with lowest fault revealing probability). Oversampling consist of randomly duplicating the data items of the minority class in order to have a more balanced data to train the classifier. In this case, we applied oversampling of the minority class for mutants which is the fault revealing class (they represent approximately 3\% of the whole data). In another attempt, we trained our classifier by only using the features that have highest information gain (those with $IG\geq0.02$ in Figure~\ref{fig:rq1-IG}). We also attempted to replace the supervised learning algorithm used in out approach by substituting the decision tree with a deep neural network classifier. We also retrained the classifier to target irrelevant mutant (mutant not killed by fault revealing tests), the motivation being that the classifier may perform better to separate irrelevant mutants than fault revealing ones. All these attempts yielded quite similar or worse results with those we report and thus, we do not detail them. %Though, we consider them as a source of interesting information for the reader. 

%% file: RelatedWork.tex
\section{Related Work}
\label{RelatedWork}

Years of research in mutation testing has shown that designing tests that are capable of revealing mutant-faults results in strong test suites that in turn reveal real faults \cite{FranklWH97, LiPO09, ChekamPTH17, Papadakis2018hi}. The technique is particularly effective and capable of revealing more faults than most of the other structural test criteria \cite{FranklWH97, LiPO09, ChekamPTH17}. Experiments using real faults, have shown that mutation testing reveals more faults than the all-uses test criterion \etal  \cite{FranklWH97}, and also that it reveals significantly more faults than the statement, branch and weak mutation \cite{ChekamPTH17} test criteria.  

Although effective, mutation requires too many mutants making the cost of generating, analysis and executing them particularly high. Recent studies have shown that only a small number of mutants is sufficient to represent them \cite{KintisPM10, AmmannDO14, PapadakisHHJT16} and that the majority of the mutants are somehow ``irrelevant'' to the underlying faults (faults that testers seek for) \cite{Papadakis2018hi}. Along the same lines, Papadakis \etal \cite{Papadakis2018Mutation} analysed different types of mutants, i.e., hard to kill, subsuming, hard to propagate and fault revealing, and demonstrated that the class of fault revealing mutants is unique and differs from the other mutant sets. These studies motivated our research by indicating that it is possible to target a specific (small) set of mutants that maximize testing effectiveness. 

Since the early days of mutation testing, researchers realised that the number of mutants is one of the most important problems of the method. Therefore, several approaches have been proposed to address this problem. Mutant random sampling was one of the first attempts~\cite{Budd80,Acree80}. Random sampling was evaluated by Wong~\cite{Wong93} who found that a sampling ratio of 10\% results in a test effectiveness loss of approximately 16\% (evaluated on Fortran programs using the Mothra mutation testing system~\cite{DeMilloGMOK88}). More recently, Papadakis and Malevris~\cite{PapadakisM10a}, using the Proteum mutation testing tool~\cite{Delamaro2001_proteum}, reported a fault loss on C operators of approximately 26\%, 16\%, 13\%, 10\%, 7\% and 6\% for sampling ratios of 10\%, 20\% ..., 60\% respectively. 

An alternative approach to reduce the number of mutants is to select them based on their types, i.e., according to the mutation operators. Mathur~\cite{Mathur91} introduced the idea of constrained mutation (also called selective mutation), using only two mutation operators. Wong \etal~\cite{WongM95b} experimented with sets of operators and found that two operators alone have a test effectiveness loss of approximately 5\%. 
 Offutt \etal~\cite{Offutt1993ICSE, OffuttLRUZ96} extended this idea and proposed a set of 5 operators, which had almost no loss on its test effectiveness. 
 This 5 mutation operator set is considered as the current standard of mutation as it has been adopted by most of the modern mutation testing tools and used in most of the recent studies \cite{PapadakisSurvey}. 
 
Many additional selective mutation approaches have been proposed. Mresa and Bottaci~\cite{MresaB99}  defined a selective mutation procedure focused on reducing the number of equivalent mutants, instead of the number of mutants alone, as done by the studies of Mathur ~\cite{Mathur91} and Offutt \etal ~\cite{OffuttLRUZ96, Offutt1993ICSE}. They report significant reductions on the numbers of equivalent mutants produced by the selected operators, with marginal effectiveness loss (evaluated on Fortran with Mothra). Later, Barbosa \etal~\cite{BarbosaMV01} defined a selective mutation procedure aimed at reducing the computational cost of mutation testing of C programs. They found that a set of 10 operators could give almost the same results with the whole set of C operators supported by Proteum (78 operators). Namin \etal ~\cite{NaminAM08} used regression analysis techniques and found that a set of 13 mutation operators of Proteum could provide substantial cost execution savings without any significant effectiveness loss (mutant reductions of approximately 93\% are reported. 

More recently, researchers have experimented with mutations involving only mutants deletion~\cite{Untch09}. Deng \etal ~\cite{DengOL13} experimented with Java programs and the MuJava mutation operators~\cite{MaOK06} and reported reductions of 80\% on the number of mutants with marginal effectiveness losses. Delamaro \etal~\cite{DelamaroOA14} defined deletion operators for C and reported that they significantly reduce the number of equivalent mutants, with again marginal effectiveness losses. 

Other attempts have explored the identification of the program locations to be mutated. The key argument 
in these research directions is that program location is among the most important factor that determines the utility of the mutants. Sun \etal ~\cite{SunXLZ17} suggested selecting mutants that are diverse in terms of static control flow graph paths that cover them. Gong \etal ~\cite{GongZYM17} used code dominator analysis in order to select mutants that, when they are covered, maximize the coverage of other mutants. This work applies weak mutation and attempts to identify dominance relations between the mutants in a static way. 

Petrovic and Ivankovic \cite{Goran} identified the arid nodes (special AST nodes) as a source of information related to utility of the mutants. Their work uses dynamic analysis (test execution) combined with static analysis (based on AST) in order to identify mutants that are helpful during code reviews. We include such features in our study with the hope that they can also capture the properties of fault revealing mutants.% However, our information gain results show that these feature play an important role, but  they are not the most discriminative ones.
 Nevertheless,  still as part of future work it is interesting to see how our features can fit within the objectives of code reviews \cite{Goran}. 

Mirshokraie \etal \cite{Mirshokraie0P15} used static (complexity) and dynamic (number of executions) analysis features to select mutants, for JavaScript programs, that reside on code parts that have low failed error propagation (they are likely to propagate to the program output). Their results show that  more than 93\% of the selected mutants are killable, and that more than 75\% of the non-trivial mutants resided in the top 30\% ranked code parts.

After several years of development of various selective mutation approaches, recent research has established that literature approaches perform similarly to random mutant sampling. Zhang \etal ~\cite{ZhangHHXM10} compared random mutant selection and selective mutation (using C programs and the Proteum mutation operators) and found that there are no significant differences between the two approaches. 
The most recent approach is that of  Kurtz \etal ~\cite{KurtzAODKG16} (using C programs and the Proteum mutation operators), which also reached the same conclusion (reporting that mutant reduction approaches, both selective mutation and random sampling, perform similarly).

From the above discussion it should be clear that despite the plethora of the selective mutation testing approaches, random sampling remains one of the most effective ones. This motivated our work, which used machine learning techniques and source code features in order to effectively tackle the problem. Moreover, as most of the methods use only one features, the mutant type, which according to our information gain results does not have relatively good prediction power, they should perform poorly. More importantly, our approach differs from the previous work in the evaluation metrics used. All previous work measured test effectiveness in terms of artificial faults (i.e., mutant kills or seeded faults found), while we used real faults. We believe that this is an important difference as our target (dependent variable) is the actual measurement of interest, i.e., the real fault revelation, and not a proxy, i.e., the number of mutants killed.

The closest work to ours is the ``predictive mutation'' one \cite{ZhangWZHZCZ16, 8304576}. Predictive mutation testing attempts to predict the mutants killed for a given test suite without any mutant execution. It employs a classification model using both static and dynamic features (both on test suite and the mutants) and achieves remarkable results with an overall 10\% error on the predicted mutation scores. Predictive mutation has a similar goal with our killable mutant prediction method. Though, predictive mutation assumes the existence of test suites, while our killable mutant prediction method does not. Nevertheless, our method targets a different problem, the prediction and prioritization of the important mutants prior to any test execution. To do so, we use only static features (on the code under test), while predictive mutation heavily relies on test code and dynamic features, and evaluate our approach using real faults (instead of mutants).

Another similar line of work is Evolutionary Mutation Testing (EMT) \cite{DELGADOPEREZ2018130}. EMT is a technique that attempts to select useful mutants based dynamic features (test execution traces) and uses them to support test augmentation. EMT learns the most interesting mutation operators and locations in the code under analysis using a search algorithm and mutant execution results. Overall, EMT achieve a 45\% reduction on the number of mutants. Although, EMT aims at the typical mutant reduction problem (while we aim at the fault revealing one), it can complement our method. Since EMT performs mutant selections after the mutant-test executions, \toolname can provide a much better starting point. Another way to combine the two techniques is to use the search engine of EMT, together with our features, to refine the mutant rankings.

A different way to reduce the mutants' number is to rank the live mutants according to their importance, so that testers can apply customised analysis according to their available budget. Along these lines, Schuler \etal ~\cite{SchulerZ13} used the mutants' impact to rank mutants according to their likelihood of being killable. Namin \etal ~\cite{NaminXRS15} introduced the MuRanker approach. MuRanker uses three features:  the differences that mutants introduce (a) on  the control-flow-graph representation (Hamming distance between the graphs), (b) on the Jimple representation (Hamming distance between the Jimple codes) and (c) on the code coverage differences produced by a given set of test cases (Hamming distance between the traces of the programs). Although our mutant prioritization scheme is similar to these approaches, we target a different problem, the static detection of valuable mutants. Thus, we do not assume the existence of test suites and mutants executions. The benefit of not making any such assumptions is that we can reduce the number of mutants to be analysed by testers, to be generated and executed by mutation testing tools.